\documentclass[reprint,twocolumn,showpacs,showkeys,preprintnumbers,amsmath,amssymb,floatfix,aps,pra,longbibliography]{revtex4-1}

\usepackage{graphicx}
\usepackage{bm}

\usepackage{physics}
\usepackage{placeins}
\usepackage{graphics}
\usepackage{color}
\usepackage{hyperref}
\usepackage{multirow}
\usepackage{blindtext}

\usepackage{pdfpages}

\makeatletter
\AtBeginDocument{\let\LS@rot\@undefined}
\makeatother  

\begin{document}


\title{Proximity effects in graphene on monolayers \\ of transition-metal phosphorus trichalcogenides MPX$_3$}

\author{Klaus Zollner}
\email{klaus.zollner@physik.uni-regensburg.de}
\affiliation{Institute for Theoretical Physics, University of Regensburg, 93040 Regensburg, Germany}
\author{Jaroslav Fabian}
\affiliation{Institute for Theoretical Physics, University of Regensburg, 93040 Regensburg, Germany}

\begin{abstract}
We investigate the electronic band structure of graphene on a series of two-dimensional magnetic transition-metal phosphorus trichalcogenide monolayers, MPX$_3$ with M=\{Mn,Fe,Ni,Co\} and X=\{S,Se\}, with first-principles calculations. A symmetry-based model Hamiltonian is employed to extract orbital parameters and sublattice resolved proximity-induced exchange couplings ($\lambda_{\textrm{ex}}^\textrm{A}$ and $\lambda_{\textrm{ex}}^\textrm{B}$) from the low-energy Dirac bands of the proximitized graphene.
Depending on the magnetic phase of the MPX$_3$ layer (ferromagnetic and three antiferromagnetic ones), completely different Dirac dispersions can be realized with exchange splittings ranging from 0 to 10~meV. Remarkably, not only the magnitude of the exchange couplings depends on the magnetic phase, but also the global sign and the type. Important, one can realize uniform ($\lambda_{\textrm{ex}}^\textrm{A} \approx \lambda_{\textrm{ex}}^\textrm{B}$) and staggered ($\lambda_{\textrm{ex}}^\textrm{A} \approx -\lambda_{\textrm{ex}}^\textrm{B}$) exchange couplings in graphene. 
From selected cases, we find that the interlayer distance, as well as a transverse electric field are efficient tuning knobs for the exchange splittings of the Dirac bands. More specifically, decreasing the interlayer distance by only about 10\%, a giant 5-fold enhancement of proximity exchange is found, while applying few V/nm of electric field, provides tunability of proximity exchange by tens of percent. 
We have also studied the dependence on the Hubbard $U$ parameter and find it to be weak. Moreover, we find that the effect of SOC on the proximitized Dirac dispersion is negligible compared to the exchange coupling. 
\end{abstract}

\maketitle

\section{Introduction}

Monolayer magnets are key ingredients for novel spintronics applications, such as in tunneling magnetoresistance and spin-orbit torque device architectures \cite{Zollner2019:PRR,Dolui2020:NL,Song2018:SC,Alghamdi2019:NL,Ostwal2020:AM,Zhao2021:arxiv}. One step towards next generation devices is to achieve highly spin polarized currents and long-distance transfer of spin information in all two dimensional (2D) van der Waals (vdW) heterostructures without the need of conventional ferromagnets (FM), where, for example, conductivity mismatch occurs at the interface \cite{Han2010:PRL,Schmidt2000:PRB,Rashba2000:PRB,Han2014:NN,Fert2001:PRB}. 
Remarkably, recent experiments have shown that in graphene/CrSBr heterostructures, where CrSBr is a layered antiferromagnet, high spin polarization of the graphene conductivity arises solely due to proximity-induced exchange coupling \cite{Ghiasi2021:NN,Kaverzin2022:arxiv}. Therefore, it is essential to study proximity effects in graphene, since it already intrinsically provides long-distance spin communication \cite{Khokhriakov2021:arxiv,Khokhriakov2020:C}, but lacks an efficient spin manipulation knob.  

Proximity effects in graphene can significantly alter the spin transport. For example, heterostructures with transition-metal dichalcogenides (TMDCs) provide a giant enhancement of the spin-orbit coupling (SOC) in graphene by proximity \cite{Gmitra2015:PRB,Gmitra2016:PRB}. Moreover, the proximity SOC can be efficiently tuned by gating and twisting in graphene/TMDC heterostructures \cite{Li2019:PRB,David2019:arxiv,Naimer2021:arxiv,Pezo2021:2DM,Peterfalvi2021:arxiv,Veneri2022:arxiv,Lee2022:arxiv}, allowing to tailor, for example, the interconversion between spin and charge currents \cite{Benitez2020:NM,Offidani2017:PRL,Ghiasi2019:NL,Khokhriakov2020:NC,Herling2020:APL,Lee2022:arxiv,Veneri2022:arxiv,Safeer2022:2DM,Offidani2017:PRL,Monaco2021:PRR,Ferreira2021:JP,Milletari2017:PRL,Dyrdal2014:PRB,Garcia2017:NL,Hoque2021:CP,Galceran2021:APL} or the spin-relaxation anisotropy \cite{Pezo2021:2DM,Cummings2017:PRL,Ghiasi2017:NL,Zihlmann2018:PRB,Leutenantsmeyer2018:PRL,Omar2019:arxiv,Offidani2018:PRB}. 
Placing graphene on a ferro- or antiferromagnet results in proximity-induced exchange coupling.
Particularly interesting are magnetic semiconductors such as Cr$_2$Ge$_2$Te$_6$ \cite{Zollner2019:PRR,Zhang2015:PRB,Karpiak2019:arxiv} or CrI$_3$ \cite{Zhang2018:PRB,Farooq2019:NPJ,Cardoso2018:PRL,Seyler2018:NL}, providing spin splitting for Dirac electrons but without contributing to transport. Moreover, the proximity-induced exchange coupling can also be tailored by gating and twisting \cite{Zollner2022:arxiv,Yan2021:PE}, important in, for example, the realization and engineering of topological states in graphene \cite{Zhang2018:PRB,Hogl2020:PRL,Vila2021:arxiv}.

Unlike ferromagnets, antiferromagnetic monolayers 
have not yet been systematically investigated for proximity effects in graphene. Transition-metal phosphorus trichalcogenides (MPX$_3$) are particularly interesting: they are semiconducting, stable in air, and can offer different magnetic configurations \cite{Wiedenmann1981:SSC,Ramos2021:2DM,Lancon2018:PRB,Lu2020:ASS,Sivadas2015:PRB,LeFlem1982:JPCS,Lee2016:NL,Chittari2020:PRB,Chittari2016:PRB,Joy1992:PRB,Li2019:PRB2,Klingen1968:NW,Pei2018:FP,Li2013:PNAS,Yang2020:RSC,Nauman2021:2DM,Autieri2022:JPC,Budniak2022:AOM,Basnet2022:arxiv,Samal2021:JMCA,Calder2021:PRB,Mai2021:SA}.
In general, MPX$_3$ monolayers are formed by transition-metal atoms, M=\{Mn,Fe,Ni,Co\}, arranged in a honeycomb lattice and surrounded by octahedrally coordinated chalcogen atoms, X=\{S,Se\}. Each honeycomb hexagon additionally has two vertically-stacked P atoms at the center.
Remarkably, the magnetic properties of the MPX$_3$ crystals can be tuned by gating and straining \cite{Chittari2020:PRB}. 
In addition, these materials  exhibit giant exciton binding energies of several hundred meV \cite{Birowska2021:PRB, Dirnberger2022:arxiv}, even larger than in TMDCs, making them canditates for optospintronics applications. 
In Ni-based transition-metal phosphorus trichalcogenides, there is even evidence of topological superconductivity \cite{Li2019:PRB2}.
Heterostructures of MnPSe$_3$ and MoS$_2$ show a type-II band alignment, important for optics, and a stacking dependent lifting of spin and valley degeneracy in MnPSe$_3$ \cite{Qi2017:SR}, potentially interesting for valleytronics applications \cite{Schaibley2016:NRM,Vitale2018:S}.
Moreover, it has been shown that antiferromagnetic FePS$_3$ increases the coercive field and Curie temperature of the 2D itinerant ferromagnet Fe$_3$GeTe$_2$ \cite{Zhang2020:AM}, while NiPS$_3$ can be important to tailor interfacial spin-orbit torques \cite{Schippers2020:PRM}.

Graphene on MPX$_3$ monolayers is expected to exhibit strong proximity exchange, depending on the magnetic configuration, stacking, twisting, and gating.
For example, MnPSe$_3$ is an Ising-type antiferromagnetic semiconductor (out-of-plane and alternating magnetic moments on Mn atoms),
with the potential to induce staggered exchange coupling in graphene, as recently demonstrated \cite{Hogl2020:PRL}.
In this study, we consider monolayer graphene in proximity to various MPX$_3$ monolayers, which can be either in the ferromagnetic, antiferromagnetic N{\'e}el, antiferromagnetic zigzag, or antiferromagnetic stripy phase. 
In order to unveil the proximity-induced exchange coupling in graphene, we calculate the low-energy Dirac dispersions of the heterostructures by density functional theory (DFT), and fit them to a symmetry-based model Hamiltonian. The effective model can be used for investigating spin transport, spin dynamics, and topologies of proximitized
Dirac electrons in graphene/MPX$_3$ structures. Such effective models are transferable due to the short-range character of the proximity effect: if a graphene/MPX$_3$ interface is part of a vdW heterostructure, the model Hamiltonian we introduce can be applied directly to it.

Depending on the transition metal M=\{Mn,Fe,Ni,Co\}, the chalcogen atom X=\{S,Se\}, and the magnetic phase, the proximity-induced exchange coupling in graphene can be markedly different. In particular, we find proximity exchange couplings that range from about 0 to 10~meV, varying in sign and type. Most important, one can realize uniform (ferromagnetic) and staggered (antiferromagnetic) proximity exchange in graphene. 
From the selected case of graphene/MnPS$_3$ in the antiferromagnetic N{\'e}el phase (ground state), we find that the interlayer distance, as well as a transverse electric field are efficient tuning knobs for the exchange splittings of the Dirac bands. More specifically, decreasing the interlayer distance by only about 10\%, a giant 5-fold enhancement of proximity exchange is found, while applying just a few V/nm of electric field can tune proximity exchange by tens of percent. 
We have also studied the dependence on the Hubbard $U$ parameter and find it to be weak. Moreover, we find that the effect of SOC on the proximitized Dirac dispersion is negligible compared to the exchange coupling. 
Our findings should be useful for spin transport, spin relaxation, optospintronics applications, as well as for tailoring topological phases of the Dirac electrons in graphene \cite{Hogl2020:PRL,Birowska2021:PRB,Qi2017:SR}.

The paper is organized as follows. In Sec. \ref{Sec:Comp_Details}, we first discuss our computational methodology and the geometry setup. In Sec. \ref{Sec:Hamiltonian}, we then introduce the symmetry-based model Hamiltonian capturing the proximity-induced exchange coupling in graphene. In Sec. \ref{Sec:Results}, we show and discuss the electronic structure and fit results of the different magnetic phases on the example of the graphene/MnPS$_3$ heterostructure. In addition, we report on the effect of the interlayer distance, a transverse electric field, the Hubbard $U$, and SOC on the proximitized Dirac dispersion. Finally, in Sec. \ref{Sec:Summary} we summarize and conclude the paper.

\section{Computational Details and Geometry}
\label{Sec:Comp_Details}

    \begin{figure*}[htb]
     \includegraphics[width=.9\textwidth]{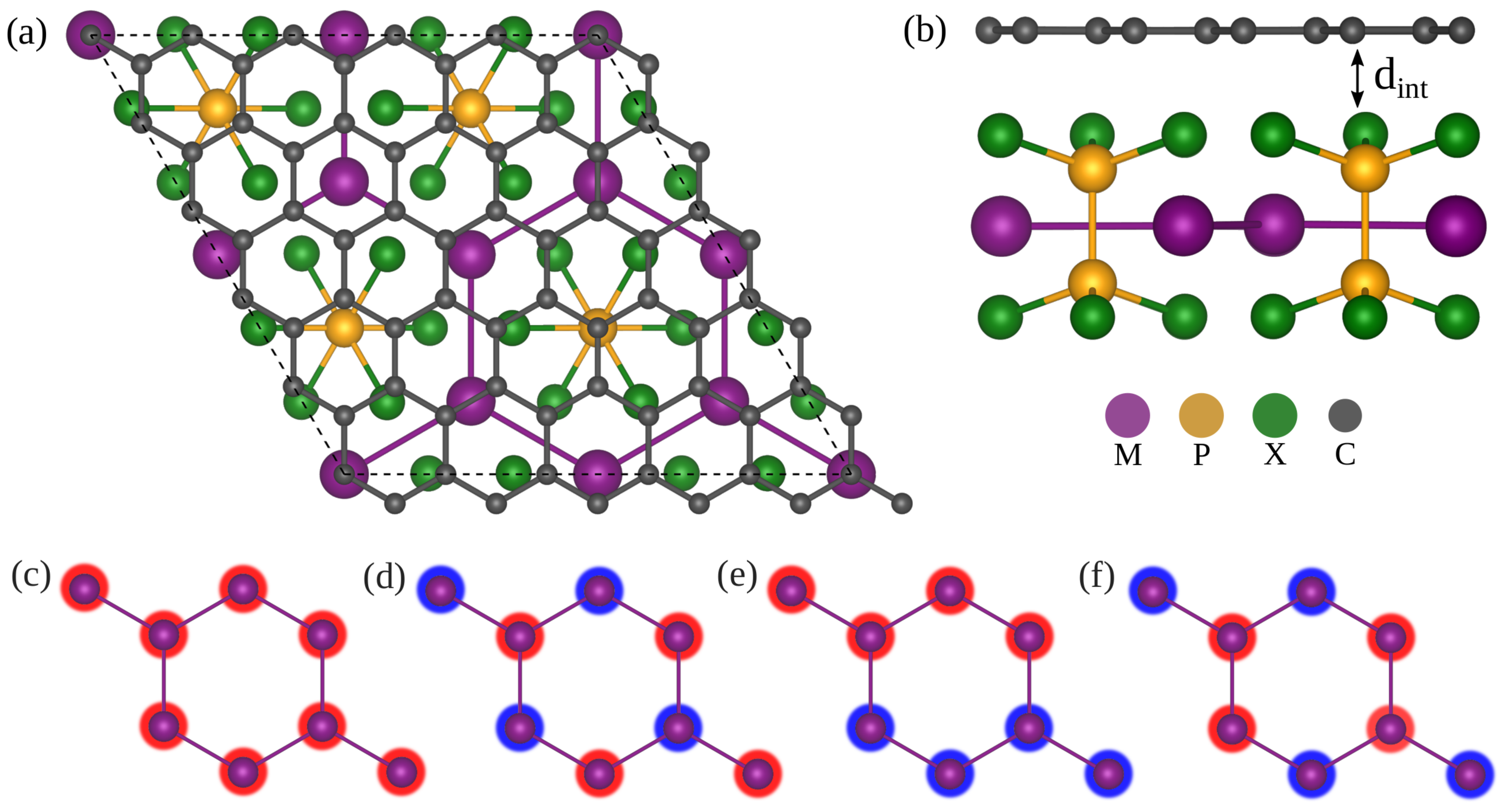}
     \caption{  (a,b) Top and side view of the graphene/MPX$_3$ heterostructure.
      Dashed line in (a) defines the heterostructure unit cell and in (b) we define the interlayer distance $\mathrm{d}_{\mathrm{int}}$. Different colors correspond to different atom types.
      Depending on the MPX$_3$ crystal, the hexagonal M-lattice can show the (c) ferromagnetic, (d) antiferromagnetic N{\'e}el, (e) antiferromagnetic zigzag, and (f) antiferromagnetic stripy phase. Red (blue) spheres indicate 
      magnetization of the M atom along $z$ ($-z$) direction. 
     }\label{Fig:crystal_structure}
    \end{figure*}

	The electronic structure calculations and structural relaxation of graphene on monolayers of the layered magnets MPX$_3$ 
	are performed by density functional theory (DFT)~\cite{Hohenberg1964:PRB} 
	with {\tt Quantum ESPRESSO}~\cite{Giannozzi2009:JPCM}.
	Self-consistent calculations are carried out with the $k$-point sampling of 
	$24\times 24\times 1$ to get converged results
	for the proximity exchange splittings.
	We perform open shell calculations that provide the 
	spin-polarized (magnetic) states of the MPX$_3$ monolayers. 
	A Hubbard parameter of $U = 4.5$~eV is used for M=\{Mn,Fe,Co,Ni\} $d$-orbitals, 
	as in recent calculations~\cite{Qi2017:SR, Pei2018:Nano}.
	We use an energy cutoff for charge density of $700$~Ry and
	the kinetic energy cutoff for wavefunctions is $70$~Ry for the scalar relativistic pseudopotential 
	with the projector augmented wave method~\cite{Kresse1999:PRB} with the 
	Perdew-Burke-Ernzerhof exchange correlation functional~\cite{Perdew1996:PRL}. 
	When SOC is included, we use the relativistic version of the pseudopotentials.  
	For the relaxation of the heterostructures, we add
	van der Waals corrections~\cite{Grimme2006:JCC,Grimme2010:JCP,Barone2009:JCC} and use 
	quasi-Newton algorithm based on trust radius procedure. 
	In order to simulate quasi-2D systems, we add a vacuum of $20$~\AA~to avoid interactions between periodic images in our slab geometry. To determine the interlayer distances, 
	the atoms of graphene are allowed to relax only along $z$-direction 
	(vertical to the layers) and the atoms of MPX$_3$ are allowed to move in all directions,
	until every component of each force is reduced below $10^{-4}$~[Ry/$a_0$], where $a_0$ is the Bohr radius.

	For some of our heterostructures, we cross-check the results with the {\tt WIEN2k} DFT-code~\cite{Wien2k}, using the relaxed structures obtained with {\tt Quantum ESPRESSO}. We use a $k$-point sampling of
  $18\times 18\times 1$ with the cutoff parameter $RK_{\textrm{max}}= 4.6$.
  The muffin-tin radii are $R_{\textrm{M}} = 2.5$,
  $R_{\textrm{P}} = 1.87~(1.86)$, $R_{\textrm{X}} = 1.96~(2.27)$,
  and $R_{\textrm{C}} = 1.32~(1.34)$ for X = S (Se). Van der Waals corrections and a Hubbard $U = 4.5$~eV are also included \cite{Grimme2010:JCP, Anisomov1993:PRB}. 
  
    \begin{table*}[!htb]
    \caption{\label{Tab:strain} Investigated crystallographic and magnetic information collected from  
    Refs. \cite{Baskin1955:PR,Gu2019:PRB, Wiedenmann1981:SSC, Wildes1998:JP, Kurosawa1983:JPSJ, Lancon2018:PRB, Lancon2016:PRB,Susner2017:AM,Wang2018:AFM,Chittari2016:PRB,Brec1986:SSI,Ouvrard1985:MRB,Sivadas2015:PRB, Rao1992:JPCS,LeFlem1982:JPCS,Li2019:PRB2,Lee2016:NL,Klingen1968:NW,Pei2018:FP, Chittari2020:PRB, Wildes2017:JP,Joy1992:PRB,Zhang2021:NL,Olsen2021:JPD}, for 
    the considered monolayers used in the graphene/MPX$_3$ heterostructures.
    The experimental (exp.) and the employed heterostructure (het.) lattice constants $a$. The strain for each subsystem, is calculated as $(a_{\textrm{het}}-a_{\textrm{exp}})/{a_{\textrm{exp}}}$. We also list the N{\'e}el temperature $T_\mathrm{N}$, the experimentally determined magnetic ground state for each MPX$_3$, as shown in Figs.~\ref{Fig:crystal_structure}(c)-(e), and the DFT-relaxed interlayer distance $\mathrm{d}_{\mathrm{int}}$ between graphene and the MPX$_3$. }
    \begin{ruledtabular}
    \begin{tabular}{lccccccccc}
     & graphene & MnPSe$_3$ & MnPS$_3$ & FePSe$_3$ & FePS$_3$ & NiPSe$_3$ & NiPS$_3$ & CoPSe$_3$ & CoPS$_3$ \\
     \hline
     $a$ (exp.) [\AA] & 2.46 & 6.39 & 6.08 & 6.26 & 5.94 & 6.13 & 5.82 & 6.19 & 5.90 \\
     $a$ (het.) [\AA] & 2.48 (2.44)\footnotemark[1] & 6.20 & 6.10 & 6.20 & 6.10 & 6.20 & 6.10 & 6.20 & 6.10 \\
     strain [\%] &	0.8 (-0.8) &  -2.9 & 0.3 & -1.0 & 2.7 & 1.1 & 4.8 & 0.2 & 3.4\\
     $T_\mathrm{N}$ [K] & - & 74 & 78 & 119 & 120 & 206 & 154 & - & 120\\
     ground state & - & N{\'e}el & N{\'e}el & Zigzag & Zigzag & Zigzag & Zigzag & - & Zigzag\\
     $\mathrm{d}_{\mathrm{int}}$ [\AA] & - & 3.457 & 3.422 & 3.471 & 3.448 & 3.474 & 3.440 & 3.488 & 3.430
    \end{tabular}
    \end{ruledtabular}
    \footnotetext[1]{For graphene/MPSe$_3$ (graphene/MPS$_3$) structures we stretch (compress) the graphene lattice constant to 2.48~\AA~(2.44~\AA).}
    \end{table*}
	
	The heterostructures of graphene/MPX$_3$ contain a $5 \times 5$ supercell 
	of graphene on a $2 \times 2$ MPX$_3$ supercell, resulting in 90 atoms in the unit cell. 
	We consider cases for M =\{Mn,Fe,Ni,Co\} and X=\{S,Se\}. 
	In Fig.~\ref{Fig:crystal_structure}(a,b) we show the exact geometry of the heterostructures. 
	The M atoms of the MPX$_3$ monolayers form a hexagonal lattice. Depending on the M atom, the MPX$_3$ monolayer can have a different magnetic ground state. 
	We consider all of our MPX$_3$ substrates to be either in the ferromagnetic, 
	antiferromagnetic N{\'e}el, antiferromagnetic zigzag, or antiferromagnetic stripy phase, as shown in Fig.~\ref{Fig:crystal_structure}(c)-(f), but with magnetizations to be collinear with the $z$-direction. Note, that structural relaxation was performed only for the ferromagnetic phase. The relaxed geometries are then employed for all other magnetic phases as well. We find that the forces in the other magnetic phases are below $10^{-3}$~[Ry/$a_0$], justifying this approach. In addition, we consider only geometries with $0^{\circ}$ relative twist angle and the stacking as shown in Fig.~\ref{Fig:crystal_structure}(a)

	In Table~\ref{Tab:strain} we summarize the experimental lattice constants of all considered monolayers, as well as their strained lattice constants used for the heterostructure calculations.
	We also list the N{\'e}el temperature $T_\mathrm{N}$, the experimentally determined magnetic ground state for each MPX$_3$, and the DFT-relaxed interlayer distance $\mathrm{d}_{\mathrm{int}}$ between graphene and the MPX$_3$. 
	For example, MnPSe$_3$ is a semiconductor, with an optical gap of about 2.3~eV \cite{Grasso1999:OSA,Yang2020:RSC}. 
    According to experiments, the lattice constant is $a = 6.39$~\AA, it has a N{\'e}el temperature of $T_\mathrm{N} = 74$~K, and the magnetic ground state is the antiferromagnetic N{\'e}el one \cite{Wiedenmann1981:SSC}, see Fig.~\ref{Fig:crystal_structure}(d). 
    Due to lattice mismatch with graphene, we strain the MnPSe$_3$ layer by about $-2.9$\% to get the mentioned supercell with commensurate lattices. The relaxed interlayer distance with graphene is $\mathrm{d}_{\mathrm{int}}=3.457$~\AA.

\section{Low energy model Hamiltonian}
\label{Sec:Hamiltonian}
	
 From our first-principles calculations we obtain the low 
    energy Dirac band structure of the proximitized graphene and extract 
    realistic parameters for an effective Hamiltonian describing the low-energy
    bands. The goal is to find an effective description for the low-energy physics, which is relevant for studying transport \cite{Fulop2021:arxiv, Veneri2022:arxiv, Lee2022:arxiv,Offidani2017:PRL,Karpiak2019:arxiv}, topology \cite{Hogl2020:PRL,Frank2018:PRL}, spin relaxation \cite{Cummings2017:PRL, Zollner2021:PRB, Zollner2019:PRB, Offidani2018:PRB}, or emergent long-range order \cite{Lin2022:S}. Due to the short-range nature of the proximity effects in van der Waals heterostructures, the effective models are transferable \cite{Zollner2020:PRL,Zollner2021:arxiv,Zollner2022:PRB}. 
    For example, a bilayer-graphene encapsulated from two sides by different materials, can be described by combining the effective models for two proximitized graphene sheets, coupled by bilayer-graphene interlayer couplings \cite{Zollner2020:PRL,Zollner2021:arxiv}.

    The systems we consider have broken time-reversal
    symmetry and either $C_3$ or no symmetry, depending on the 
    magnetic phase of the underlying MPX$_3$. The following Hamiltonian,
    derived from symmetry~\cite{Kochan2017:PRB,Phong2017:2DM, Zollner2016:PRB}, is able to describe 
    the graphene bands in the vicinity of the Dirac points when proximity exchange is present
    \begin{flalign}
    \label{Eq:Hamiltonian}
    &\mathcal{H} = \mathcal{H}_{0}+\mathcal{H}_{\Delta}+\mathcal{H}_{\textrm{ex}}+\mathrm{E_D},\\
    &\mathcal{H}_{0} = -\sum_{s} f_{s}(\textbf{q})\ket{\Psi_{\mathrm{A}},s}\bra{\Psi_{\mathrm{B}},s}+h.c. \\
    &\mathcal{H}_{\Delta} =\Delta \sigma_z \otimes s_0,\\
    &\mathcal{H}_{\textrm{ex}} =  (-\lambda_{\textrm{ex}}^\textrm{A} \sigma_{+}	+\lambda_{\textrm{ex}}^\textrm{B} \sigma_{-}) \otimes s_z.
    \end{flalign}
    The orbital structural function we use is \mbox{$f_{s}(\textbf{q})= t_{1,s}+t_{2,s}\mathrm{e}^{\mathrm{i}\textbf{q}\textbf{R}_2} +t_{3,s}\mathrm{e}^{-\mathrm{i}\textbf{q}\textbf{R}_3}$}, where we denote $t_{\alpha,s}$ as the spin (\mbox{$s=\{\uparrow,\downarrow\}$}) and direction 
    (\mbox{$\alpha=\{1,2,3\}$}) dependent hopping parameters, along direction 
    \mbox{$\textbf{R}_{\alpha}=\textrm{a}(\cos{\frac{2\pi(\alpha-1)}{3}},\sin{\frac{2\pi(\alpha-1)}{3}})$}. 
    The wavevector is \mbox{$\textbf{q}=\tau\textbf{K}+\textbf{k}$}, for the valley index $\tau = \pm 1$ and \mbox{$\tau\textbf{K}=\pm(\frac{4\pi}{3\textrm{a}},0)$} with the lattice constant a.
    
    This orbital structural function is a generalized case for the orbital Hamiltonian as defined in Ref. \cite{Kochan2017:PRB}.
    Typically, the standard linearized orbital Hamiltonian is used \cite{Kochan2017:PRB,Zollner2019:PRB}, but
    with our definition, we can describe cases where the nearest-neighbor hoppings are spin dependent and can be different along different spatial directions.
    This is necessary, because the magnetic phase of the MPX$_3$ monolayer substrate determines   the symmetry of the system and the dependence of hoppings on spatial directions. Additionaly, because the substrate is magnetic, hoppings can be spin dependent. 
    
    The Pauli spin
    matrices are $s_j$, acting on spin space ($\uparrow,
    \downarrow$), and pseudospin matrices are $\sigma_j$,
    acting on sublattice space (C$_\textrm{A}$, C$_\textrm{B}$), with
    $j = \{ 0,x,y,z \}$ where $j = 0$ denotes the $2 \times 2$
    unit matrix. For shorter notation, we use $\sigma_{\pm} =
    \frac{1}{2}(\sigma_z \pm \sigma_0)$. The staggered
    potential gap is $\Delta$ and the parameters,
    $\lambda_{\textrm{ex}}^\textrm{A}$ and
    $\lambda_{\textrm{ex}}^\textrm{B}$, represent the sublattice-resolved proximity-induced exchange couplings. The four basis states are
    $|\Psi_{\textrm{A}}, \uparrow\rangle$,  $|\Psi_{\textrm{A}},
    \downarrow\rangle$,  $|\Psi_{\textrm{B}}, \uparrow\rangle$,  and
    $|\Psi_{\textrm{B}}, \downarrow\rangle$.  The model
    Hamiltonian is centered around the Fermi level at zero energy.
    Since first-principles results capture charge-transfer effects, we also introduce the 
    parameter $\mathrm{E_D}$ (termed Dirac point energy) which shifts the Dirac bands.
    
   For each considered heterostructure, we calculate the proximitized low energy Dirac bands in the vicinity of the K point.
    To extract the fit parameters from the first-principles data, we employ a least-squares routine \cite{lmfit}, taking into account band energies, splittings, and spin expectation values (spin up and spin down).

\section{Band Structure and Fit Results}
\label{Sec:Results}
In the following, we first neglect SOC in the calculations, since we are mainly interested in the proximity-induced exchange couplings. The effects of SOC are discussed at a later point. 
For our analysis of proximity effects, we consider graphene on the magnetic monolayers MPX$_3$, for
different magnetic phases, as shown in Fig.~\ref{Fig:crystal_structure}(c)-(f). 
Depending on the magnetic phase of the MPX$_3$, the system shows either 3-fold symmetry for ferromagnetic and antiferromagnetic N{\'e}el phases or lacks all symmetries in the antiferromagnetic zigzag and stripy phases. 
In the following, we explicitly show and discuss the band structure and model calculations (with fitted parameters) for graphene on MnPS$_3$ as an exemplary case. 
In the Supplemental Material we show the band structure and the fitted results for the remaining MPX$_3$ substrates \footnotemark[3].
Furthermore, for graphene on MnPS$_3$ in the antiferromagnetic N{\'e}el phase (the ground state), we discuss the influence of the interlayer distance, the Hubbard $U$, and a transverse electric field on the fitted parameters.

\subsection{Ferromagnetic phase}

    \begin{figure}[htb]
     \includegraphics[width=.99\columnwidth]{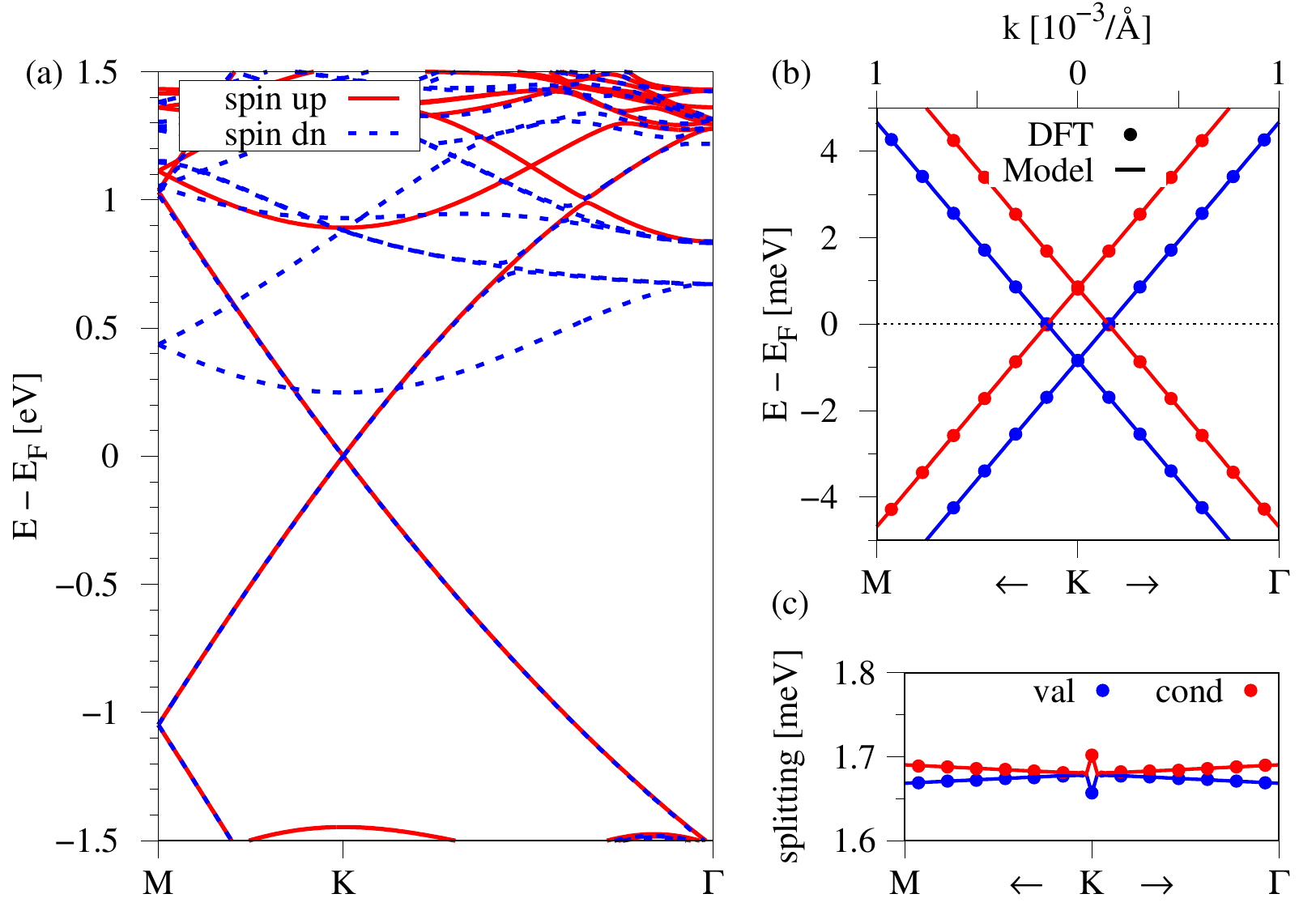}
     \caption{  (a) DFT-calculated band structure of the graphene/MnPS$_3$ heterostructure along the high-symmetry path M-K-$\Gamma$ without SOC, when the MnPS$_3$ is in the ferromagnetic phase. Red solid (blue dashed) lines correspond to spin up (spin down).
     (b) Zoom to the DFT-calculated (symbols) low energy Dirac bands near the K point with a fit to the model Hamiltonian (solid line). 
     (c) The splitting of valence and conduction band.
     }\label{Fig:bands_FM_MnPS3}
    \end{figure}

First, we consider the MnPS$_3$ monolayer to be in the ferromagnetic phase in which the heterostructure has a 3-fold symmetry and our model can be further simplified by setting 
$t_{1,s} = t_{2,s} = t_{3,s}$ for both spin species $s$, respectively. 
In Fig.~\ref{Fig:bands_FM_MnPS3}(a) we show the corresponding DFT-calculated 
band structure of the graphene/MnPS$_3$ without SOC. 
We find that the graphene Dirac states are located at the Fermi level and are well preserved. The parabolic spin down band above the Dirac cone originates from the magnetic substrate. 
In Fig.~\ref{Fig:bands_FM_MnPS3}(b), we show a zoom to the low energy Dirac bands near the K point. We find that the model perfectly agrees with the DFT-calculated dispersion by using the parameters summarized in Table~\ref{tab:fitresults_noSOC}. 
In this ferromagnetic case, uniform exchange parameters, $\lambda_{\textrm{ex}}^\textrm{A} \approx \lambda_{\textrm{ex}}^\textrm{B} \approx -0.8$~meV, are found. The calculated positive and uniform spin polarization on graphene is consistent with this picture, see Supplemental Material~\footnotemark[3]. The staggered potential $\Delta$ is small ($\approx 10~\mu$eV), indicating that nearly no sublattice symmetry breaking is present. The orbital hopping parameters $t_{s}$ are only slightly different for the two spin species. 
Using these parameters, not only the dispersion but also the band splittings, of about 1.7~meV, can be perfectly reproduced, see Fig.~\ref{Fig:bands_FM_MnPS3}(c).

\subsection{Antiferromagnetic N{\'e}el phase} 

    \begin{figure}[htb]
     \includegraphics[width=.99\columnwidth]{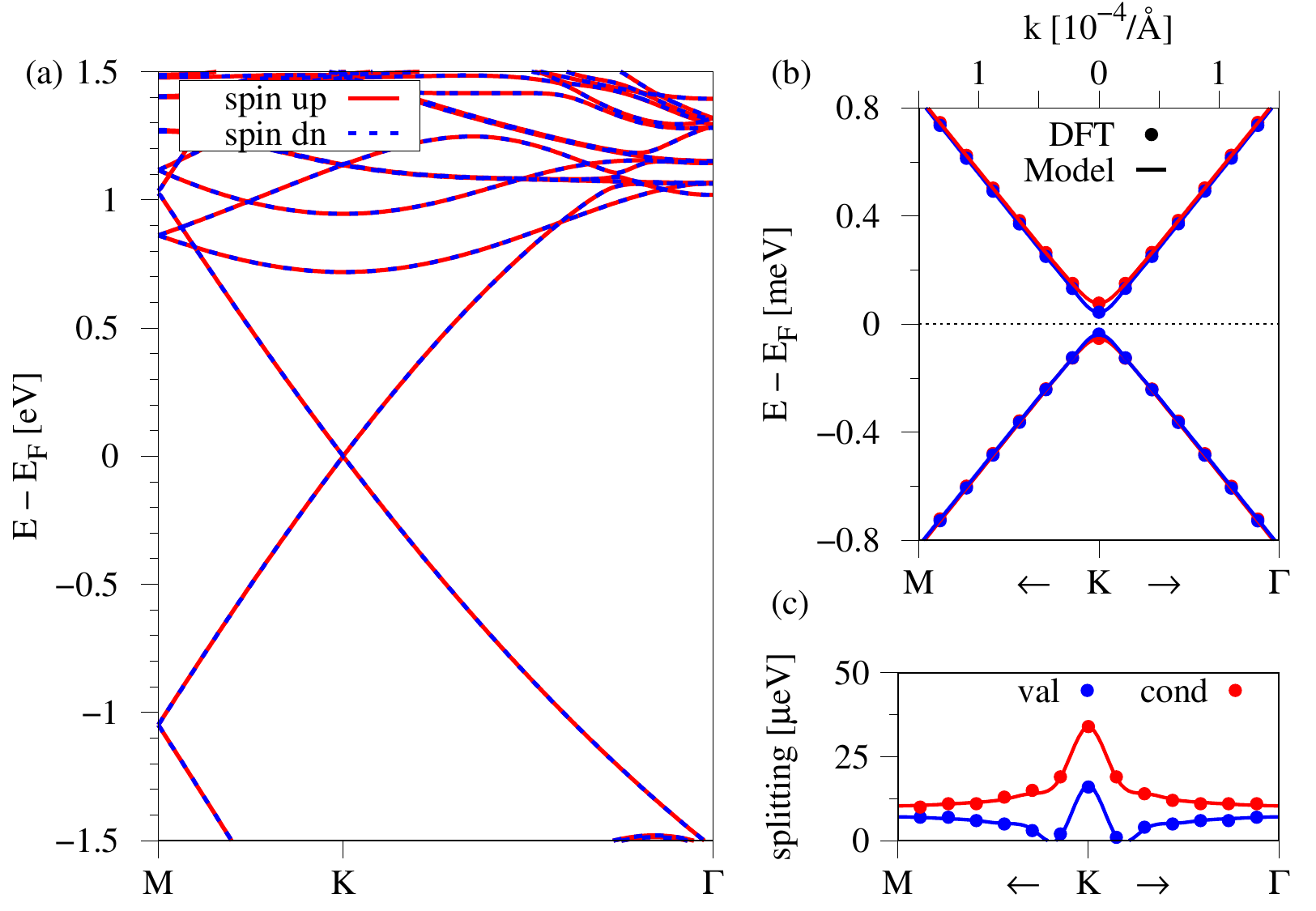}
     \caption{  Same as Fig.~\ref{Fig:bands_FM_MnPS3}, but for the  MnPS$_3$ in the antiferromagnetic N{\'e}el phase.}\label{Fig:bands_AFM_Neel_MnPS3}
    \end{figure}

Second, we consider the MnPS$_3$ monolayer to be in the antiferromagnetic N{\'e}el phase, which is the ground state for this material.  
As before, we calculate the dispersion and employ our model Hamiltonian. Again, the heterostructure has a 3-fold symmetry in the antiferromagnetic N{\'e}el phase and we can use $t_{1,s} = t_{2,s} = t_{3,s}$ for the fitting procedure.
The calculation and fitting results are summarized in Fig.~\ref{Fig:bands_AFM_Neel_MnPS3}.
The Dirac bands of graphene are again well preserved and located at the Fermi level. Also the model agrees perfectly with the DFT-calculated dispersion and band splittings, employing the parameters from Table~\ref{tab:fitresults_noSOC}. 
Compared to the ferromagnetic phase, band splittings are drastically diminished in magnitude from about 1.7~meV to 20~$\mu$eV. 
We can understand this by the following argument. 
In the antiferromagnetic N{\'e}el phase, the average proximity exchange on the graphene sublattices is reduced, because Mn atoms have alternating magnetizations. Therefore, the exchange field felt by graphene is strongly diminished. In contrast in the ferromagnetic phase, all Mn atoms have the same magnetization and proximity effects are much stronger. 
In addition, for the ferromagnetic phase, sublattice resolved proximity exchange parameters had the same sign, while for the antiferromagnetic N{\'e}el phase, we find opposite sign, 
$\lambda_{\textrm{ex}}^\textrm{A} < 0$ and $\lambda_{\textrm{ex}}^\textrm{B} > 0$.
The calculated spin polarization, being opposite on the graphene sublattices, supports this result of staggered exchange couplings \footnotemark[3].
This is a very important finding, since the magnetic phase decides not only about the magnitude, but also the type of proximity exchange coupling, which can have dramatic consequences for topological phases in proximitized graphene~\cite{Hogl2020:PRL}.

 \subsection{Antiferromagnetic Zigzag phase} 
 
    \begin{figure}[htb]
     \includegraphics[width=.99\columnwidth]{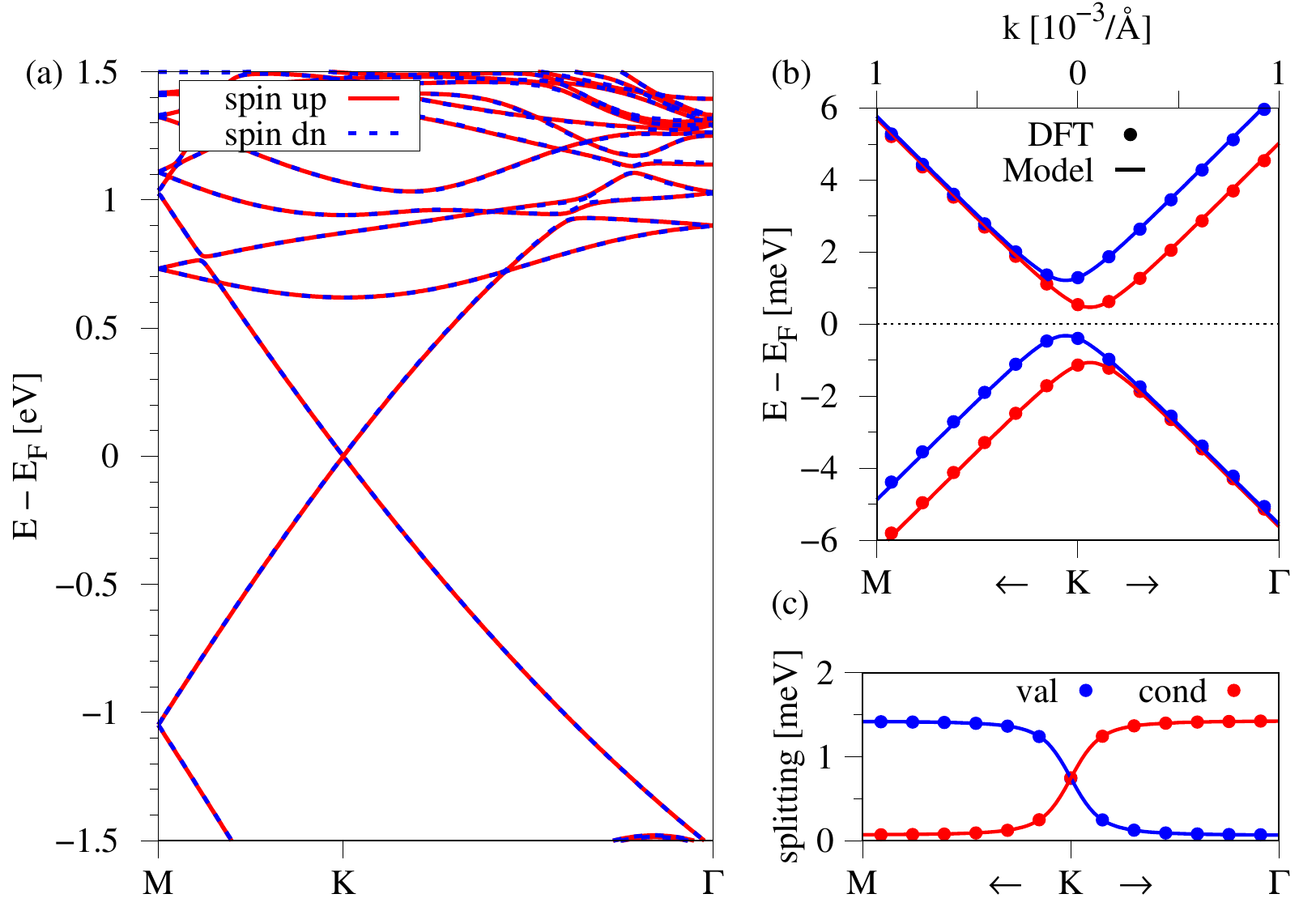}
     \caption{  Same as Fig.~\ref{Fig:bands_FM_MnPS3}, but for the MnPS$_3$  in the antiferromagnetic zigzag phase.}\label{Fig:bands_AFM_zigzag_MnPS3}
    \end{figure}
    
 Third, we consider the MnPS$_3$ monolayer to be in the antiferromagnetic zigzag phase. Now, the heterostructure has no symmetry and nearest-neighbor hopping amplitudes are becoming inequivalent $t_{1,s} \neq t_{2,s} \neq t_{3,s}$. This asymmetry in the
hopping amplitudes leads to the shift of the Dirac 
point in momentum space \cite{Pereira2009:PRB, Wunsch2008:NJP}.
 Because the system is additionally magnetic, spin up and spin down Dirac points do not need to be shifted equally away from the K point. In the Supplemental Material \footnotemark[3], we explicitly show a 2D-map of the low energy Dirac bands in momentum space, corresponding to Fig.~\ref{Fig:bands_AFM_zigzag_MnPS3}(b).
 Especially in this case we take into account DFT-calculated bands not along a high-symmetry path, but in a small $k$-space region around the K point, to obtain accurate fitting results. 
 In Fig.~\ref{Fig:bands_AFM_zigzag_MnPS3}, we summarize our calculation and fit results for the graphene/MnPS$_3$ heterostructure when the MnPS$_3$ is in the antiferromagnetic zigzag phase. 
 The global band structure of the zigzag and the N{\'e}el phases are quite similar at first glance. However, proximity-induced exchange coupling in graphene is completely different and the Dirac cones for spin up and spin down are indeed shifted in momentum space, see Fig.~\ref{Fig:bands_AFM_zigzag_MnPS3}(b). As a result, band splittings are strongly direction dependent, see Fig. \ref{Fig:bands_AFM_zigzag_MnPS3}(c).
 However, our model is also capable of describing this situation with parameters summarized in Table~\ref{tab:fitresults_noSOC}, as can be seen in Fig.~\ref{Fig:bands_AFM_zigzag_MnPS3}(b,c).
 Interestingly, sublattice-resolved proximity exchange parameters of the antiferromagnetic zigzag phase are about 0.4~meV in magnitude, half of the value and opposite in sign compared to the ferromagnetic case, see Table~\ref{tab:fitresults_noSOC}. Comparing the calculated spin polarizations on graphene in the ferromagnetic and antiferromagnetic zigzag phases~\footnotemark[3], the opposite sign of proximity exchange is reasonable.
 In addition, the staggered potential parameter $\Delta$ is drastically enhanced compared to the other phases.
 Therefore, mainly one sublattice contributes to the spin polarization in the antiferromagnetic zigzag phase, while in the ferromagnetic phase, both sublattices contribute equally.
 
 So far, we have seen all types of sublattice-resolved proximity-induced exchange coupling in graphene, uniform (positive and negative) and staggered, only by changing the magnetic phase of the MnPS$_3$ substrate. 
 \subsection{Antiferromagnetic Stripy phase} 
 
    \begin{figure}[htb]
     \includegraphics[width=.99\columnwidth]{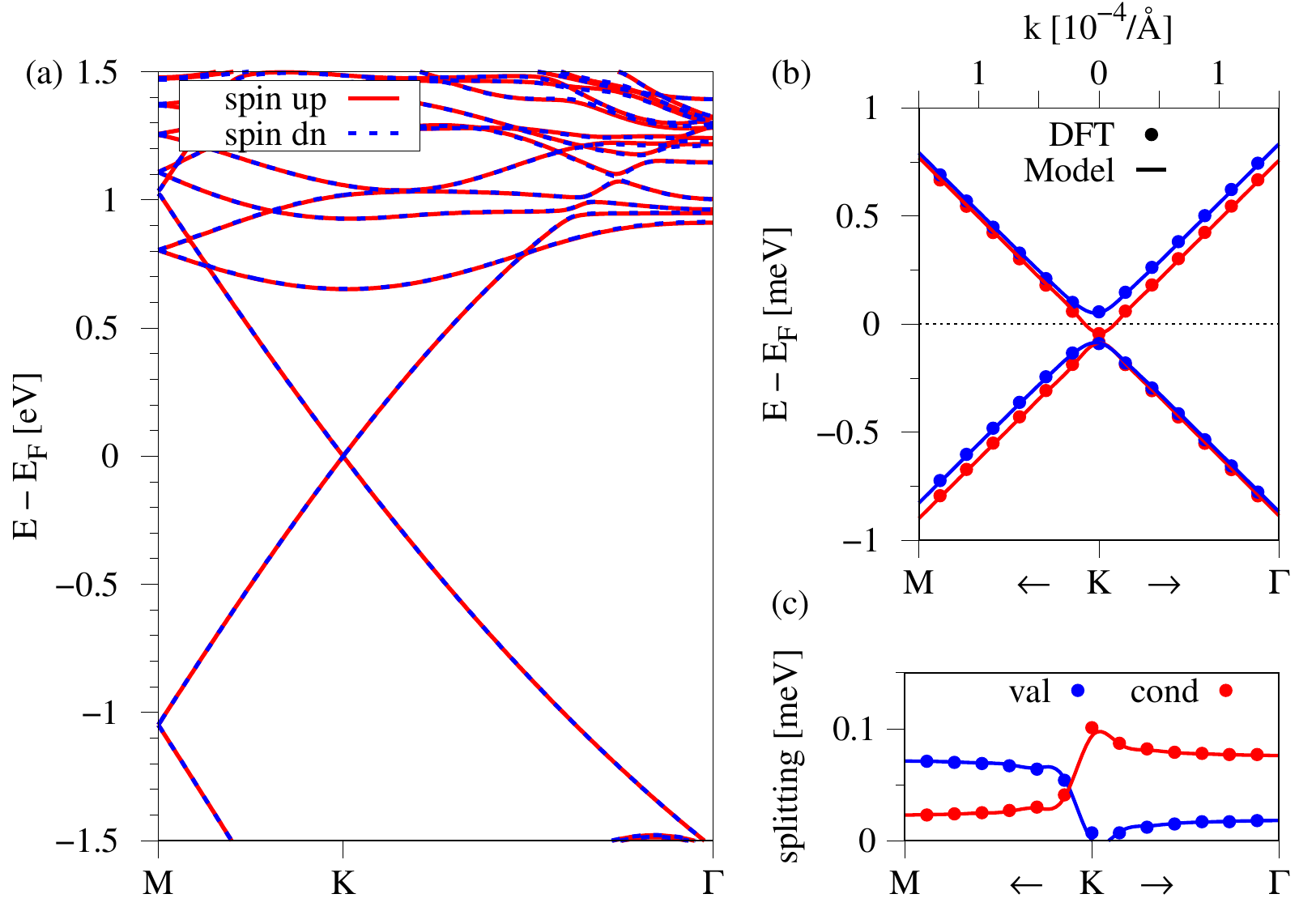}
     \caption{  Same as Fig.~\ref{Fig:bands_FM_MnPS3}, when the MnPS$_3$ is in the antiferromagnetic stripy phase.}\label{Fig:bands_AFM_stripy_MnPS3}
    \end{figure}
 
  Finally, we consider the MnPS$_3$ monolayer to be in the antiferromagnetic stripy phase. Similar to the zigzag phase, the heterostructure has no symmetry and hopping amplitudes become inequivalent $t_{1,s} \neq t_{2,s} \neq t_{3,s}$. 
  All relevant features of the dispersion are nearly identical to the antiferromagnetic zigzag phase. 
  The main difference is that sublattice resolved proximity exchange couplings and Dirac band splittings are much smaller in magnitude. Looking at the calculated spin polarization in real space~\footnotemark[3], we find a similar picture as for the antiferromagnetic N{\'e}el phase, with opposite polarizations on the graphene sublattices.
  This scenario arises due to the small sublattice symmetry breaking $\Delta$ and small but uniform exchange parameters $\lambda_{\textrm{ex}}$ in combination with slightly asymmetric hoppings.

\begin{table*}[htb]
\caption{\label{tab:fitresults_noSOC} 
Fit parameters of Hamiltonian $\mathcal{H}$ for the graphene/MPX$_3$ heterostructures 
for different magnetic phases (FM = ferromagnetic, AFM = antiferromagnetic) of the MPX$_3$. 
Calculations were performed without SOC. We list the hopping parameters t$_{\alpha,s}$, the
staggered potential $\Delta$, proximity exchange parameters $\lambda_{\textrm{ex}}^\textrm{A}$ 
and $\lambda_{\textrm{ex}}^\textrm{B}$, the Dirac point energy $\mathrm{E_D}$, 
and the total energy $\mathrm{E}_{\textrm{tot}}$, with respect to the ground state. }
\begin{ruledtabular}
\begin{tabular}{l c c c c c c c }
MPX$_3$ & magnetic & t$_{\mathrm{1},\uparrow}$,t$_{\mathrm{2},\uparrow}$,t$_{\mathrm{3},\uparrow}$/t$_{\mathrm{1},\downarrow}$,t$_{\mathrm{2},\downarrow}$,t$_{\mathrm{3},\downarrow}$  & $\Delta$   & $\lambda_{\textrm{ex}}^\textrm{A}$   & $\lambda_{\textrm{ex}}^\textrm{B}$   & $\mathrm{E_D}$  & $\mathrm{E}_{\textrm{tot}}$  \\ 
& phase & [eV] & [$\mu$eV] & [$\mu$eV] & [$\mu$eV] & [meV] & [meV] \\\hline
          & FM  & 2.5038/2.5030   & 19.2  & $-219.6$ & $-220.1$ & $-0.047$ & 231.4 \\
 & AFM N{\'e}el & 2.5032/2.5032   & 42.3  & $-11.4$  & $-13.0$  & 0.029     & 0    \\
MnPSe$_3$ & AFM Zigzag & 2.5057,2.5032,2.5008/2.4997,2.5021,2.5046   &    629.2   &   469.5 & 463.1  &   0.060    & 91.8     \\
        & AFM Stripy  & 2.5035,2.5034,2.5033/2.5030,2.5032,2.5035  & 112.3 & 41.3 & 43.8  & 0.007 &  116.4 \\
         & FM\footnotemark[1]  & 2.5083/2.5070   & 68.9  & $-215.0$ & $-232.9$ & $-0.004$ &  164.9\\
         & AFM N{\'e}el\footnotemark[1] &  2.5088/2.5086  & 67.7  &  20.2 & $-38.5$   &  $-0.001$  & 0    \\\hline
          & FM   & 2.6144/2.6093   & 13.2  & $-850.9$ & $-828.5$ & $-0.004$  & 238.8 \\
  & AFM N{\'e}el & 2.6126/2.6126   & 52.9  & $-17.0$  &  8.2     &  0.007    & 0    \\
 MnPS$_3$     & AFM Zigzag &  2.6136,2.6116,2.6097/2.6100,2.6120,2.6140   & 496.6  &  362.4   & 383.2 & 0.073       & 103.5     \\
         & AFM Stripy  & 2.6123,2.6122,2.6122/2.6120,2.6122,2.6124  & 32.9 & 35.1  & 12.0 & $-0.041$  &  113.7\\
          & FM\footnotemark[1]  & 2.6192/2.6109  &  24.5 & $-954.3$ & $-1016.3$ & 0.031 & 212.4 \\
         & AFM N{\'e}el\footnotemark[1] &  2.6173/2.6173  & 35.4  & 14.6  & $-19.2$   &  0.001  & 0    \\\hline
          & FM & 2.5050/2.3475  & 3261.2 & 992.2 & $-5926.5$ & $-8.6$ & 1511.1\\
FePSe$_3$ & AFM N{\'e}el \footnotemark[2]  & -  & - & - & - & -      & 1040.1    \\
          & AFM Zigzag &  2.5054,2.5027,2.5015/2.5003,2.5026,2.5039     & 289.3   &  196.4   &  428.3     & $-0.001$     &   0   \\
         & AFM Stripy  & 2.5028,2.5028,2.5031/2.4988,2.4997,2.5006  & 196.7 & 99.2 & $-94.1$ & 0.880  & 275.5 \\\hline
          & FM           &  2.6151/2.5824   & 364.7  & 430.8 & $-786.4$ & 331.5 & 1735.0\\
FePS$_3$  & AFM N{\'e}el &    2.6110/2.6109      &  126.9     & $-14.7$    &   $-63.2$       &   550.9       &   1451.7  \\
          & AFM Zigzag &    2.5996,2.5936,2.5891/2.6003,2.6019,2.6051  &  570.7    &   668.5    & 458.4    &  $-0.229$  &   0   \\
          & AFM Stripy  & 2.5452,2.5343,2.5470/2.5347,2.5429,2.5446  & 342.1 & $-183.9$ & 455.0 & $-0.034$ & 134.5 \\\hline
          & FM           &  2.5036/1.9911 & 53.7 & 10764.4 & 10575.8 & 85.0 & 479.0 \\
NiPSe$_3$ & AFM N{\'e}el &  2.4797/2.4795  & 75.7 & $-36.0$ &  $-0.7$   &    0.018      & 82.0     \\
          & AFM Zigzag &     2.4851,2.4814,2.4779/2.4858,2.4883,2.4907 &  655.3  &  549.4 &  609.2 &    $-0.126$  &   0 \\
             & AFM Stripy\footnotemark[2]  & -  & - & - & - & - & 535.7  \\\hline
          & FM           &  2.6130/2.3893 & 860.3 & 1538.5  & 115.1 & 387.2 & 291.1\\
NiPS$_3$  & AFM N{\'e}el &   2.5732/2.5755  &    114.6   &   53.1       &    $-65.9$      &    265.6      &   37.8  \\
          & AFM Zigzag & 2.1102,2.1146,2.1169/2.0470,2.0545,2.0610         &  79.3     &   144.9     &  303.6    &   293.7 &  0    \\
                  & AFM Stripy\footnotemark[2]  & -  & - & - & - & -  & 341.6 \\\hline
                    & FM & 2.4221/2.5022  & 0.9 & $-3097.7$ & $-3069.9$ & 364.1 & 127.8 \\
CoPSe$_3$ & AFM N{\'e}el &  2.4786/2.4797   & 87.6    &     $-91.7$ &   39.5    &  363.4 &  1120.1   \\
          & AFM Zigzag &   2.4506,2.4461,2.4415/2.4526,2.4509,2.4495   &   346.8    &  $-12.2$ &   37.4 & 370.4    &  1306.7    \\
                  & AFM Stripy  & 2.4784,2.4781,2.4769/2.4848,2.4854,2.4847  & 65.2 & 55.9 & $-60.2$ & 362.2 & 0  \\\hline
          & FM           & 2.6103/2.5989    & 43.7 & $-1337.2$ & $-1261.7$  & 562.7 & 534.0 \\
CoPS$_3$  & AFM N{\'e}el &    2.5280/2.5269      &   115.2    &  $-63.8$        &  25.3        &    0.103      &  19.2   \\
          & AFM Zigzag &   2.4576,2.4494,2.4414/2.5166,2.5237,2.5304    & 1371.3  & 1349.9     & 1442.2     &    0.067      &  0    \\
        & AFM Stripy\footnotemark[2]  & -  & - & - & - & - & 229.9 \\
\end{tabular}
\end{ruledtabular}
\footnotetext[1]{Calculated with {\tt WIEN2k}~\cite{Wien2k}.}
\footnotetext[2]{A reasonable fit could not be performed, as can be seen from the DFT-calculated band structure in the Supplemental Material.}
\end{table*}

\begin{figure}[htb]
\includegraphics[width=.9\columnwidth]{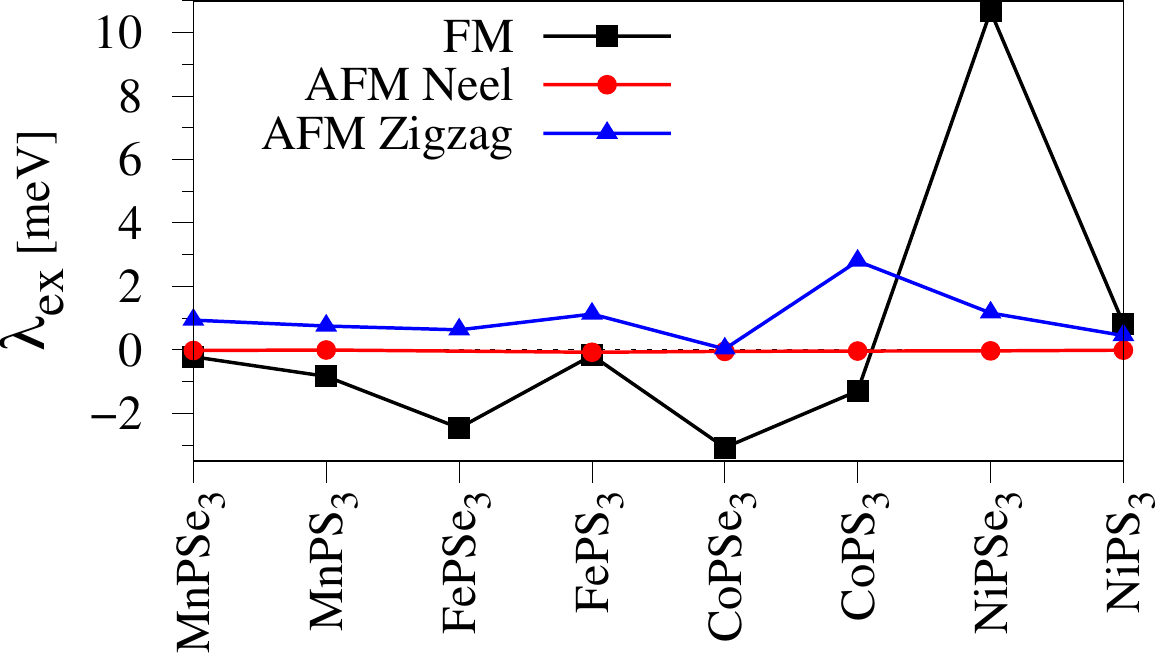}
\caption{  The average proximity exchange coupling, $\lambda_{\textrm{ex}}$, as function of the substrate material and the magnetic phase (FM = ferromagnetic, AFM = antiferromagnetic).  }\label{Fig:PEX_graph}
\end{figure}

In order to unveil specific trends on how the proximity-induced exchange depends on the specific material and magnetic phase, we plot in Fig.~\ref{Fig:PEX_graph} the average coupling, \mbox{$\lambda_{\textrm{ex}} = (\lambda_{\textrm{ex}}^\textrm{A}+\lambda_{\textrm{ex}}^\textrm{B}$)/2} (listed in Tab.~\ref{tab:fitresults_noSOC}). We only show the results for the ferromagnetic, antiferromagnetic N\'{e}el, and antiferromagnetic zigzag phases, since in the antiferromagnetic stripy phase we do not have data for all material combinations. In addition, the stripy phase is only energetically favorable in the case of CoPSe$_3$. 

Let us first discuss the dependence in the ferromagnetic phase (black data points). 
From Fig.~\ref{Fig:PEX_graph} it is evident that $\lambda_{\textrm{ex}}$ increases in magnitude in ascending order of the transition metal (Mn, Fe, Co, Ni), for Se-based MPX$_3$ materials. In addition, Mn, Fe, and Co provide negative $\lambda_{\textrm{ex}}$, while Ni provides positive $\lambda_{\textrm{ex}}$. 
The reason for the increasing magnitude can be seen from the density of states (DOS), see Supplemental Material~\footnotemark[3]. Focusing on Se-based materials, the transition metal DOS contribution near the Dirac point continuously increases from Mn to Ni. Since these states from the transition metals provide magnetism and can couple to the Dirac electrons, the proximity exchange increases. 
In general it also seems that Se-based materials provide stronger proximity exchange compared to S-based ones. 
Focusing on the two antiferromagnetic phases (red and blue data points), we find that proximity exchange is much more suppressed compared to the ferromagnetic phase. This can be understood from the fact that an antiferromagnetic substrate has alternating magnetizations and therefore the exchange field that graphene can pick-up is diminished. 
In general, the antiferromagnetic N{\'e}el phase provides the smallest proximity exchange in graphene, while the antiferromagnetic zigzag phase can still provide moderate exchange coupling. This may be attributed to the different arrangements of magnetic metal ions below graphene. In the N{\'e}el phase, magnetic ions with different magnetization orientations form a regular pattern (considering one metal ion, the three nearest neighbors have opposite magnetizations). In contrast, in the zigzag phase, there are at least small "domains" (zigzag-stripes), where the magnetization is uniform (considering one metal ion, two nearest neighbors have the same magnetization).

\subsection{Distance Study}

Proximity exchange coupling is a short range effect and can be strongly enhanced by diminishing the interlayer distance between materials, as experimentally demonstrated for proximity SOC in graphene/TMDC heterostructures \cite{Fulop2021:arxiv,Fulop2021:arxiv2}. In addition, in first-principles calculations, the equilibrium interlayer distance depends on the exchange-correlation functional and the vdW corrections, as for example demonstrated for graphite and hexagonal boron nitride \cite{Graziano2012:JP}.

How does the interlayer distance affect proximity exchange effects in our bilayers?
As an exemplary case we consider MnPS$_3$ in the antiferromagnetic N{\'e}el phase (ground state) and tune the distance $\mathrm{d}_{\mathrm{int}}$ between graphene and the MnPS$_3$ monolayer.
Especially this structure is interesting, since sublattice resolved proximity exchange parameters have opposite sign, $\lambda_{\textrm{ex}}^\textrm{A} < 0$ and $\lambda_{\textrm{ex}}^\textrm{B} > 0$, see Table~\ref{tab:fitresults_noSOC}, potentially important for the realization of novel topological states \cite{Hogl2020:PRL}.

    \begin{figure}[htb]
     \includegraphics[width=.99\columnwidth]{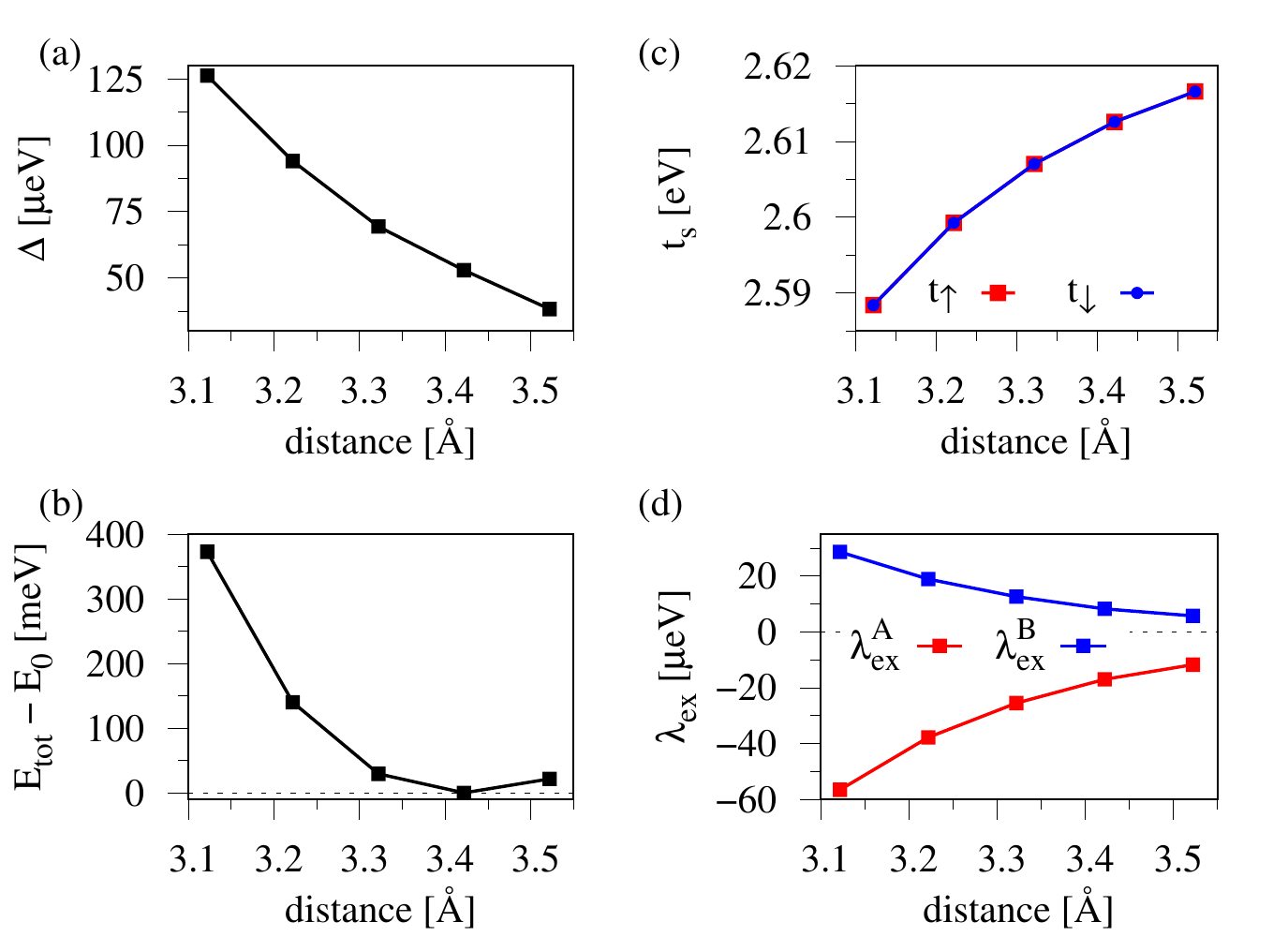}
     \caption{  Interlayer distance dependence of (a) the staggered potential $\Delta$, (b) the total energy $\mathrm{E}_{\mathrm{tot}}$, (c) the spin dependent hopping parameters t$_{\uparrow}$ and t$_{\downarrow}$, and (d) the sublattice resolved proximity exchange parameters $\lambda_{\textrm{ex}}^\textrm{A}$ and $\lambda_{\textrm{ex}}^\textrm{B}$ for the graphene/MnPS$_3$ heterostructure  when the MnPS$_3$ is in the antiferromagnetic N{\'e}el phase.
     }\label{Fig:distance_AFM_Neel_MnPS3}
    \end{figure}

In Fig.~\ref{Fig:distance_AFM_Neel_MnPS3}, we show the evolution of the fit parameters and the total energy as function of the interlayer distance for the graphene/MnPS$_3$ heterostructure. 
We notice that the staggered potential $\Delta$ increases with decreasing distance, because the sublattice-symmetry breaking in graphene gets stronger.
The total energy is minimized for the relaxed interlayer distance given in Table~\ref{Tab:strain}. 
The two spin dependent hopping parameters, t$_{\uparrow}$ and t$_{\downarrow}$, have nearly the same value for this system, as can be seen in Table~\ref{tab:fitresults_noSOC}, and decrease with decreasing van der Waals gap.
The proximity exchange parameters $\lambda_{\textrm{ex}}^\textrm{A}$ and $\lambda_{\textrm{ex}}^\textrm{B}$ increase in magnitude with decreasing interlayer distance.
We find that by decreasing $\mathrm{d}_{\mathrm{int}}$ by about 10\%, one can achieve a significant 5-fold enhancement of proximity exchange couplings.
Important, the sign of both parameters, $\lambda_{\textrm{ex}}^\textrm{A} < 0$ and $\lambda_{\textrm{ex}}^\textrm{B} > 0$, is not affected.

In the Supplemental Material~\footnotemark[3], we additionally show the evolution of the fit parameters as function of the interlayer distance for the graphene/MnPSe$_3$ heterostructure in the antiferromagnetic N{\'e}el phase. There, we even find a crossover from uniform to staggered exchange couplings, when decreasing the interlayer distance.
In conclusion, we expect similar behavior for the other material combinations and the different magnetic phases.

\subsection{Hubbard U}

    \begin{figure}[htb]
     \includegraphics[width=.99\columnwidth]{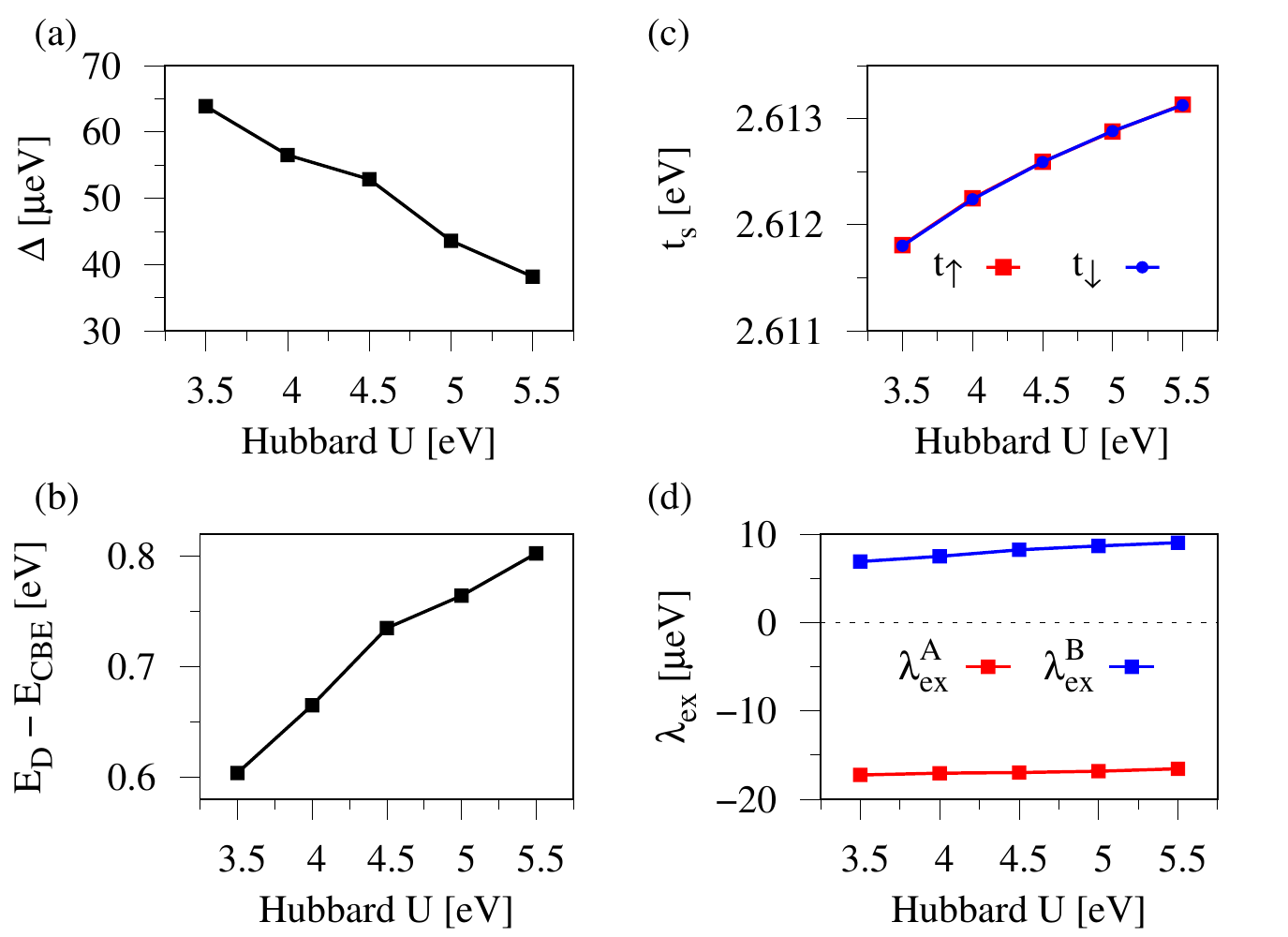}
     \caption{  Hubbard $U$ dependence of (a) the staggered potential $\Delta$, (b) the band offset of the Dirac point, $\mathrm{E_D}$, with respect to the MnPS$_3$ conduction band edge, $\mathrm{E_{CBE}}$, at the K point, (c) the spin dependent hopping parameters t$_{\uparrow}$ and t$_{\downarrow}$, and (d) the sublattice resolved proximity exchange parameters $\lambda_{\textrm{ex}}^\textrm{A}$ and $\lambda_{\textrm{ex}}^\textrm{B}$ for the graphene/MnPS$_3$ heterostructure  when the MnPS$_3$ is in the antiferromagnetic N{\'e}el phase.
     }\label{Fig:HubbardU_AFM_Neel_MnPS3}
    \end{figure}
    
The Hubbard $U$ parameter is variable for the DFT calculations. 
How does the $U$ affect proximity exchange splittings?
Again, we consider graphene on MnPS$_3$ in the antiferromagnetic N{\'e}el phase and tune the $U$ parameter between 3.5 to 5.5~eV, which is a typical range of values for $d$ orbitals in strongly correlated materials. 
The results are summarized in Fig.~\ref{Fig:HubbardU_AFM_Neel_MnPS3}. 
Overall, the fit parameters $\Delta$, t$_{\uparrow}$, t$_{\downarrow}$,  $\lambda_{\textrm{ex}}^\textrm{A}$, and $\lambda_{\textrm{ex}}^\textrm{B}$, barely change because for the graphene/MnPS$_3$ heterosturcutre, the Dirac bands are located at the Fermi level and well isolated from the substrate's bands, see Fig.~\ref{Fig:bands_AFM_Neel_MnPS3}(a). 
Even though the Hubbard $U$ shifts the Mn $d$ orbital levels in energy and tunes the band offset, the graphene states are nearly unaffected. 

However, the effect of the Hubbard $U$ on the fitted parameters can be also quite strong, as we demonstrate in the case of NiPSe$_3$ in the antiferromagnetic zigzag phase (see Supplemental Material~\footnotemark[3]). There, we find that proximity exchange couplings can be tuned from about 400 to 700~$\mu$eV within our considered range of the $U$ parameter.

\subsection{Transverse Electric Field}

\begin{figure}[htb]
     \includegraphics[width=.99\columnwidth]{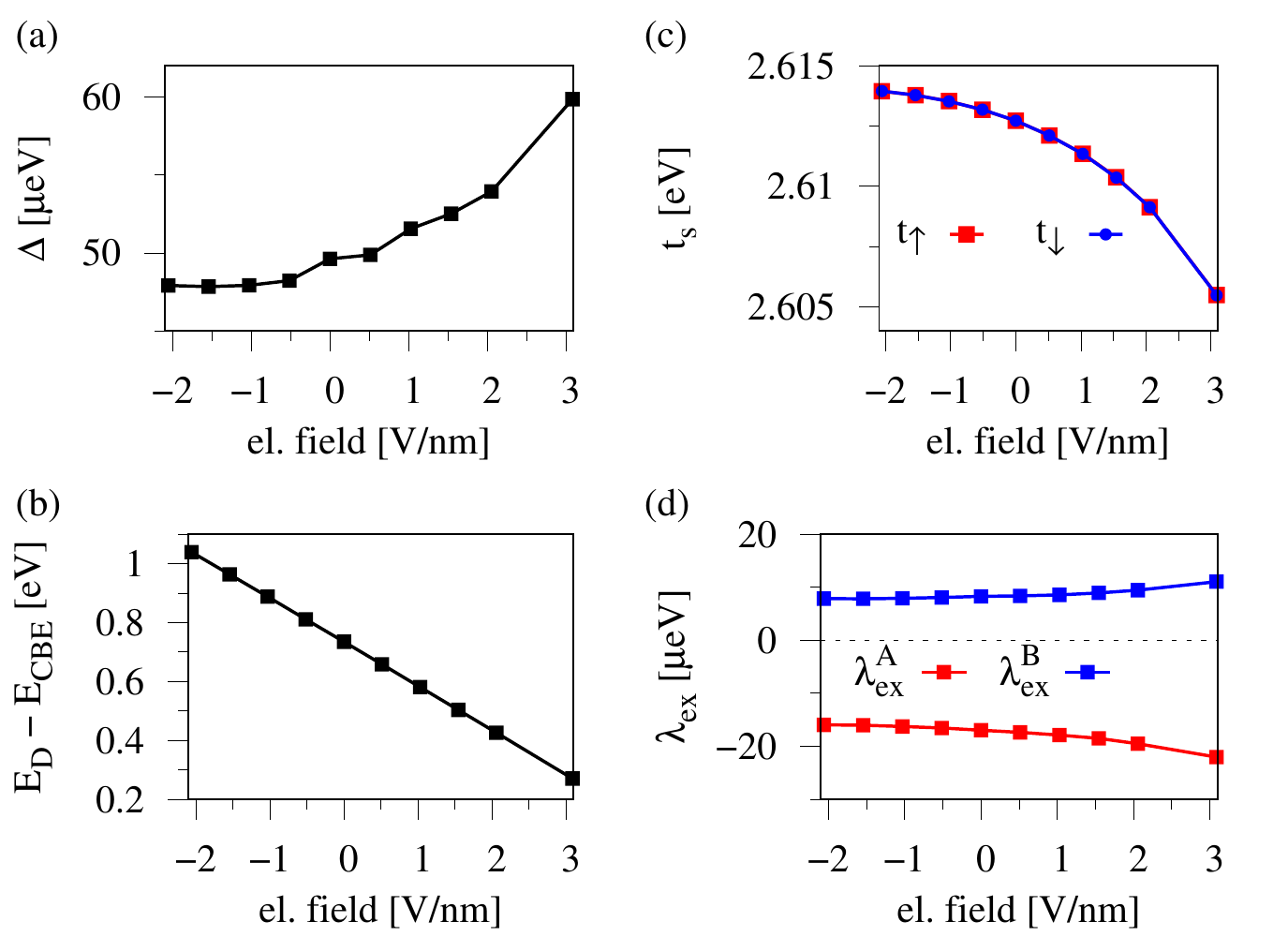}
     \caption{  Transverse electric field  dependence of (a) the staggered potential $\Delta$, (b) the band offset of the Dirac point, $\mathrm{E_D}$, with respect to the MnPS$_3$ conduction band edge, $\mathrm{E_{CBE}}$, at the K point, (c) the spin dependent hopping parameters t$_{\uparrow}$ and t$_{\downarrow}$, and (d) the sublattice resolved proximity exchange parameters $\lambda_{\textrm{ex}}^\textrm{A}$ and $\lambda_{\textrm{ex}}^\textrm{B}$ for the graphene/MnPS$_3$ heterostructure  when the MnPS$_3$ is in the antiferromagnetic N{\'e}el phase.
     }\label{Fig:Efield_AFM_Neel_MnPS3}
    \end{figure}

In experiment, gating is a tool to control proximity-induced exchange coupling as well as the doping level. 
How does a transverse electric field affect the Dirac bands in our scenario?
Similar to before, we consider graphene on MnPS$_3$ in the antiferromagnetic N{\'e}el phase and apply an electric field across the heterostructure. 
The results are summarized in Fig.~\ref{Fig:Efield_AFM_Neel_MnPS3}. 
Within our calculated field range of $-2$ to 3~V/nm, we find that the fitted parameters barely change. 
Most important, we can strongly tune the band offset of the Dirac point, $\mathrm{E_D}$, with respect to the MnPS$_3$ conduction band edge, $\mathrm{E_{CBE}}$, at the K point, see Fig.~\ref{Fig:Efield_AFM_Neel_MnPS3}(b).
For a field of 3~V/nm, we can bring the Dirac point at about 0.25~eV below the conduction band edge of the MnPS$_3$. 
In addition, it seems that the closer the Dirac cone gets to the conduction band edge, staggered potential and exchange couplings increase, while the hopping amplitudes decrease.
Even though, the proximity exchange couplings are small in value, their relative tunability is giant, since they can be enhanced by about 30\% by tuning the field from $-2$ to 3~V/nm.
In general, the tunability should hold also for other magnetic phases, where proximity exchange is stronger.

\subsection{Effects of SOC}

So far, we have not included SOC in the calculations.
Since the number of possible heterostructures in different magnetic phases is too large, we limit ourselves to a few selected cases and check how the inclusion of SOC influences the low energy graphene Dirac bands. 
For simplicity, we only consider Mn-based MPX$_3$ monolayers in the ferromagnetic phase.
This should be enough to find out about the magnitude of the proximity-induced SOC, which we can compare to the induced exchange coupling. 
In addition, when SOC is included, we employ constrained magnetization calculations (forcing the magnetization parallel to $z$ direction), since {\tt Quantum ESPRESSO} implements noncollinear magnetism.
Note, that the plane wave and pseudopotential method, implemented in {\tt Quantum ESPRESSO}, does not correctly reproduce the intrinsic SOC values of bare monolayer graphene (12~$\mu$eV), since the relevant $d$-orbitals are not present \cite{Gmitra2009:PRB,Konschuh2010:PRB}. However, proximity-induced SOC (if present) is captured.

    \begin{figure}[htb]
     \includegraphics[width=.99\columnwidth]{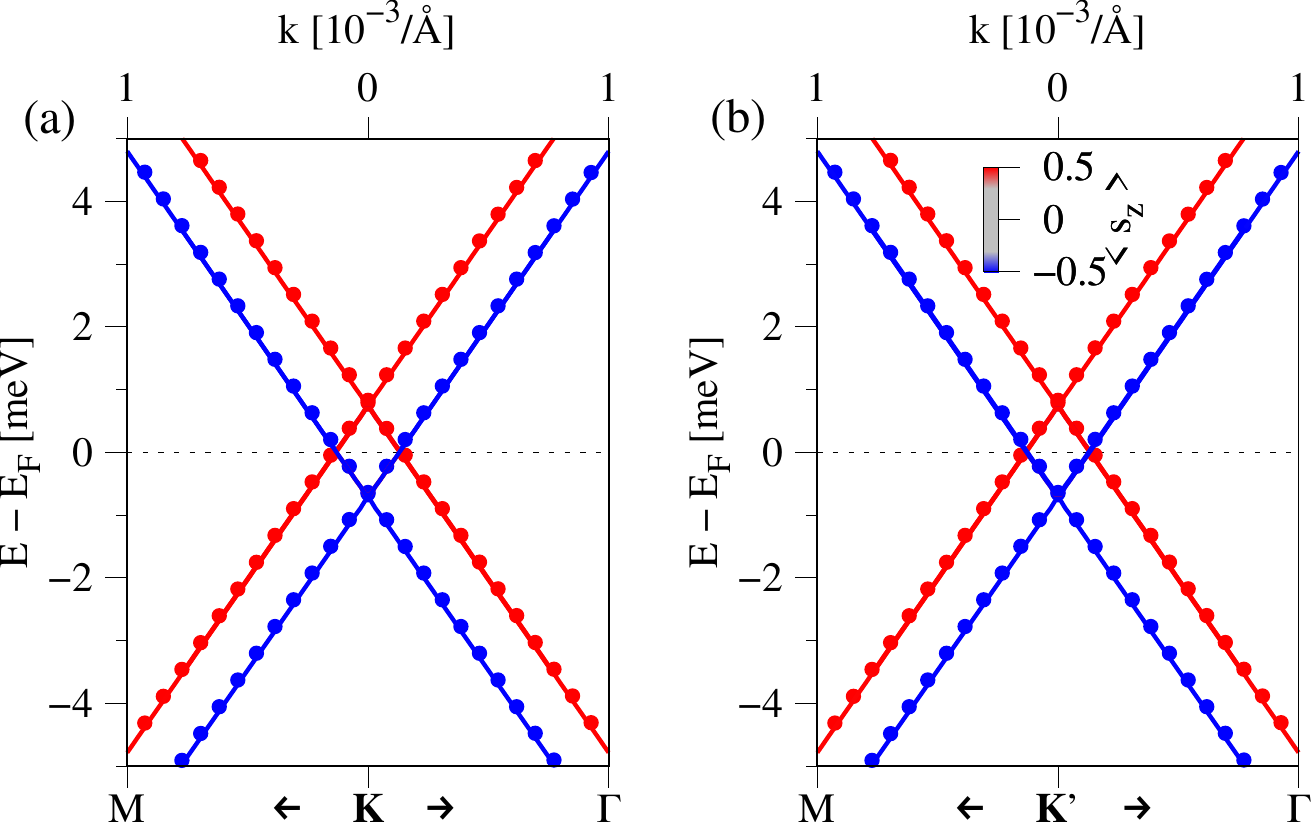} 
     \caption{  (a) DFT-calculated low energy Dirac bands near the K point for the graphene/MnPS$_3$ heterostructure, when MnPS$_3$ is in the ferromagnetic phase and SOC is included. The symbols are color coded by their $s_z$ spin expectation value. (b) Same as (a) but near the K$^\prime$ point. 
     }\label{Fig:MnPS3_FM_SOC}
    \end{figure}

In Fig.~\ref{Fig:MnPS3_FM_SOC}, we show the DFT-calculated low energy bands near the K and K$^\prime$ points of the graphene/MnPS$_3$ heterostructure, when MnPS$_3$ is in the ferromagnetic phase and SOC is included. 
We find that bands near the K and K$^\prime$ valleys are nearly identical and fully $s_z$ polarized, indicating that proximity-induced SOC is negligible compared to the exchange one. Indeed, comparing the calculated low energy bands from Fig.~\ref{Fig:MnPS3_FM_SOC} (with SOC) and Fig.~\ref{Fig:bands_FM_MnPS3} (without SOC), we find no marked change by inclusion of SOC. 
This result is not surprising, because recent calculations have shown that SOC effects are small in bare monolayer MnPS$_3$ in the antiferromagnetic N{\'e}el phase \cite{Yang2020:RSC}. Therefore, also proximity-induced SOC is small, even though we consider the ferromagnetic phase here. 
Similar results hold for MnPSe$_3$ in the ferromagnetic phase (low-energy dispersions with SOC are not shown).

\subsection{Further Discussion}

In some cases, band hybridization (anticrossings) additionally affect the Dirac bands, leading to renormalized proximity-induced exchange splittings, as previously shown for graphene/hBN/Co heterostructures \cite{Zollner2016:PRB}. Similar things can happen here, for example in the case of graphene/NiPSe$_3$ in the ferromagnetic phase, spin down bands originating from the NiPSe$_3$ are located near the Dirac point energy and hybridize with the spin down Dirac states, see the band structure in the Supplemental Material \footnotemark[3]. 
As a consequence, the effective proximity exchange parameters are much larger (10~meV) than for any other case we consider here. Also the spin down hopping parameter t$_{\downarrow}$ differs by about 0.5~eV from the spin up one t$_{\uparrow}$, see Table~\ref{tab:fitresults_noSOC}.
For the graphene/NiPS$_3$ structure in the antiferromagnetic zigzag phase, the band hybridization is even more pronounced, since NiPS$_3$ bands from both spin species are present near the Dirac point. 

Whenever bands, which originate from the magnetic MPX$_3$ substrate, are located near the Dirac point energy, band hybridization can occur. As a consequence, the spin splitting of the graphene Dirac states is affected, in addition to the bare proximity-induced exchange coupling.
By comparing the band structures and fitted parameters from all of our structures, this happens only in a few cases we consider here. 
However, from previous studies \cite{Zollner2016:PRB, Zollner2019b:PRB}, we know that band offsets can be influenced by the interlayer distance, a transverse electric field, or the Hubbard $U$. 
In our exemplary case of MnPS$_3$ in the antiferromagnetic N{\'e}el phase, we have seen that the interlayer distance strongly affects proximity exchange, while the Hubbard $U$ does not, since band hybridization effects are absent (given the large band offset). This behavior can be very different in other heterostructures, especially in cases where band hybridization already occurs.

Does another DFT code give similar results?
We additionally employ the full-potential linearized augmented plane wave code {\tt WIEN2k} \cite{Wien2k} and recalculate the dispersion for some of our structures. The fit results are also summarized in Table~\ref{tab:fitresults_noSOC}.
We consider graphene on MnPS$_3$ and MnPSe$_3$ in the ferromagnetic and antiferromagnetic N{\'e}el phases, respectively. Because of the broken symmetry in the antiferromagnetic zigzag and stripy phases, {\tt WIEN2k} calculations are computationally too demanding and thus not included in this study. 
The purpose to employ two different codes is to test the robustness of our predictions (magnitudes and signs of proximity exchange parameters).

Both codes predict the antiferromagnetic N{\'e}el phase to be the ground state for MnPSe$_3$ and MnPS$_3$, in agreement with experimental data. 
For both substrates in both magnetic phases, {\tt WIEN2k} predicts similar proximity exchange splittings as {\tt Quantum ESPRESSO}.
For MnPSe$_3$, results of the antiferromagnetic N{\'e}el phase differ. The different DFT codes predict small exchange parameters, but of different type ({\tt WIEN2k} predicts staggered ones, while {\tt Quantum ESPRESSO} predicts uniform ones). 
For MnPS$_3$ in the antiferromagnetic N{\'e}el phase, the sublattice resolved proximity exchange parameters are staggered for both codes, but with a global sign change. 
The results for the ferromagnetic phases coincide for both materials and both codes. 
Indeed, when proximity exchange is larger than about 100 $\mu$eV, both codes essentially agree, as we can see from the ferromagnetic phase results. However, when the proximity effects become rather small (few to tens of $\mu$eV), the codes reach their limits in numerical accuracy and discrepancies may arise, as we can see from the antiferromagnetic N{\'e}el phase results.
We conclude that, whenever exchange parameters are small (few to tens of $\mu$eV), a precise prediction of results is computationally very challenging and the extracted parameters should be treated with caution.
However, we do not see this as a critical problem, since the magnitudes of the proximity exchange in the different magnetic phases are the same, independent of the DFT code. The proximity effects depend on the hybridization of the orbitals across the interface, and on the energies of the orbitals that take part of the hybridizations. The energetics of higher lying orbitals depends on the code, which is most likely causing the discrepancies.

Finally, we want to comment on the magnetic easy axis of the substrate MPX$_3$ materials. In our calculations, we have considered only out-of-plane (collinear to $z$ direction) magnetic configurations. 
Otherwise, the calculations would be computationally very demanding since noncollinear magnetism needs to be treated, which is accompanied by inclusion of SOC.  
However, for some MPX$_3$ monolayers, experiments and theory indicate in-plane antiferromagnetism, such as for NiPS$_3$ \cite{Lancon2018:PRB,Olsen2021:JPD}. 
Therefore, one should keep in mind that our presented results of proximity exchange may be different in real samples. 
With our DFT approach, we can give at most rough predictions and indications. 
Nevertheless, we find that the class of magnetic MPX$_3$ monolayers is highly interesting in terms of proximity effects, as different types and magnitudes of exchange coupling can be induced in graphene. We also hope that future experiments can confirm at least some of our findings and that the potential of these materials will be recognized for future spintronics applications.

\section{Summary}
\label{Sec:Summary}
In this work, we have considered heterostructures of graphene and the magnetic transition-metal phosphorus trichalcogenides MPX$_3$. From first-principles calculations, we have extracted the proximitized Dirac bands of graphene, which we fitted to a model Hamiltonian, including orbital and proximity exchange parameters. We found that, depending on the magnetic phase of the MPX$_3$ substrate layer, the metal atom M=\{Mn,Fe,Ni,Co\}, and the chalcogen atom X=\{S,Se\}, very different proximity-induced exchange couplings can be realized in graphene. Not only that the exchange splittings range from about 0 to 10~meV, also the sign and type of the exchange couplings differ from case to case. We demonstrated, that the MPX$_3$ family of magnetic monolayers are a platform to induce uniform ($\lambda_{\textrm{ex}}^\textrm{A} \approx \lambda_{\textrm{ex}}^\textrm{B}$), as well as staggered ($\lambda_{\textrm{ex}}^\textrm{A} \approx -\lambda_{\textrm{ex}}^\textrm{B}$) exchange couplings in graphene, leading to very different spin polarizations.
In addition, decreasing the interlayer distance between graphene and the substrate is a very efficient tool to enhance proximity-induced exchange couplings. An applied transverse electric field can potentially result in a similar tunability, but was not observed in our exemplary case. The Hubbard $U$, which is a parameter in the DFT calculation, can have a huge impact on Dirac band splittings, since the $U$ determines the band alignment and to some extent the coupling of the monolayers. Finally, we found that for Mn-based materials in the ferromagnetic phase, the proximity-induced SOC is negligible compared to the exchange parameters. 

\acknowledgments

This work was funded by the Deutsche Forschungsgemeinschaft (DFG, German Research Foundation) SFB 1277 (Project No. 314695032), SPP 2244 (Project No. 443416183), and the European Union Horizon 2020 Research and Innovation Program under contract number 881603 (Graphene Flagship).
    
    \footnotetext[3]{See Supplemental Material, where we show band structures, density of states, and fit results for the remaining graphene/MPX$_3$ heterostructures. For the MnPS$_3$ case we also show the calculated spin polarizations on graphene. }

\bibliography{references}

\cleardoublepage


\section*{Supplementary material}

# Supplemental Material: Proximity effects in graphene on monolayers of transition-metal phosphorus trichalcogenides MPX₃

Klaus Zollner[1, *] and Jaroslav Fabian[1]

[1]*Institute for Theoretical Physics, University of Regensburg, 93040 Regensburg, Germany*

In the Supplemental Material we show the band structures, density of states, and fit results for the remaining heterostructures not shown in the main text. More specifically we show additional results for graphene on $MnPS_3$, $NiPS_3$, $FePS_3$, $CoPS_3$, $MnPSe_3$, $NiPSe_3$, $FePSe_3$, and $CoPSe_3$, where each $MPX_3$ monolayer can be in the ferromagnetic, the antiferromagnetic Néel, the antiferromagnetic zigzag, and the antiferromagnetic stripy phase.

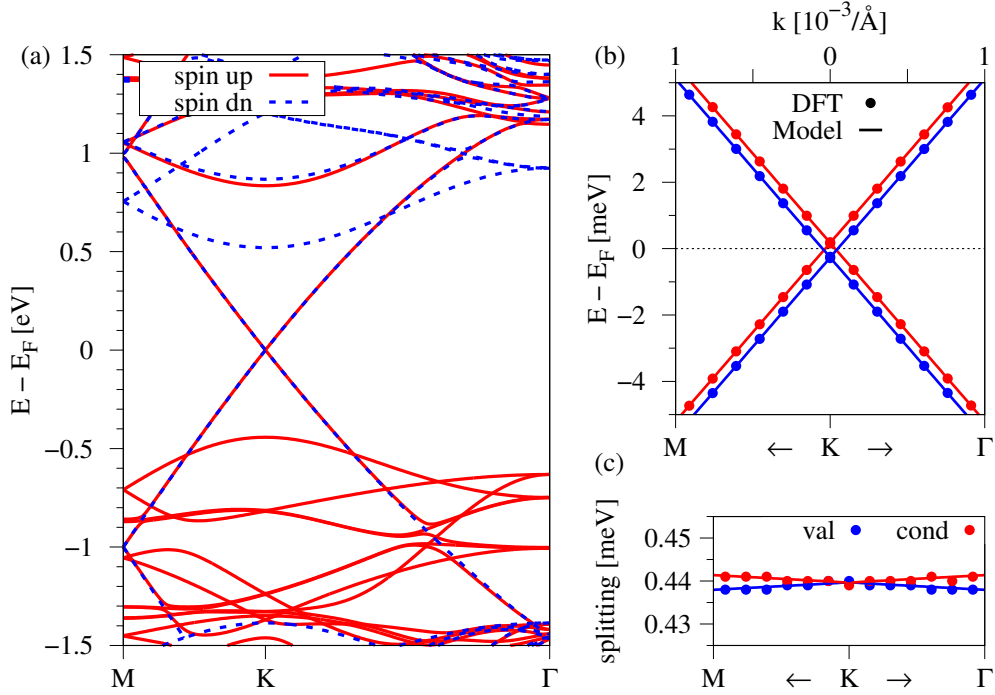

FIG. S1. (a) DFT-calculated band structure of the graphene/$MnPSe_3$ heterostructure along the high-symmetry path M-K-$\Gamma$ without SOC, when the $MnPSe_3$ is in the ferromagnetic phase. Red solid (blue dashed) lines correspond to spin up (spin down). (b) Zoom to the DFT-calculated (symbols) low energy Dirac bands near the K point with a fit to the model Hamiltonian (solid line). (c) The splitting of valence and conduction band.

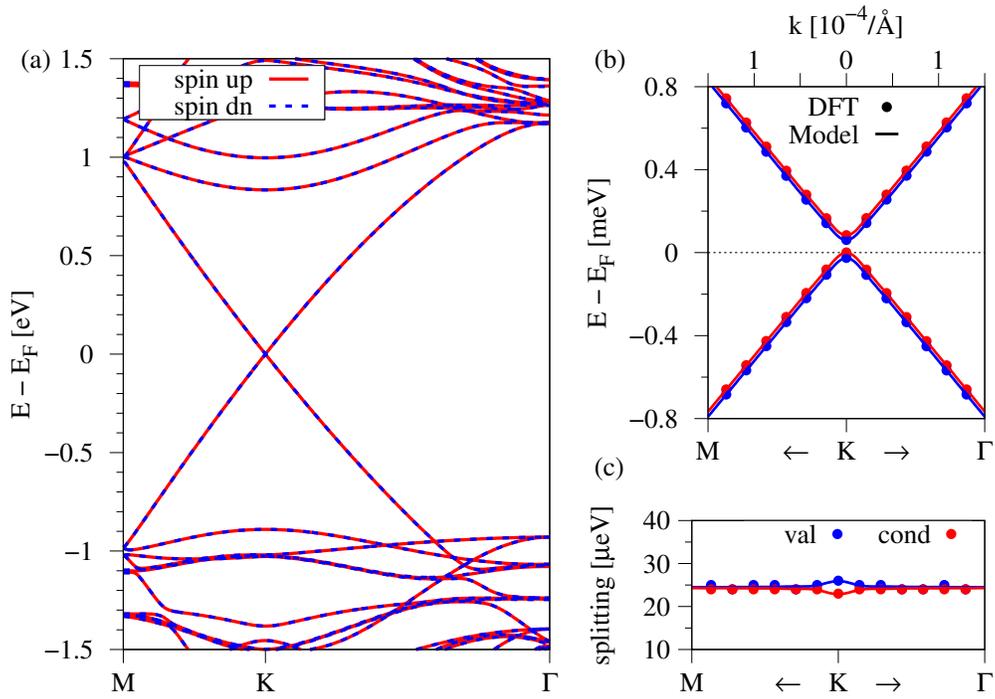

FIG. S2. (a) DFT-calculated band structure of the graphene/MnPSe$_3$ heterostructure along the high-symmetry path M-K-Γ without SOC, when the MnPSe$_3$ is in the antiferromagnetic Néel phase. Red solid (blue dashed) lines correspond to spin up (spin down). (b) Zoom to the DFT-calculated (symbols) low energy Dirac bands near the K point with a fit to the model Hamiltonian (solid line). (c) The splitting of valence and conduction band.

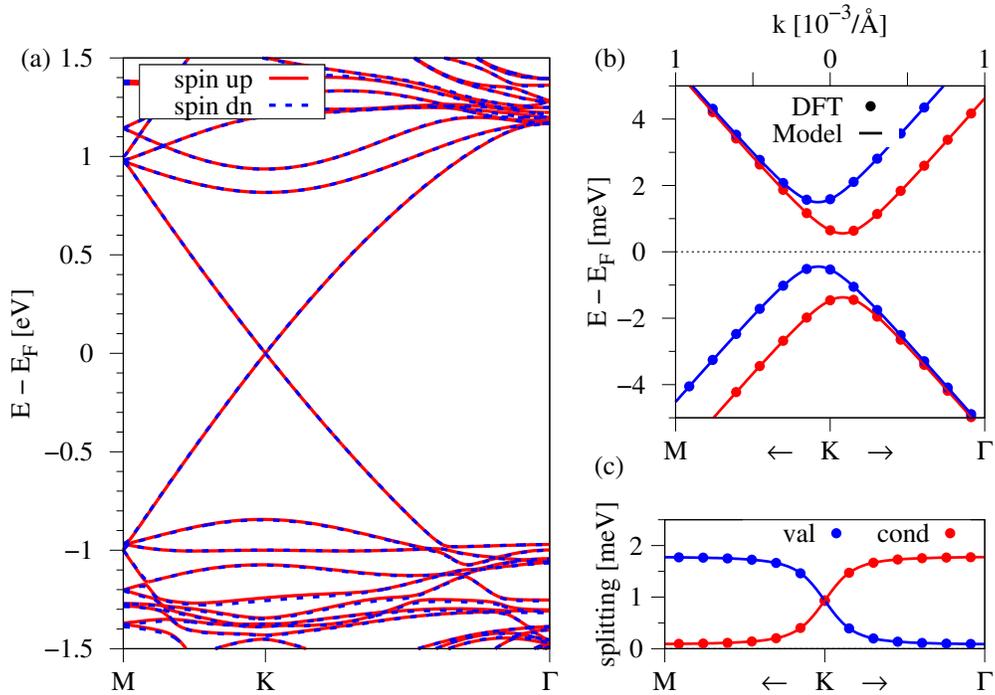

FIG. S3. (a) DFT-calculated band structure of the graphene/MnPSe$_3$ heterostructure along the high-symmetry path M-K-Γ without SOC, when the MnPSe$_3$ is in the antiferromagnetic zigzag phase. Red solid (blue dashed) lines correspond to spin up (spin down). (b) Zoom to the DFT-calculated (symbols) low energy Dirac bands near the K point with a fit to the model Hamiltonian (solid line). (c) The splitting of valence and conduction band.

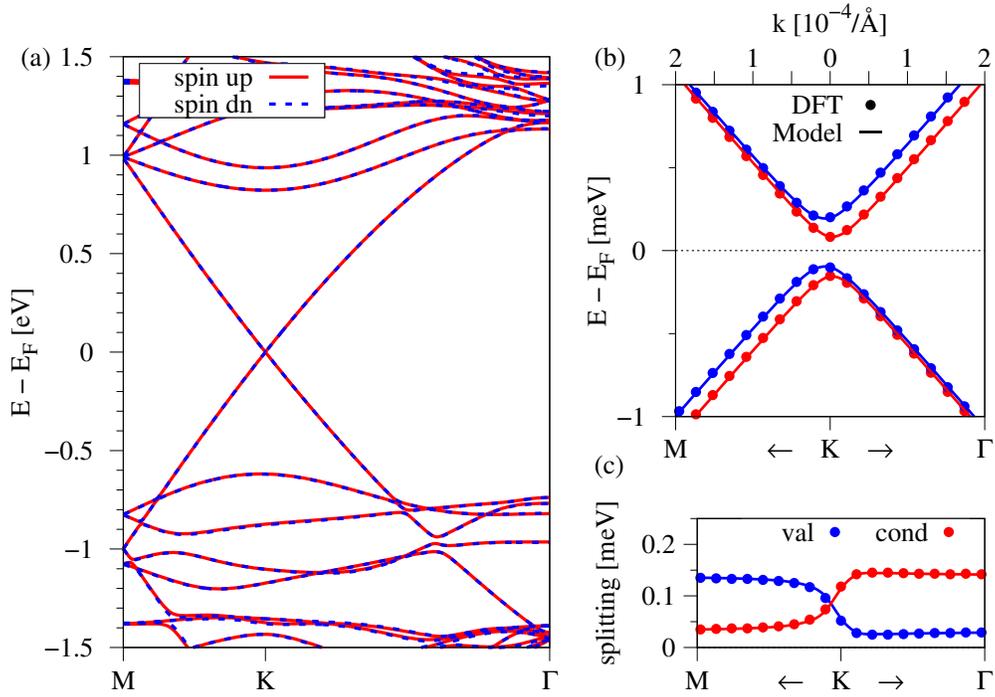

FIG. S4. (a) DFT-calculated band structure of the graphene/MnPSe$_3$ heterostructure along the high-symmetry path M-K-Γ without SOC, when the MnPSe$_3$ is in the antiferromagnetic stripy phase. Red solid (blue dashed) lines correspond to spin up (spin down). (b) Zoom to the DFT-calculated (symbols) low energy Dirac bands near the K point with a fit to the model Hamiltonian (solid line). (c) The splitting of valence and conduction band.

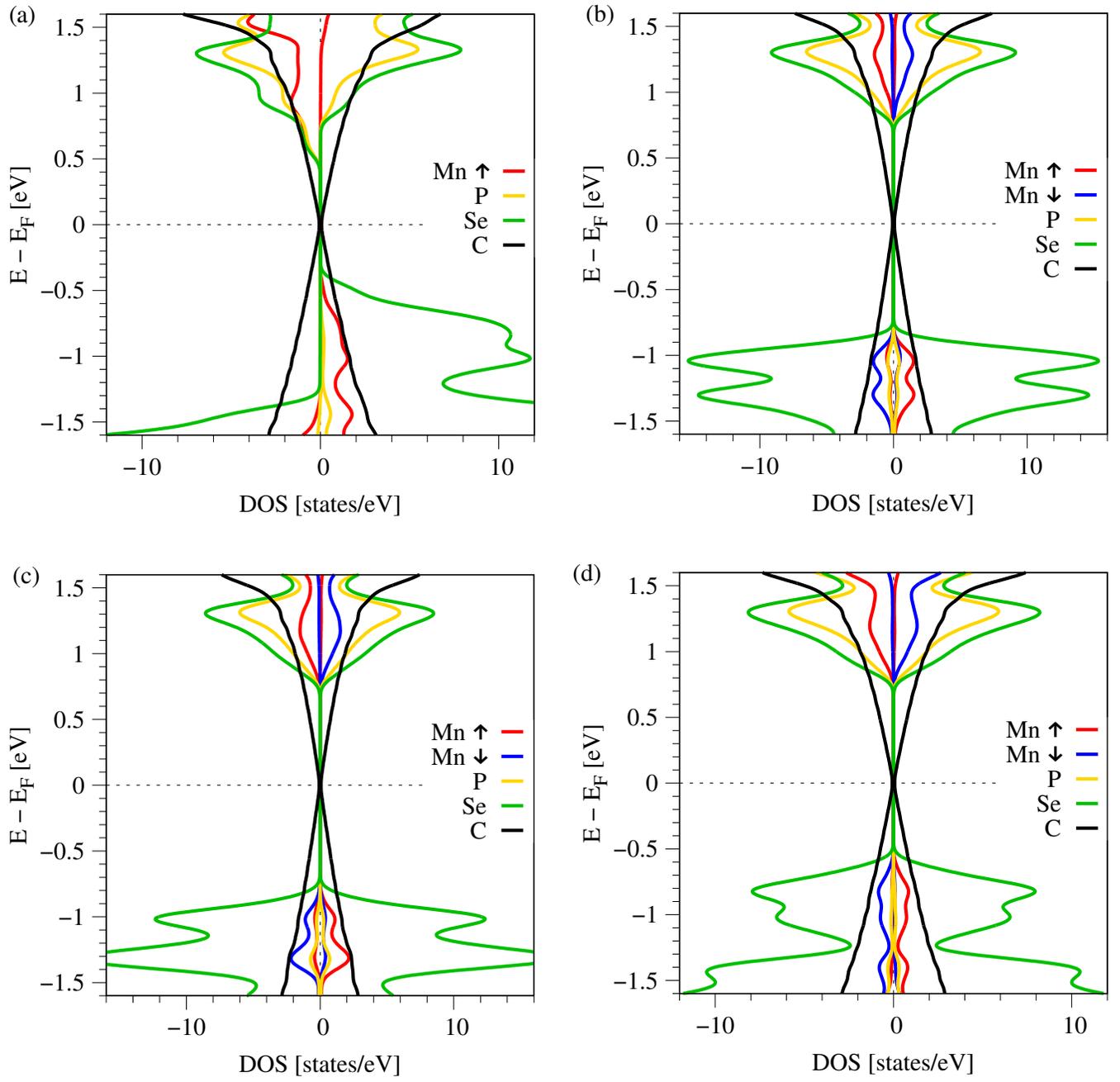

FIG. S5. Spin and atom resolved Density of States (DOS) for graphene on MnPSe₃ when MnPSe₃ is in the (a) ferromagnetic, (b) antiferromagnetic Néel, (c) antiferromagnetic zigzag, or (d) antiferromagnetic stripy phase.

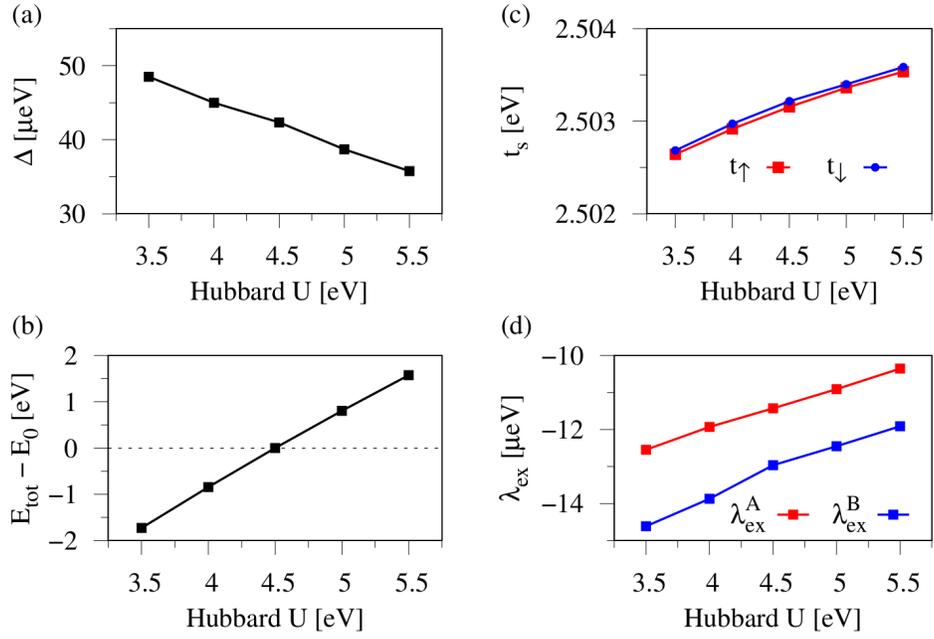

FIG. S6. Hubbard $U$ dependence of (a) the staggered potential $\Delta$, (b) the total energy $E_{tot}$, (c) the spin dependent hopping parameters $t_\uparrow$ and $t_\downarrow$, and (d) the sublattice resolved proximity exchange parameters $\lambda_{ex}^A$ and $\lambda_{ex}^B$ for the graphene/MnPSe₃ heterostructure when the MnPSe₃ is in the antiferromagnetic Néel phase.

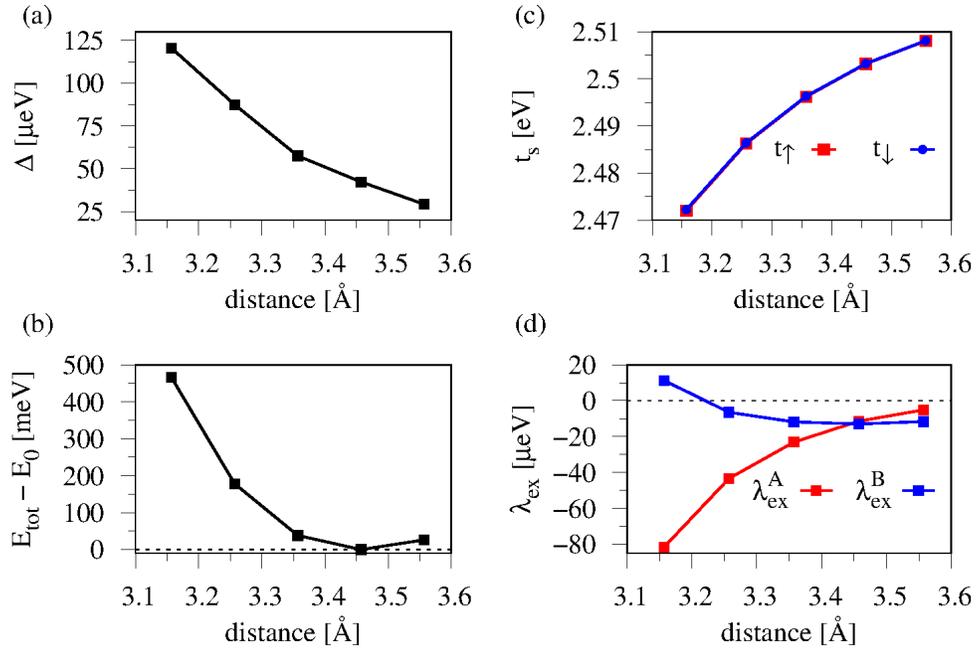

FIG. S7. Interlayer distance dependence of (a) the staggered potential $\Delta$, (b) the total energy $E_{tot}$, (c) the spin dependent hopping parameters $t_\uparrow$ and $t_\downarrow$, and (d) the sublattice resolved proximity exchange parameters $\lambda_{ex}^A$ and $\lambda_{ex}^B$ for the graphene/MnPSe₃ heterostructure when the MnPSe₃ is in the antiferromagnetic Néel phase.

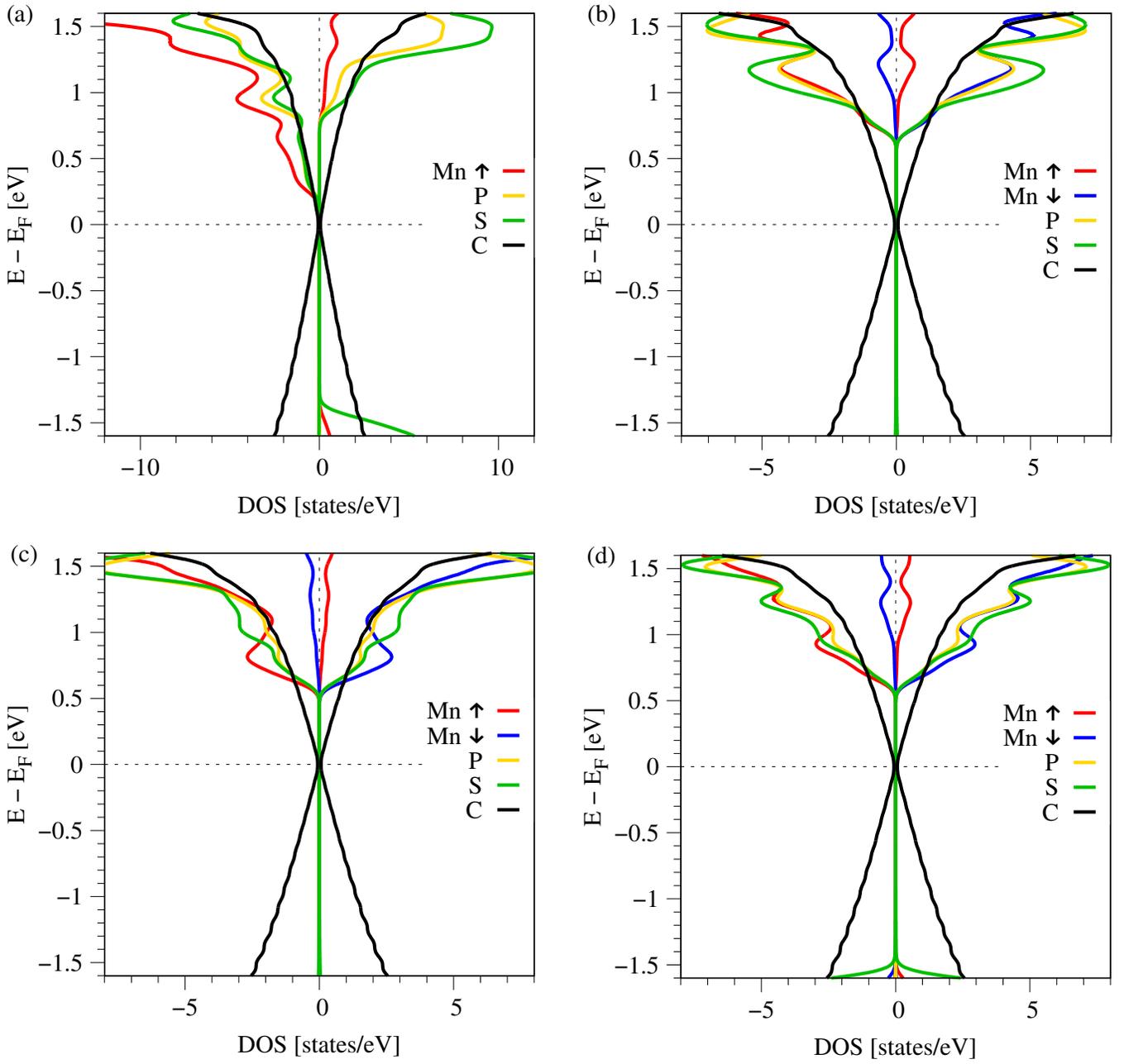

FIG. S8. Spin and atom resolved Density of States (DOS) for graphene on MnPS$_3$ when MnPS$_3$ is in the (a) ferromagnetic, (b) antiferromagnetic Néel, (c) antiferromagnetic zigzag, or (d) antiferromagnetic stripy phase.

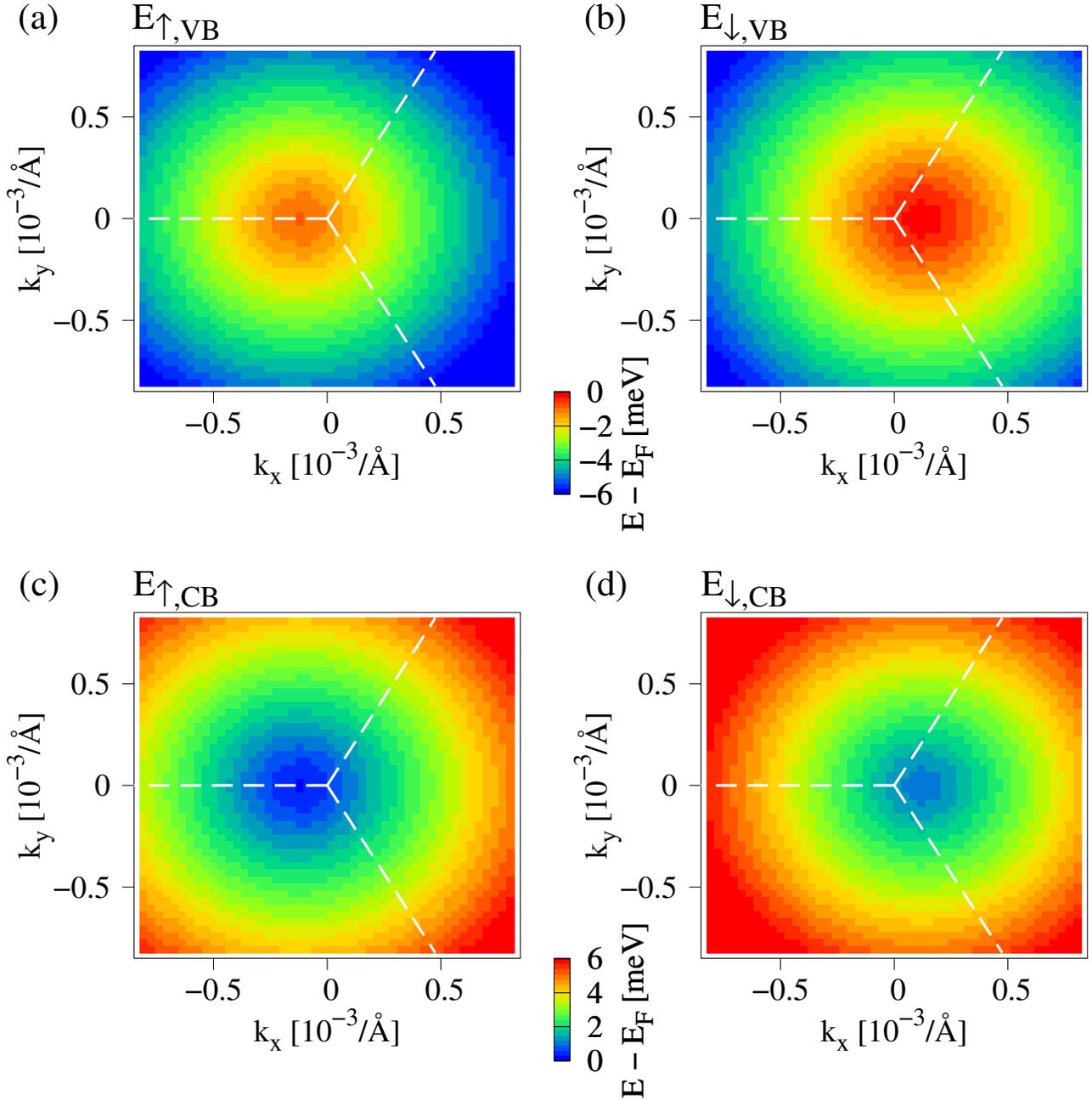

FIG. S9. 2D-map of the DFT-calculated low energy Dirac bands of the graphene/MnPS$_3$ heterostructure without SOC, when the MnPS$_3$ is in the antiferromagnetic zigzag phase. (a,b) The valence band energies for spin up E$_{\uparrow,VB}$ and down E$_{\downarrow,VB}$. (c,d) Same as (a,b) but for the conduction band. The dashed white lines mark the edges of the Brillouin zone with the K point at the center.

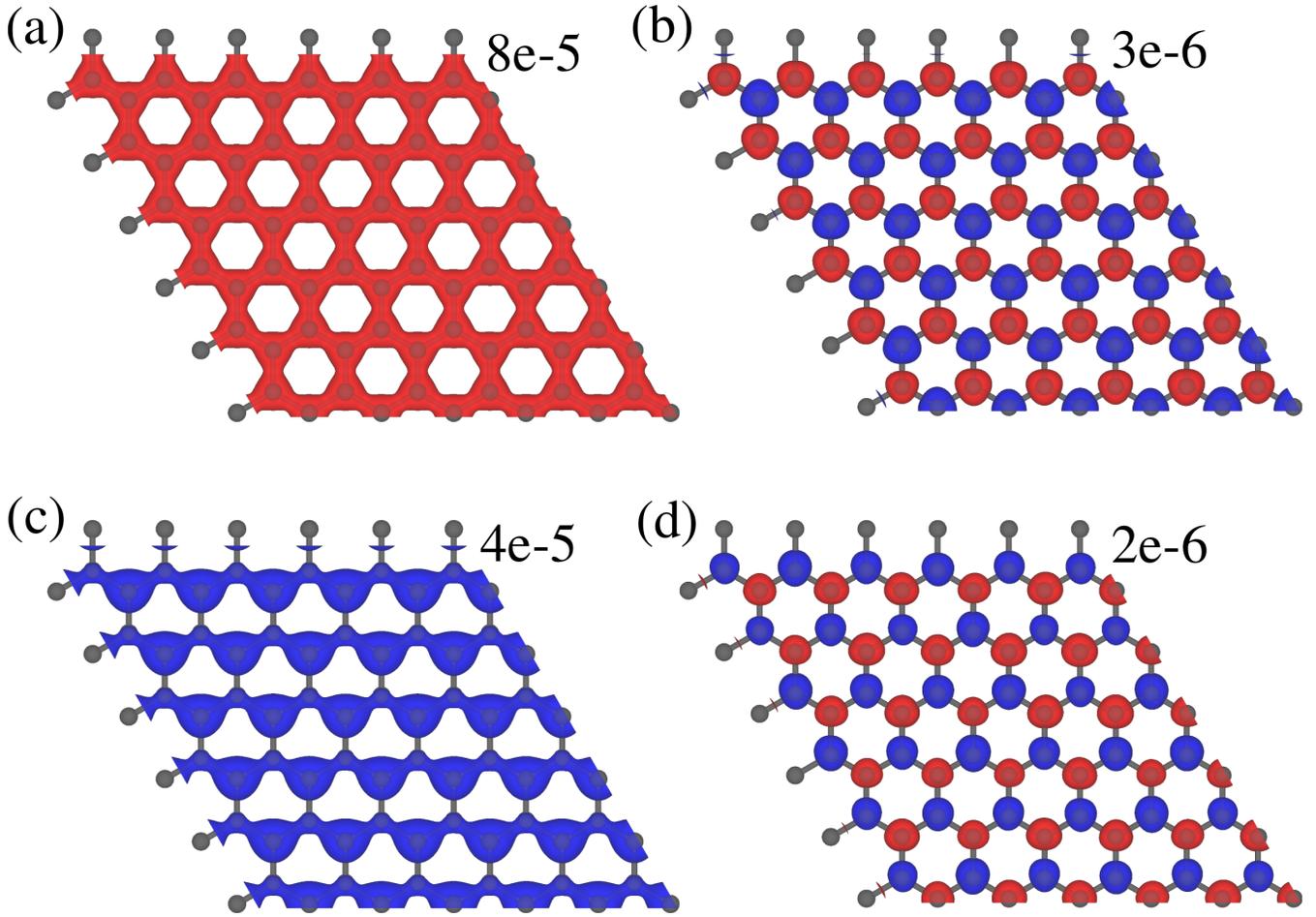

FIG. S10. Calculated spin polarizations, $\Delta\rho = \rho_\uparrow - \rho_\downarrow$, for graphene on MnPS$_3$ when MnPS$_3$ is in the (a) ferromagnetic, (b) antiferromagnetic Néel, (c) antiferromagnetic zigzag, or (d) antiferromagnetic stripy phase. For better visualization of the spin polarizations on graphene, we have deleted the MnPS$_3$ substrate in the figures. The spin densities $\rho_{\uparrow/\downarrow}(\boldsymbol{r}) = \sum_{n,\boldsymbol{k}} |\phi_{n,\uparrow/\downarrow}^{\boldsymbol{k}}(\boldsymbol{r})|^2$ are sums over eigenstates with the corresponding spin. Here, we take into account Dirac states in the energy window of $-6$ to $-4$ meV below the Fermi level. The color red (blue) corresponds to $\Delta\rho > 0$ ($\Delta\rho < 0$). The isosurfaces correspond to isovalues, in units $\text{Å}^{-3}$, as indicated next to each subfigure.

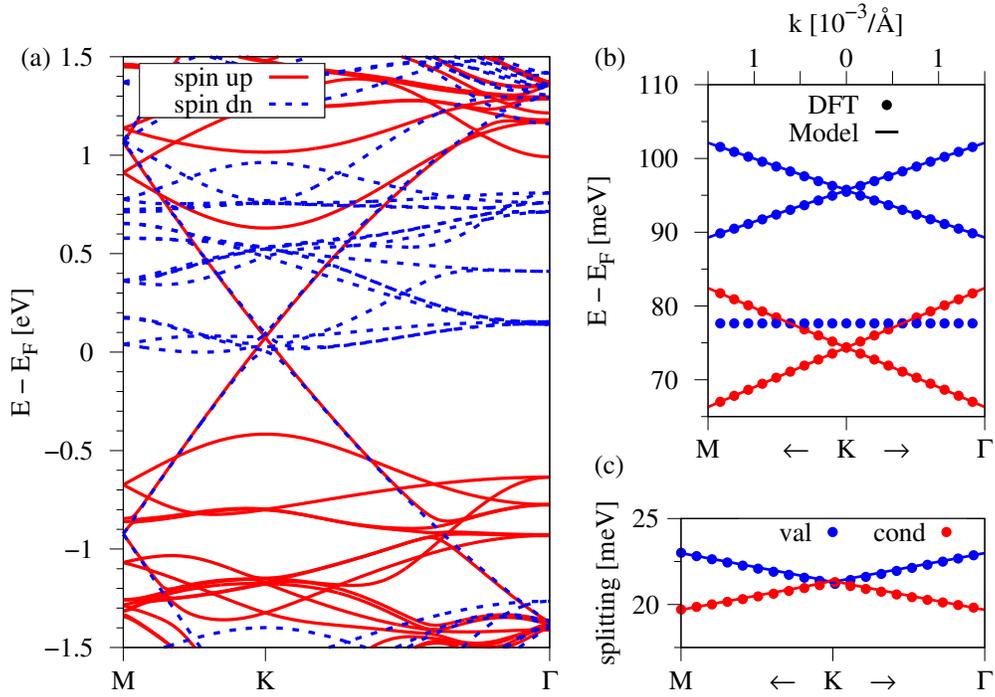

FIG. S11. (a) DFT-calculated band structure of the graphene/NiPSe₃ heterostructure along the high-symmetry path M-K-Γ without SOC, when the NiPSe₃ is in the ferromagnetic phase. Red solid (blue dashed) lines correspond to spin up (spin down). (b) Zoom to the DFT-calculated (symbols) low energy Dirac bands near the K point with a fit to the model Hamiltonian (solid line). (c) The splitting of valence and conduction band.

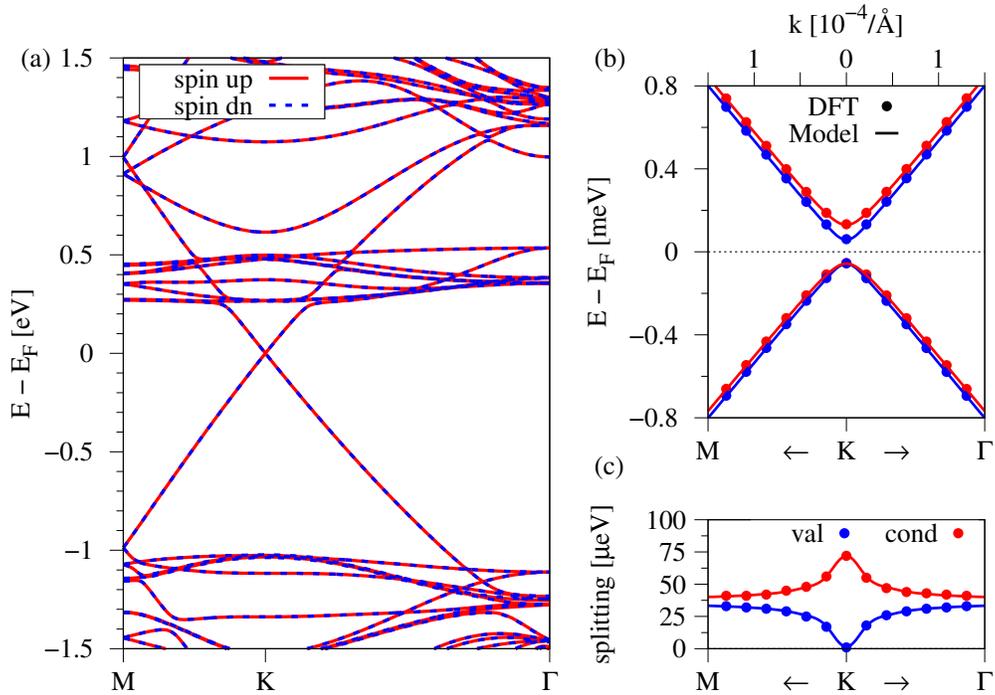

FIG. S12. (a) DFT-calculated band structure of the graphene/NiPSe₃ heterostructure along the high-symmetry path M-K-Γ without SOC, when the NiPSe₃ is in the antiferromagnetic Néel phase. Red solid (blue dashed) lines correspond to spin up (spin down). (b) Zoom to the DFT-calculated (symbols) low energy Dirac bands near the K point with a fit to the model Hamiltonian (solid line). (c) The splitting of valence and conduction band.

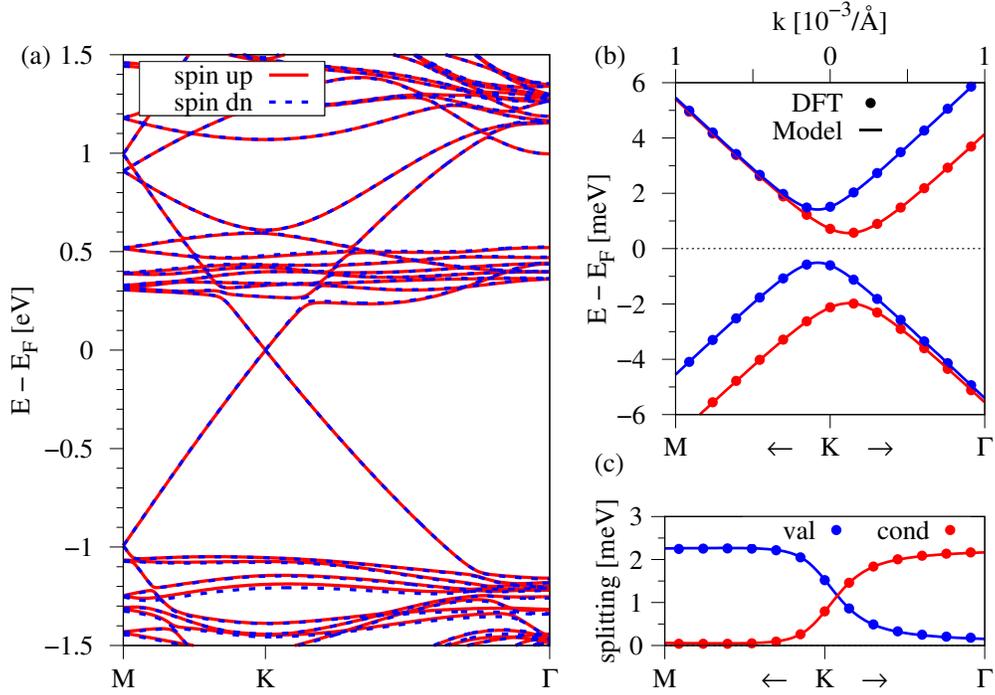

FIG. S13. (a) DFT-calculated band structure of the graphene/NiPSe₃ heterostructure along the high-symmetry path M-K-Γ without SOC, when the NiPSe₃ is in the antiferromagnetic zigzag phase. Red solid (blue dashed) lines correspond to spin up (spin down). (b) Zoom to the DFT-calculated (symbols) low energy Dirac bands near the K point with a fit to the model Hamiltonian (solid line). (c) The splitting of valence and conduction band.

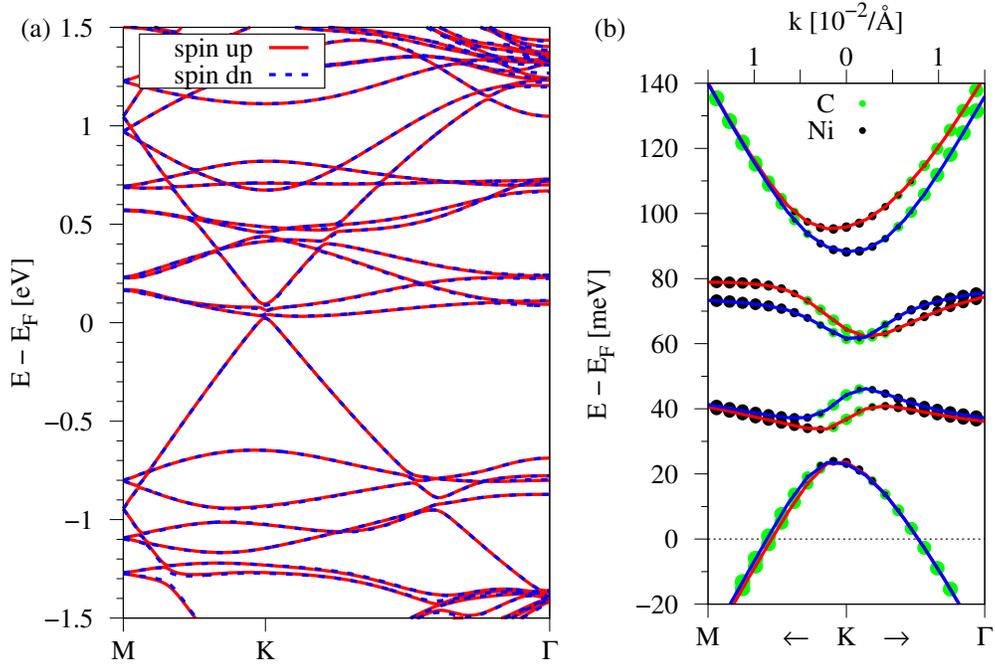

FIG. S14. (a) DFT-calculated band structure of the graphene/NiPSe₃ heterostructure along the high-symmetry path M-K-Γ without SOC, when the NiPSe₃ is in the antiferromagnetic stripy phase. Red solid (blue dashed) lines correspond to spin up (spin down). (b) Zoom to the DFT-calculated low energy Dirac bands near the K point (red and blue solid lines for spin up and down). The green and black spheres represent the projections onto C and Ni states. This emphasizes that the graphene Dirac states are strongly hybridized with Ni states from the NiPSe₃ and we cannot apply our model fit properly.

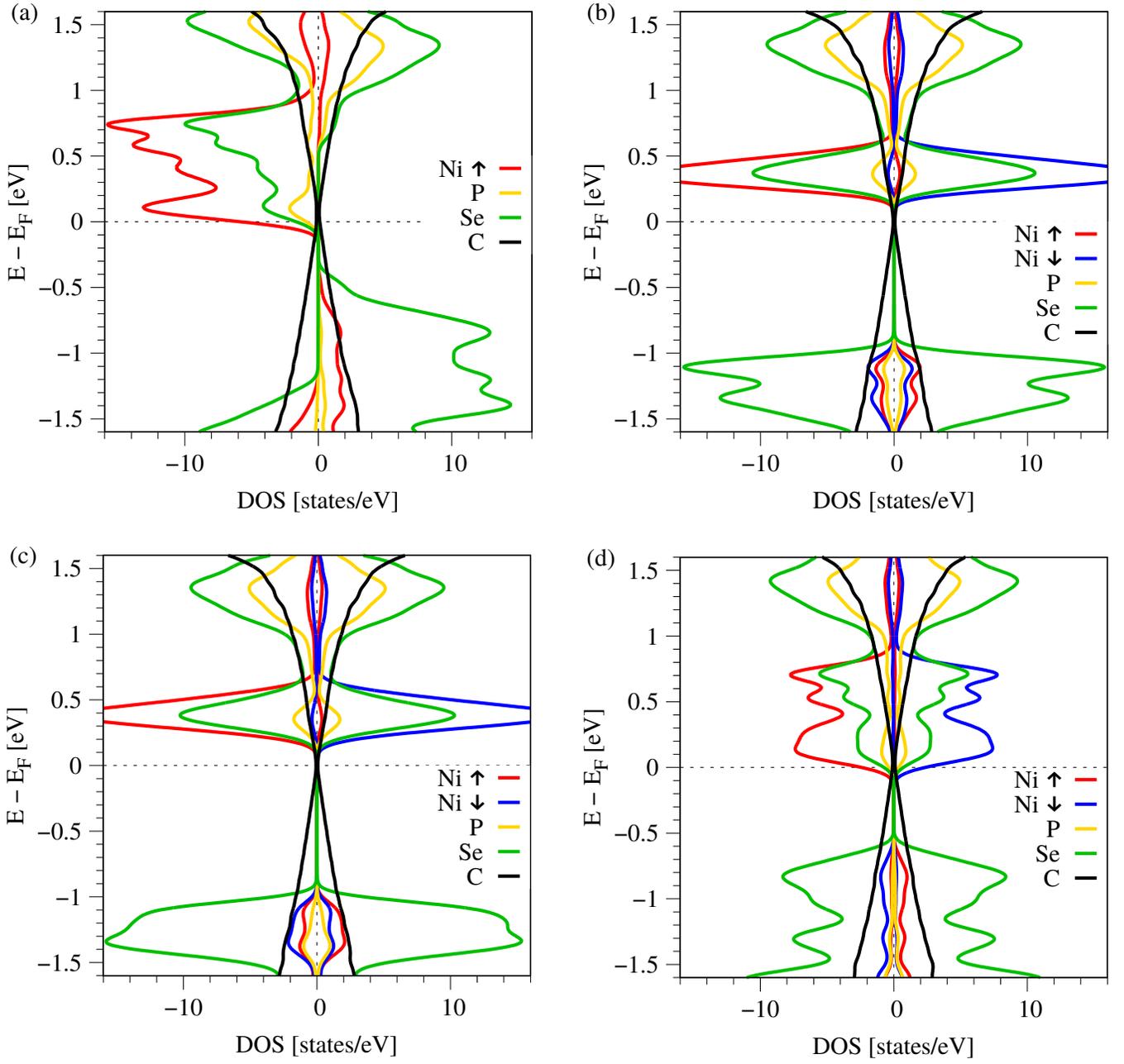

FIG. S15. Spin and atom resolved Density of States (DOS) for graphene on NiPSe₃ when NiPSe₃ is in the (a) ferromagnetic, (b) antiferromagnetic Néel, (c) antiferromagnetic zigzag, or (d) antiferromagnetic stripy phase.

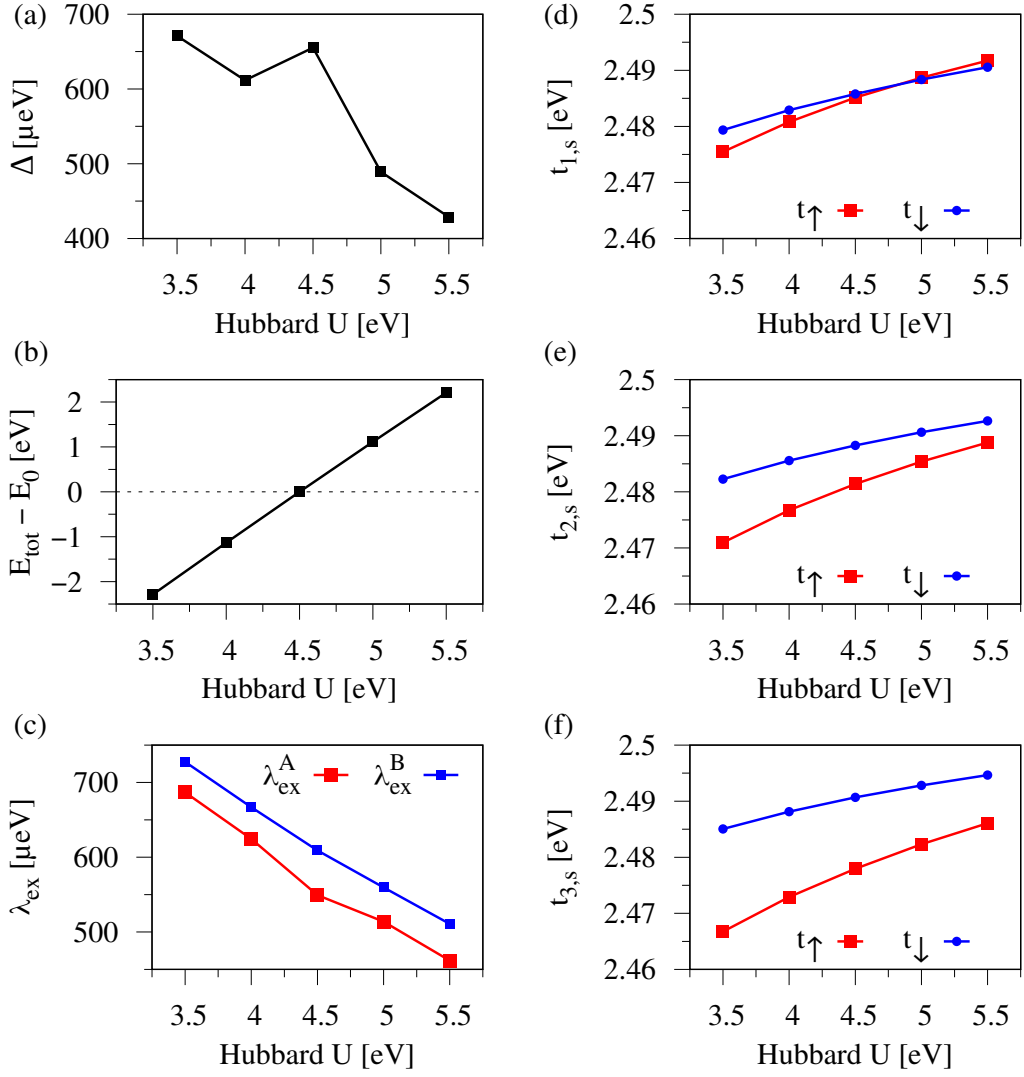

FIG. S16. Hubbard $U$ dependence of (a) the staggered potential $\Delta$, (b) the total energy $E_{tot}$, (c) the sublattice resolved proximity exchange parameters $\lambda_{ex}^A$ and $\lambda_{ex}^B$, and (d-f) the spin and direction dependent hopping parameters $t_{\alpha,\uparrow}$ and $t_{\alpha,\downarrow}$, for the graphene/NiPSe$_3$ heterostructure when the NiPSe$_3$ is in the antiferromagnetic zigzag phase.

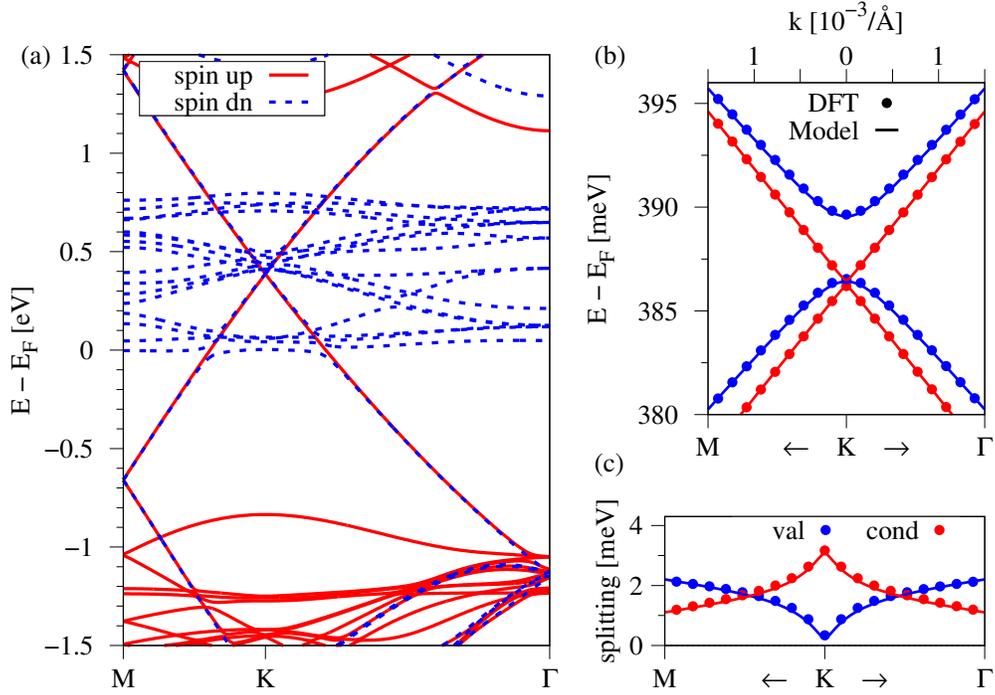

FIG. S17. (a) DFT-calculated band structure of the graphene/NiPS$_3$ heterostructure along the high-symmetry path M-K-Γ without SOC, when the NiPS$_3$ is in the ferromagnetic phase. Red solid (blue dashed) lines correspond to spin up (spin down). (b) Zoom to the DFT-calculated (symbols) low energy Dirac bands near the K point with a fit to the model Hamiltonian (solid line). (c) The splitting of valence and conduction band.

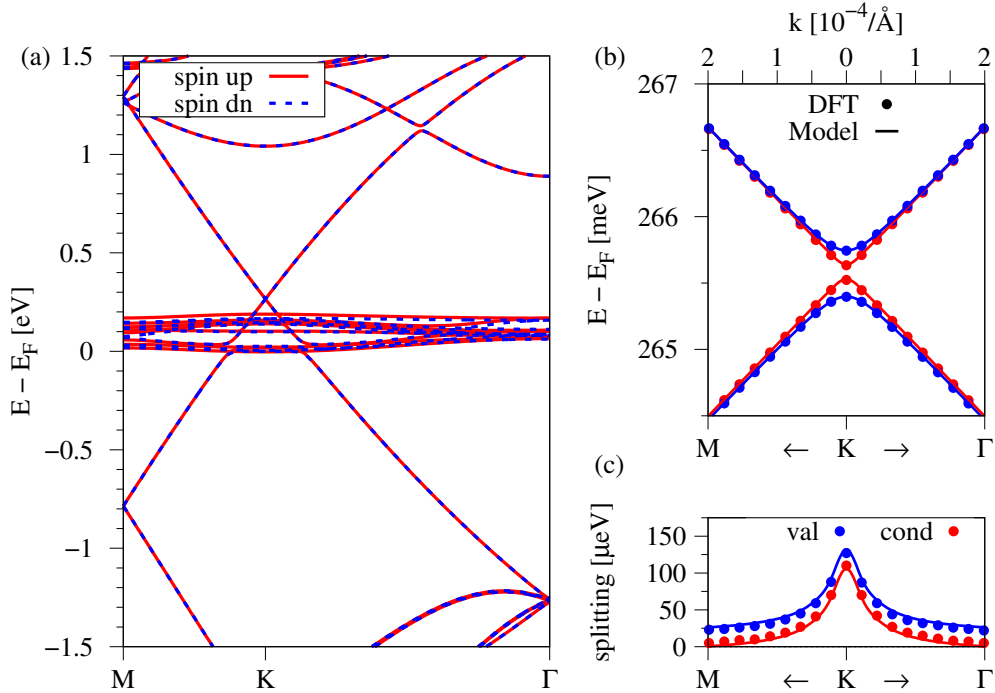

FIG. S18. (a) DFT-calculated band structure of the graphene/NiPS$_3$ heterostructure along the high-symmetry path M-K-Γ without SOC, when the NiPS$_3$ is in the antiferromagnetic Néel phase. Red solid (blue dashed) lines correspond to spin up (spin down). (b) Zoom to the DFT-calculated (symbols) low energy Dirac bands near the K point with a fit to the model Hamiltonian (solid line). (c) The splitting of valence and conduction band.

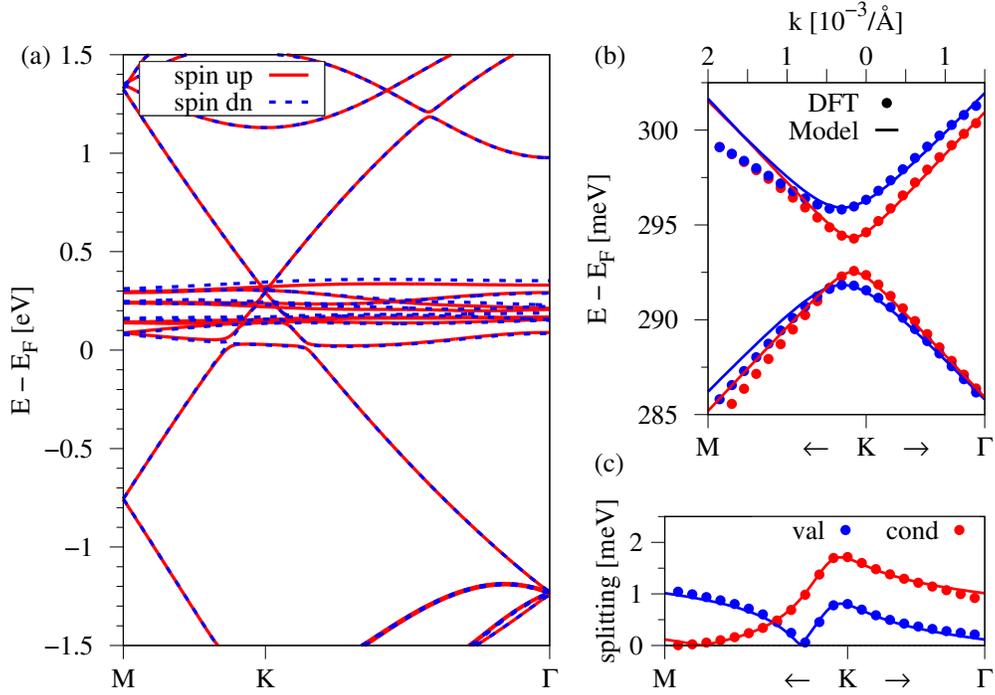

FIG. S19. (a) DFT-calculated band structure of the graphene/NiPS$_3$ heterostructure along the high-symmetry path M-K-Γ without SOC, when the NiPS$_3$ is in the antiferromagnetic zigzag phase. Red solid (blue dashed) lines correspond to spin up (spin down). (b) Zoom to the DFT-calculated (symbols) low energy Dirac bands near the K point with a fit to the model Hamiltonian (solid line). (c) The splitting of valence and conduction band.

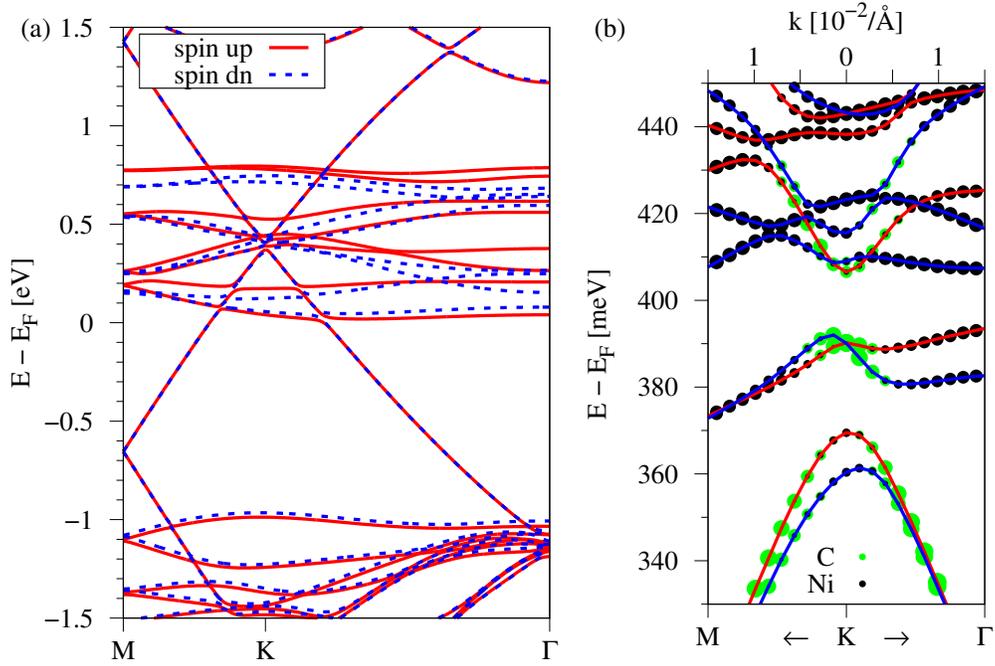

FIG. S20. (a) DFT-calculated band structure of the graphene/NiPS$_3$ heterostructure along the high-symmetry path M-K-Γ without SOC, when the NiPS$_3$ is in the antiferromagnetic stripy phase. Red solid (blue dashed) lines correspond to spin up (spin down). (b) Zoom to the DFT-calculated low energy Dirac bands near the K point (red and blue solid lines for spin up and down). The green and black spheres represent the projections onto C and Ni states. This emphasizes that the graphene Dirac states are strongly hybridized with Ni states from the NiPS$_3$ and we cannot apply our model fit properly.

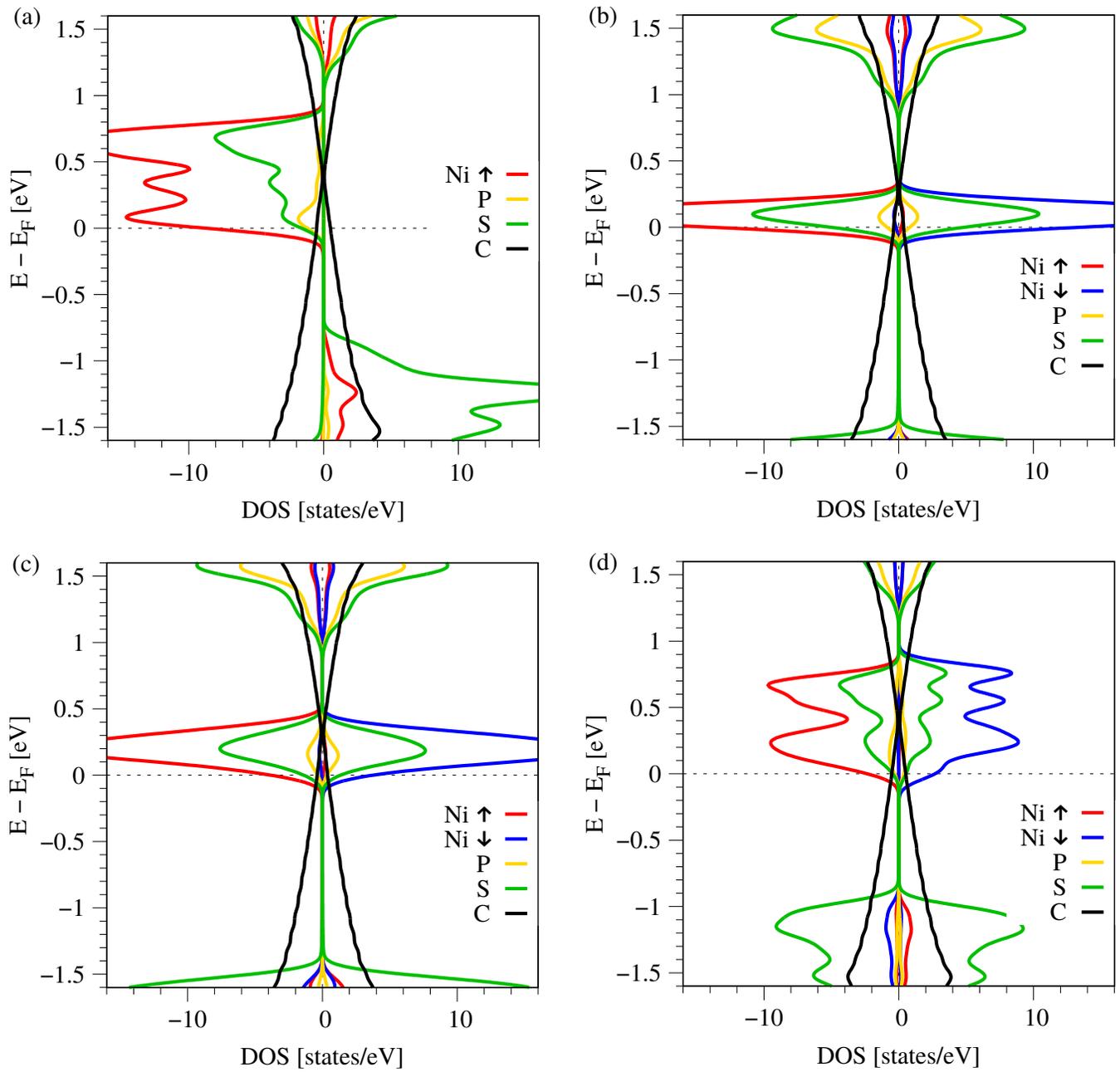

FIG. S21. Spin and atom resolved Density of States (DOS) for graphene on NiPS₃ when NiPS₃ is in the (a) ferromagnetic, (b) antiferromagnetic Néel, (c) antiferromagnetic zigzag, or (d) antiferromagnetic stripy phase.

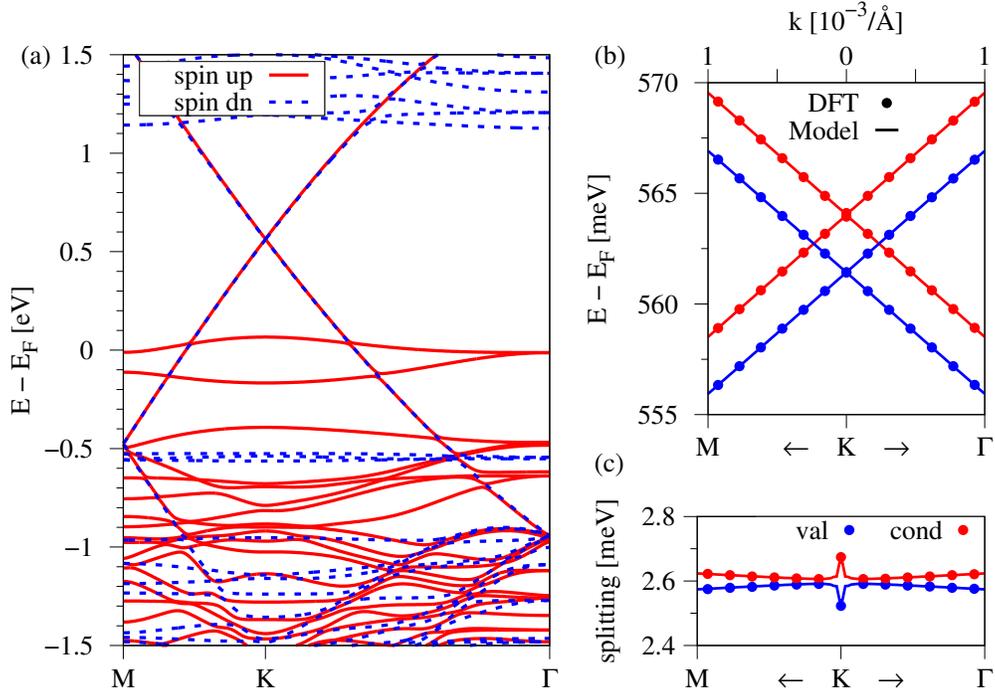

FIG. S22. (a) DFT-calculated band structure of the graphene/CoPS₃ heterostructure along the high-symmetry path M-K-Γ without SOC, when the CoPS₃ is in the ferromagnetic phase. Red solid (blue dashed) lines correspond to spin up (spin down). (b) Zoom to the DFT-calculated (symbols) low energy Dirac bands near the K point with a fit to the model Hamiltonian (solid line). (c) The splitting of valence and conduction band.

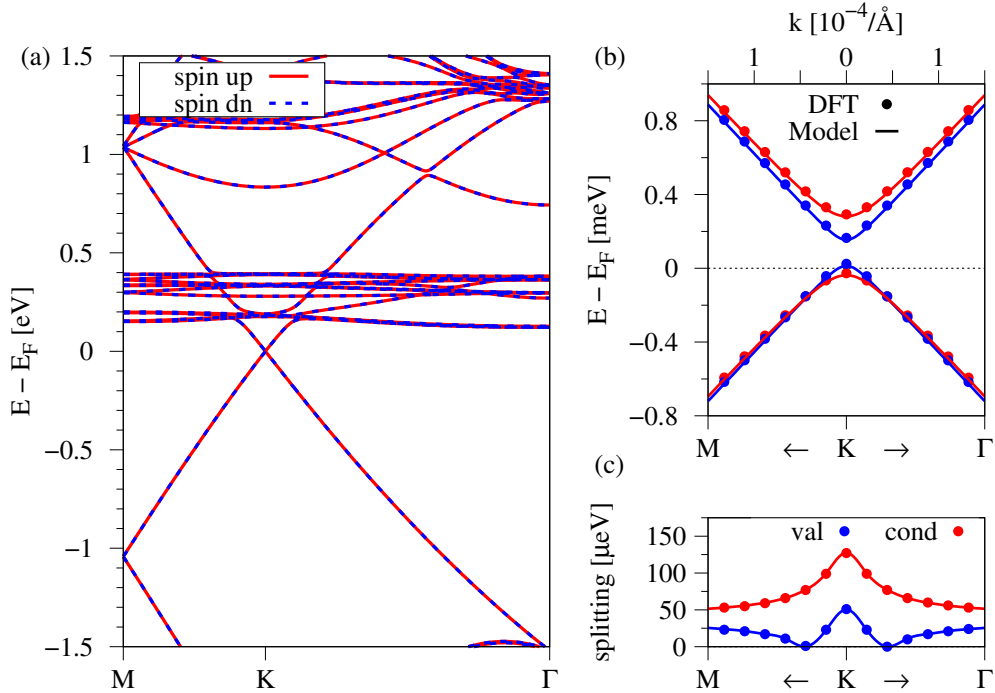

FIG. S23. (a) DFT-calculated band structure of the graphene/CoPS₃ heterostructure along the high-symmetry path M-K-Γ without SOC, when the CoPS₃ is in the antiferromagnetic Néel phase. Red solid (blue dashed) lines correspond to spin up (spin down). (b) Zoom to the DFT-calculated (symbols) low energy Dirac bands near the K point with a fit to the model Hamiltonian (solid line). (c) The splitting of valence and conduction band.

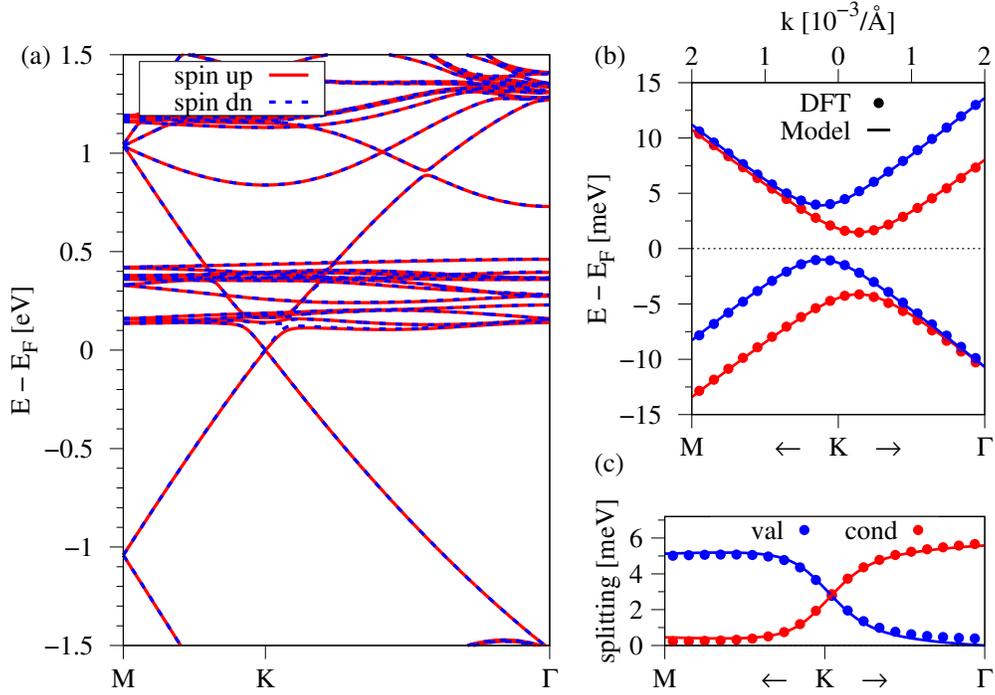

FIG. S24. (a) DFT-calculated band structure of the graphene/CoPS₃ heterostructure along the high-symmetry path M-K-Γ without SOC, when the CoPS₃ is in the antiferromagnetic zigzag phase. Red solid (blue dashed) lines correspond to spin up (spin down). (b) Zoom to the DFT-calculated (symbols) low energy Dirac bands near the K point with a fit to the model Hamiltonian (solid line). (c) The splitting of valence and conduction band.

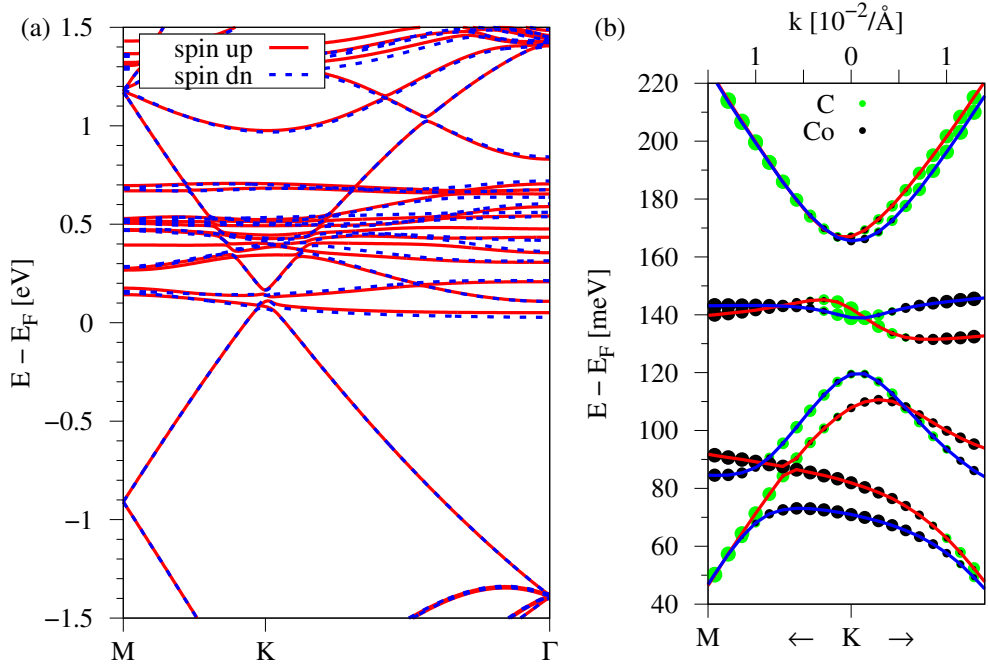

FIG. S25. (a) DFT-calculated band structure of the graphene/CoPS₃ heterostructure along the high-symmetry path M-K-Γ without SOC, when the CoPS₃ is in the antiferromagnetic stripy phase. Red solid (blue dashed) lines correspond to spin up (spin down). (b) Zoom to the DFT-calculated low energy Dirac bands near the K point (red and blue solid lines for spin up and down). The green and black spheres represent the projections onto C and Co states. This emphasizes that the graphene Dirac states are strongly hybridized with Co states from the CoPS₃ and we cannot apply our model fit properly.

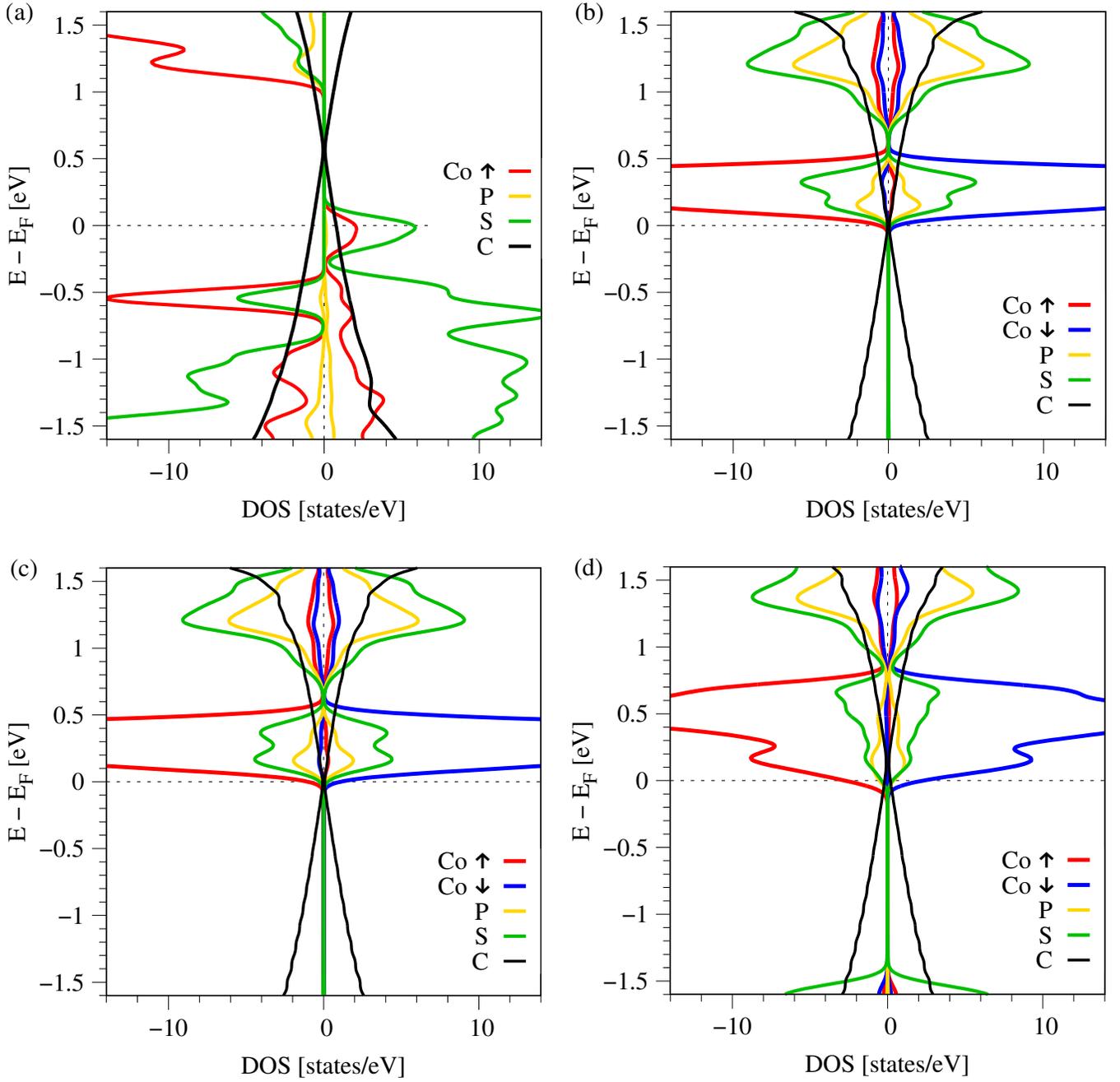

FIG. S26. Spin and atom resolved Density of States (DOS) for graphene on CoPS$_3$ when CoPS$_3$ is in the (a) ferromagnetic, (b) antiferromagnetic Néel, (c) antiferromagnetic zigzag, or (d) antiferromagnetic stripy phase.

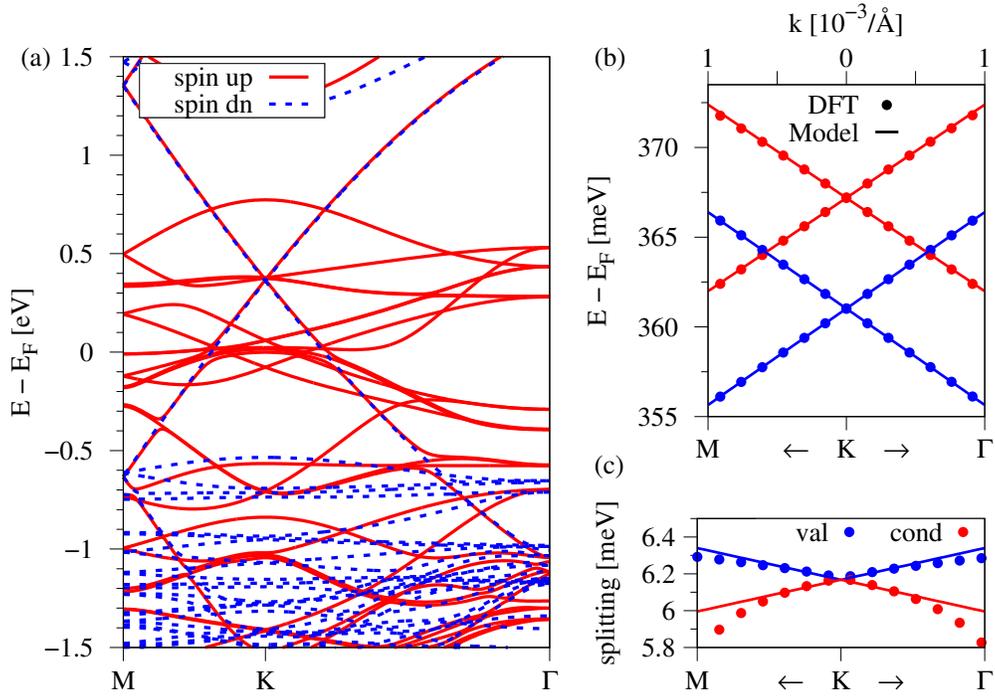

FIG. S27. (a) DFT-calculated band structure of the graphene/CoPSe₃ heterostructure along the high-symmetry path M-K-Γ without SOC, when the CoPSe₃ is in the ferromagnetic phase. Red solid (blue dashed) lines correspond to spin up (spin down). (b) Zoom to the DFT-calculated (symbols) low energy Dirac bands near the K point with a fit to the model Hamiltonian (solid line). (c) The splitting of valence and conduction band.

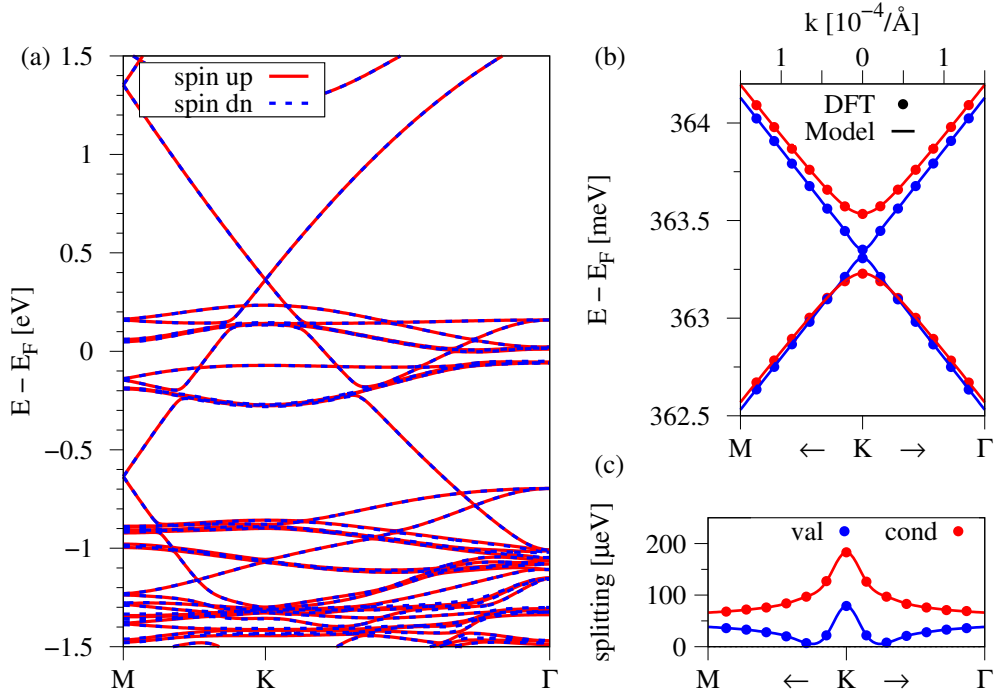

FIG. S28. (a) DFT-calculated band structure of the graphene/CoPSe₃ heterostructure along the high-symmetry path M-K-Γ without SOC, when the CoPSe₃ is in the antiferromagnetic Néel phase. Red solid (blue dashed) lines correspond to spin up (spin down). (b) Zoom to the DFT-calculated (symbols) low energy Dirac bands near the K point with a fit to the model Hamiltonian (solid line). (c) The splitting of valence and conduction band.

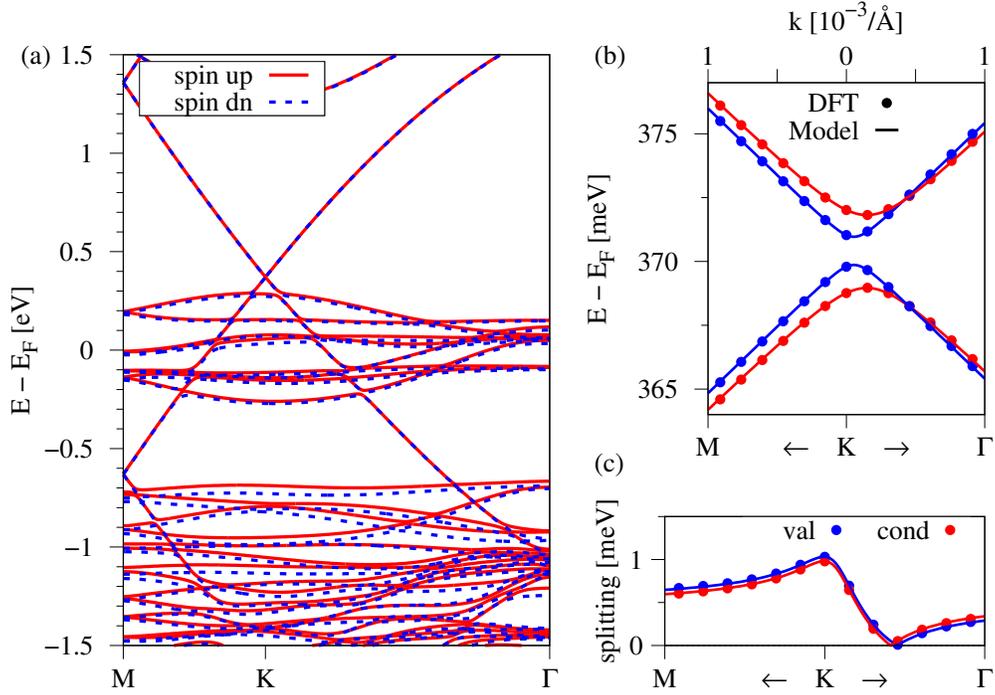

FIG. S29. (a) DFT-calculated band structure of the graphene/CoPSe₃ heterostructure along the high-symmetry path M-K-Γ without SOC, when the CoPSe₃ is in the antiferromagnetic zigzag phase. Red solid (blue dashed) lines correspond to spin up (spin down). (b) Zoom to the DFT-calculated (symbols) low energy Dirac bands near the K point with a fit to the model Hamiltonian (solid line). (c) The splitting of valence and conduction band.

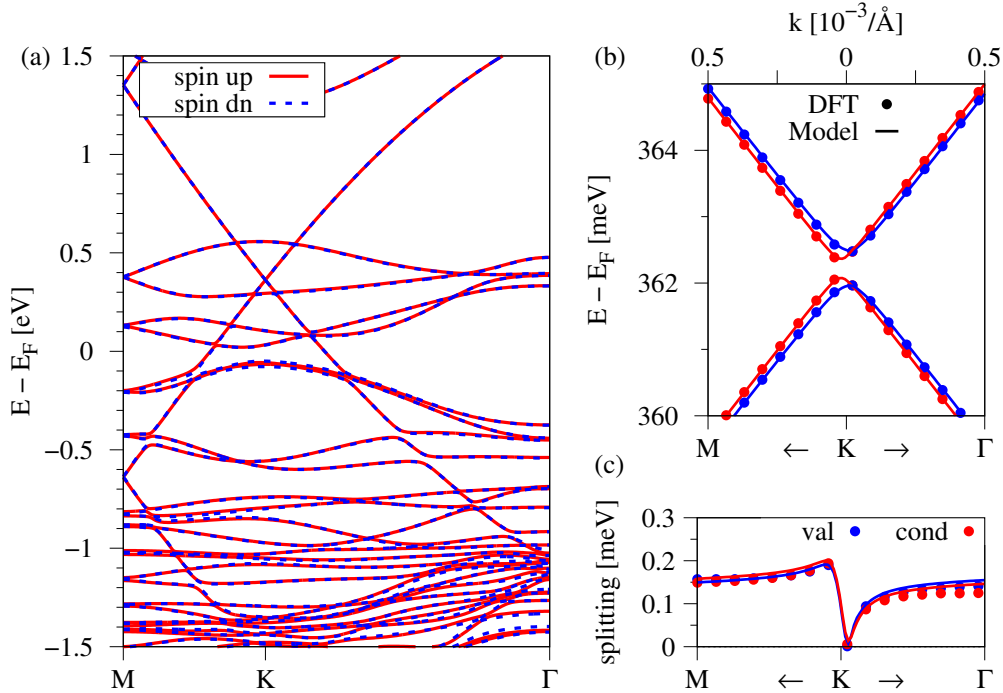

FIG. S30. (a) DFT-calculated band structure of the graphene/CoPSe₃ heterostructure along the high-symmetry path M-K-Γ without SOC, when the CoPSe₃ is in the antiferromagnetic stripy phase. Red solid (blue dashed) lines correspond to spin up (spin down). (b) Zoom to the DFT-calculated (symbols) low energy Dirac bands near the K point with a fit to the model Hamiltonian (solid line). (c) The splitting of valence and conduction band.

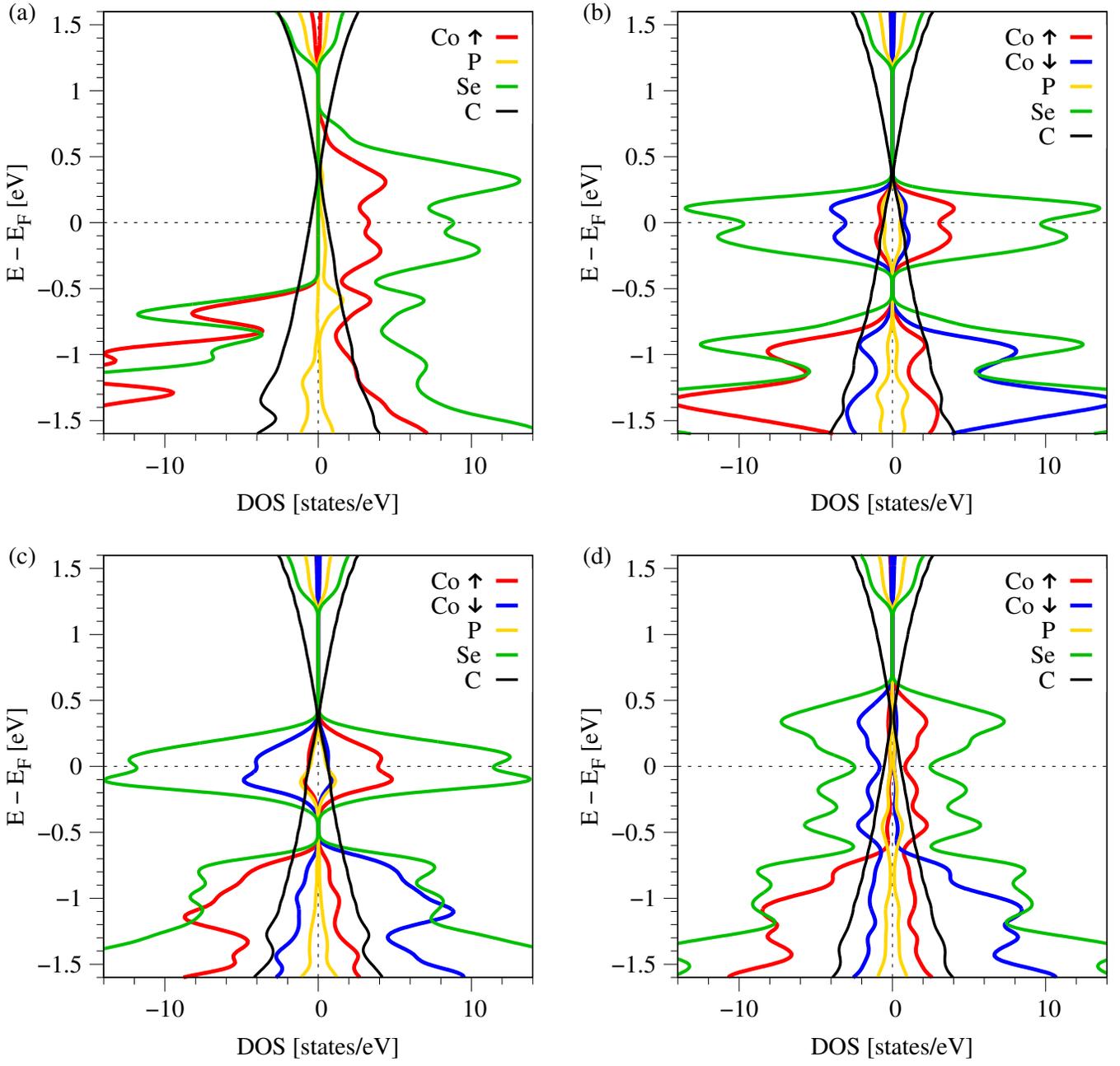

FIG. S31. Spin and atom resolved Density of States (DOS) for graphene on CoPSe₃ when CoPSe₃ is in the (a) ferromagnetic, (b) antiferromagnetic Néel, (c) antiferromagnetic zigzag, or (d) antiferromagnetic stripy phase.

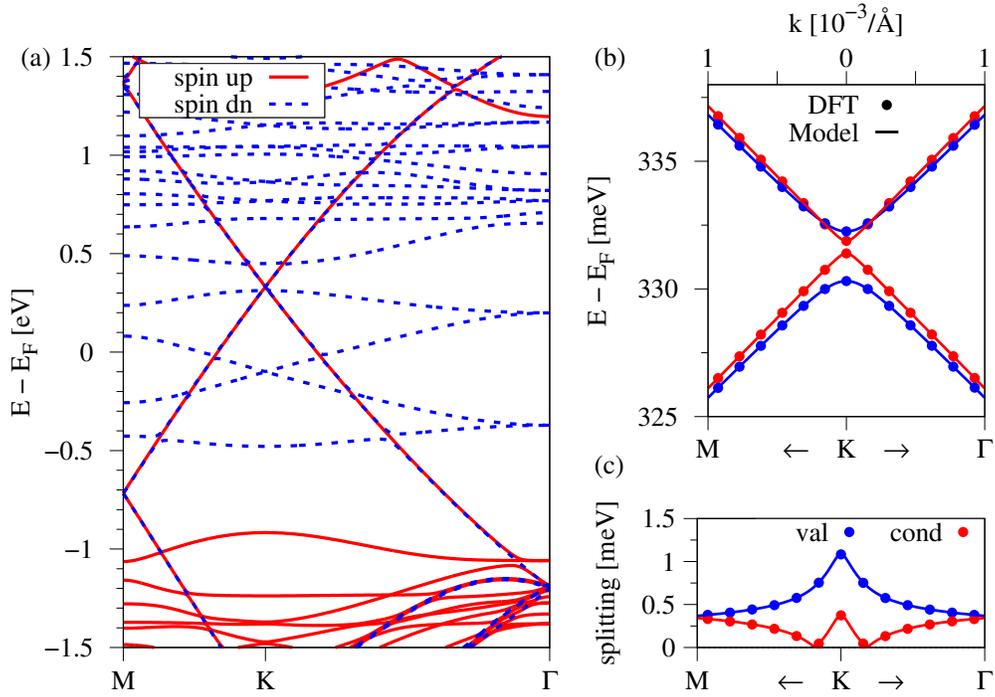

FIG. S32. (a) DFT-calculated band structure of the graphene/FePS$_3$ heterostructure along the high-symmetry path M-K-Γ without SOC, when the FePS$_3$ is in the ferromagnetic phase. Red solid (blue dashed) lines correspond to spin up (spin down). (b) Zoom to the DFT-calculated (symbols) low energy Dirac bands near the K point with a fit to the model Hamiltonian (solid line). (c) The splitting of valence and conduction band.

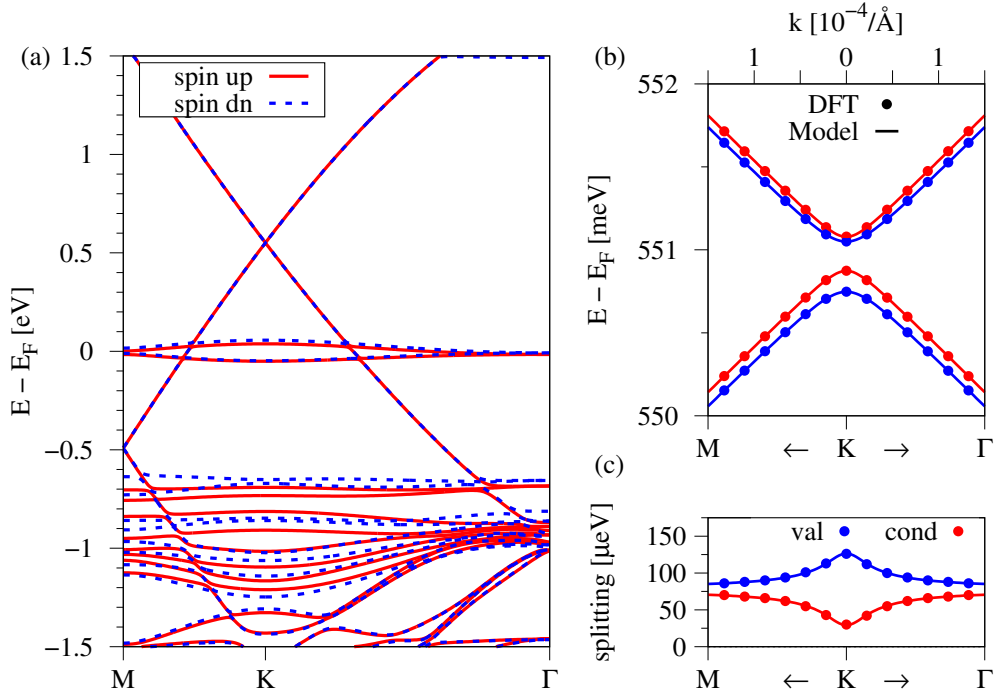

FIG. S33. (a) DFT-calculated band structure of the graphene/FePS$_3$ heterostructure along the high-symmetry path M-K-Γ without SOC, when the FePS$_3$ is in the antiferromagnetic Néel phase. Red solid (blue dashed) lines correspond to spin up (spin down). (b) Zoom to the DFT-calculated (symbols) low energy Dirac bands near the K point with a fit to the model Hamiltonian (solid line). (c) The splitting of valence and conduction band.

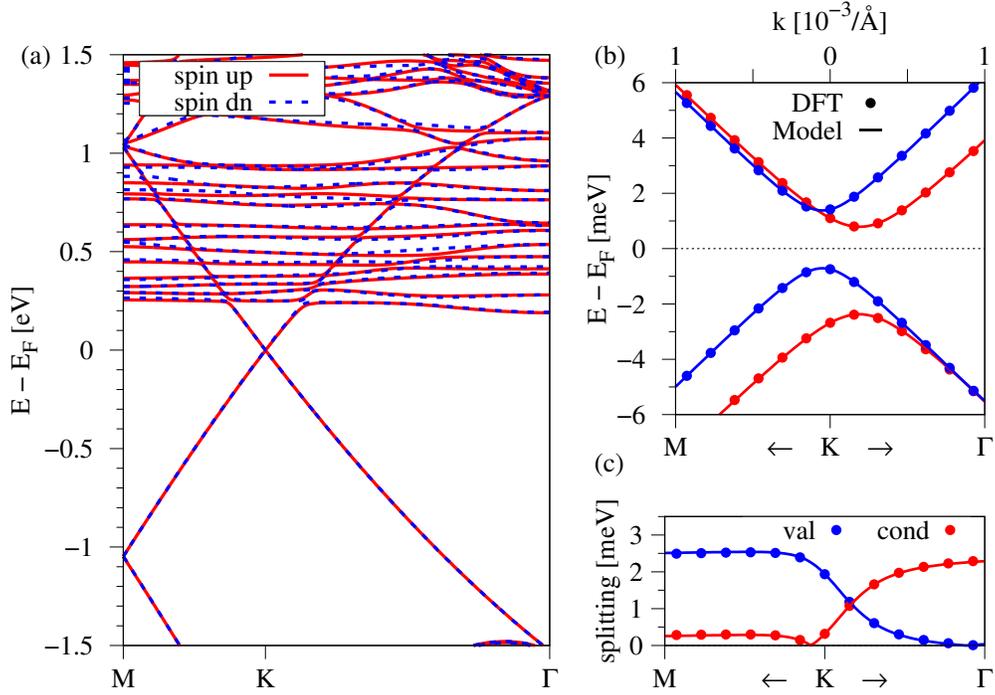

FIG. S34. (a) DFT-calculated band structure of the graphene/FePS$_3$ heterostructure along the high-symmetry path M-K-Γ without SOC, when the FePS$_3$ is in the antiferromagnetic zigzag phase. Red solid (blue dashed) lines correspond to spin up (spin down). (b) Zoom to the DFT-calculated (symbols) low energy Dirac bands near the K point with a fit to the model Hamiltonian (solid line). (c) The splitting of valence and conduction band.

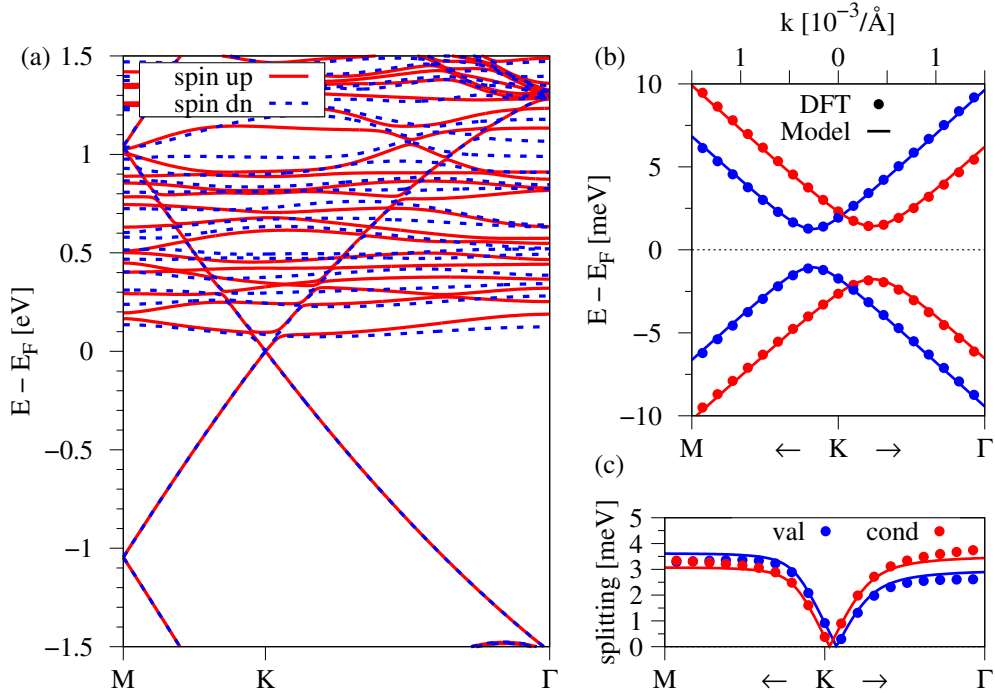

FIG. S35. (a) DFT-calculated band structure of the graphene/FePS$_3$ heterostructure along the high-symmetry path M-K-Γ without SOC, when the FePS$_3$ is in the antiferromagnetic stripy phase. Red solid (blue dashed) lines correspond to spin up (spin down). (b) Zoom to the DFT-calculated (symbols) low energy Dirac bands near the K point with a fit to the model Hamiltonian (solid line). (c) The splitting of valence and conduction band.

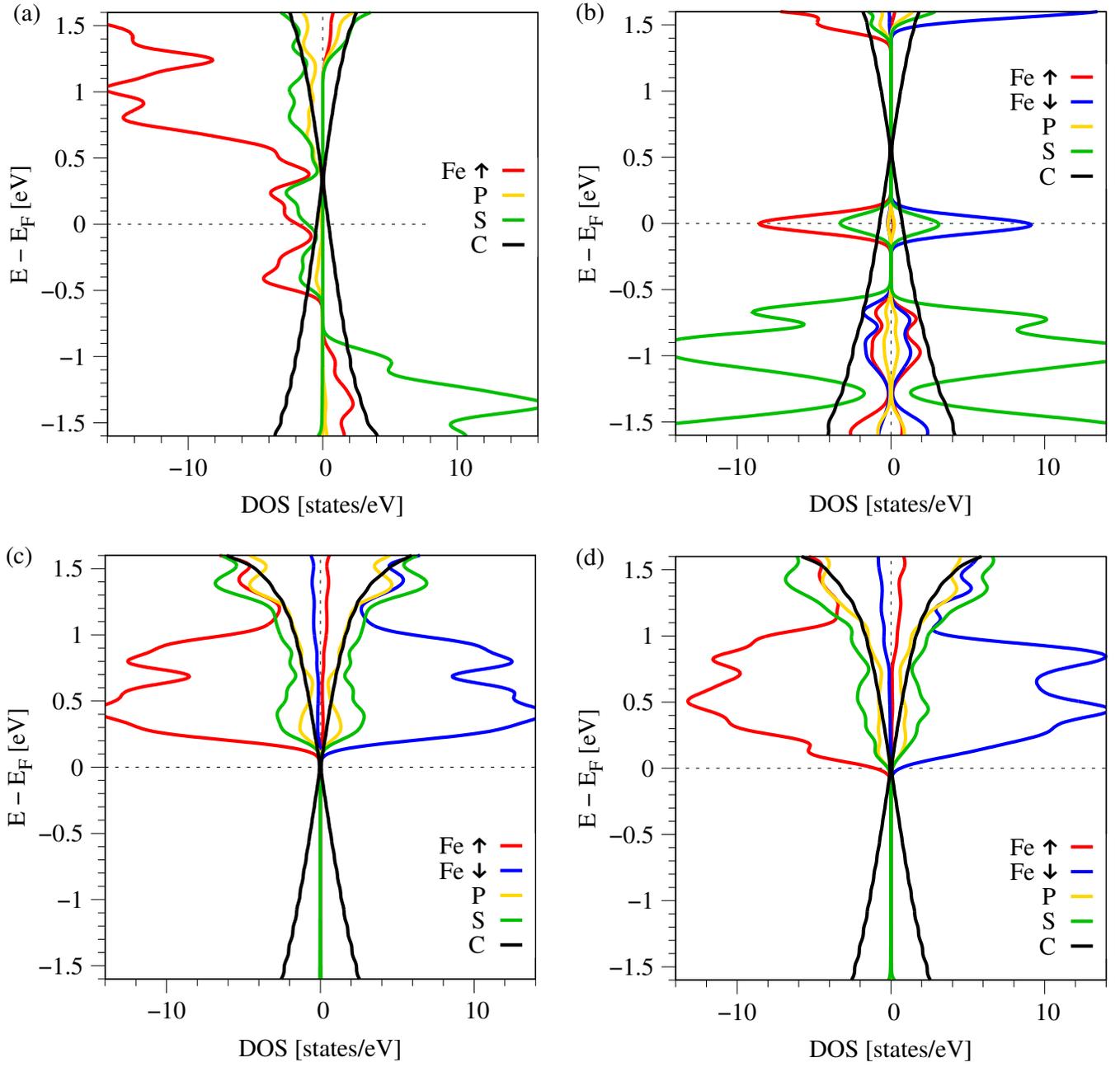

FIG. S36. Spin and atom resolved Density of States (DOS) for graphene on FePS$_3$ when FePS$_3$ is in the (a) ferromagnetic, (b) antiferromagnetic Néel, (c) antiferromagnetic zigzag, or (d) antiferromagnetic stripy phase.

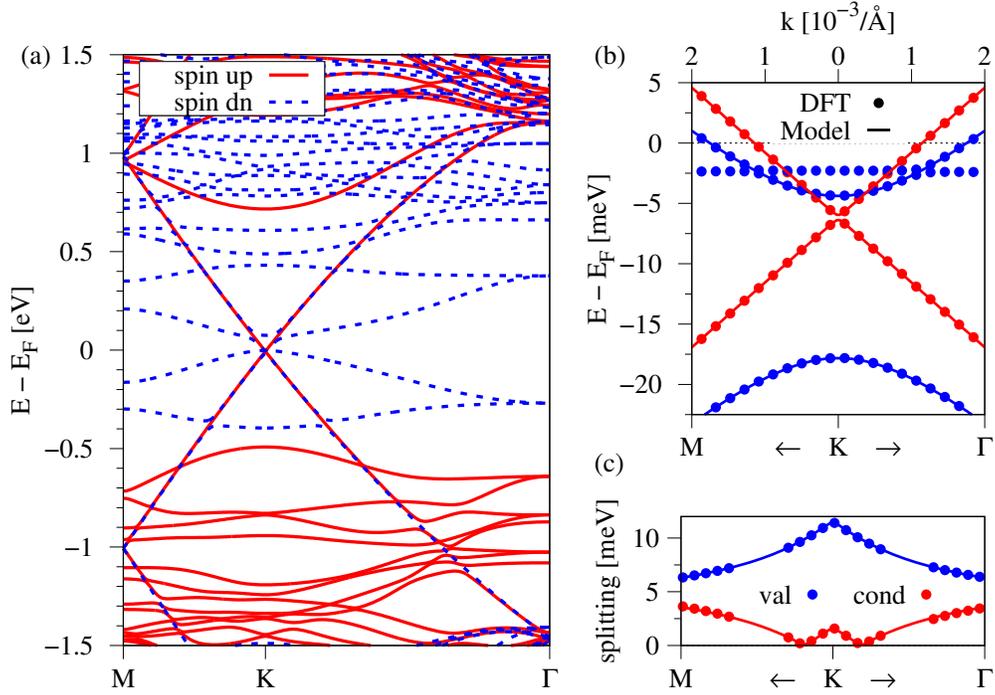

FIG. S37. (a) DFT-calculated band structure of the graphene/FePSe$_3$ heterostructure along the high-symmetry path M-K-Γ without SOC, when the FePSe$_3$ is in the ferromagnetic phase. Red solid (blue dashed) lines correspond to spin up (spin down). (b) Zoom to the DFT-calculated (symbols) low energy Dirac bands near the K point with a fit to the model Hamiltonian (solid line). (c) The splitting of valence and conduction band.

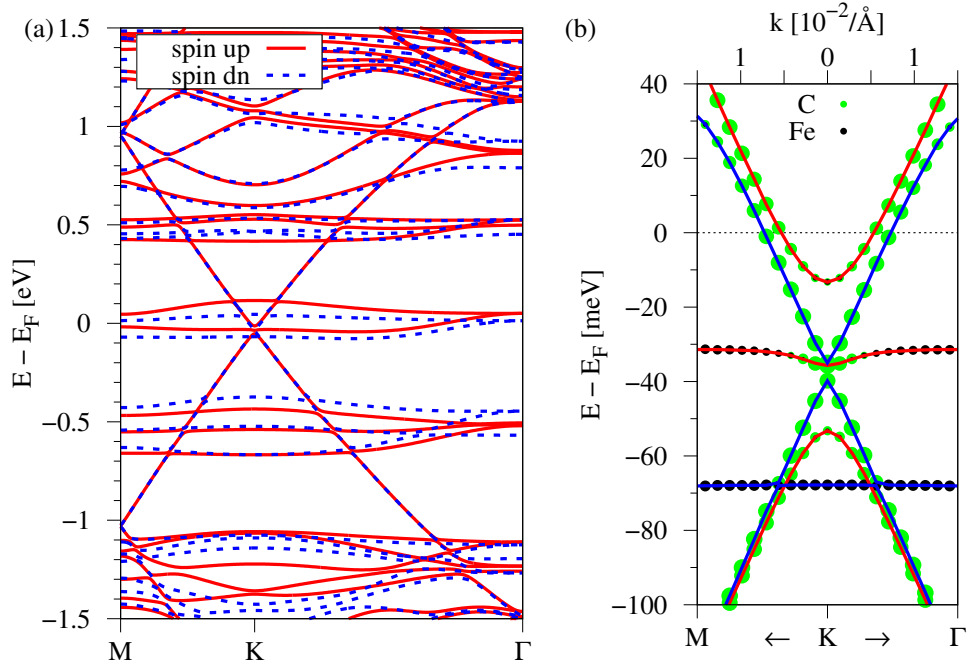

FIG. S38. (a) DFT-calculated band structure of the graphene/FePSe$_3$ heterostructure along the high-symmetry path M-K-Γ without SOC, when the FePSe$_3$ is in the antiferromagnetic Néel phase. Red solid (blue dashed) lines correspond to spin up (spin down). (b) Zoom to the DFT-calculated (symbols) low energy Dirac bands near the K point with a fit to the model Hamiltonian (solid line). (c) The splitting of valence and conduction band.

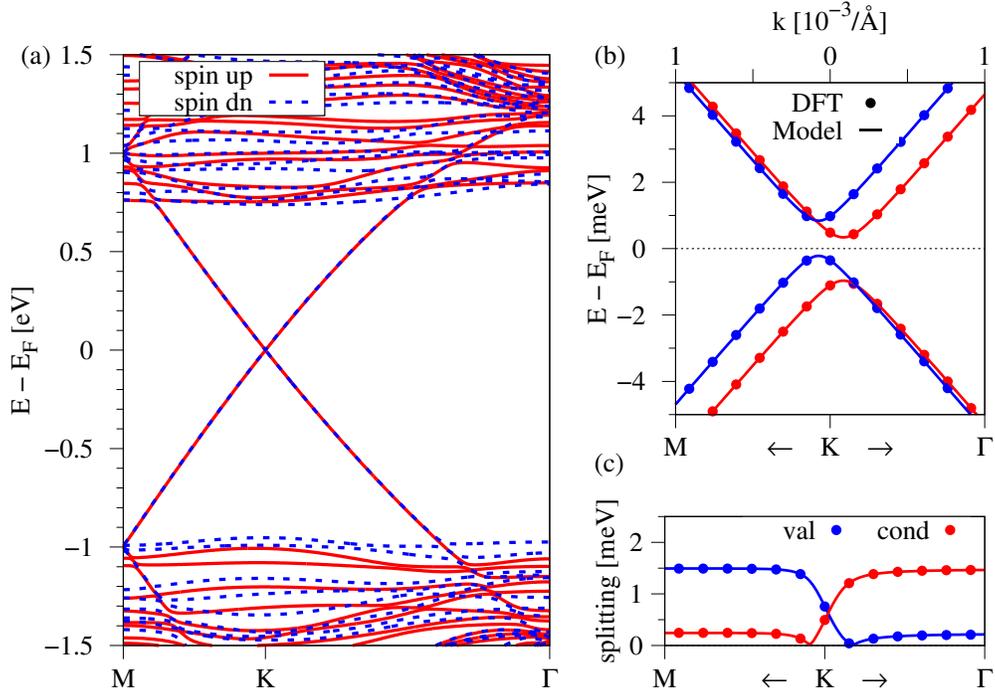

FIG. S39. (a) DFT-calculated band structure of the graphene/FePSe₃ heterostructure along the high-symmetry path M-K-Γ without SOC, when the FePSe₃ is in the antiferromagnetic zigzag phase. Red solid (blue dashed) lines correspond to spin up (spin down). (b) Zoom to the DFT-calculated (symbols) low energy Dirac bands near the K point with a fit to the model Hamiltonian (solid line). (c) The splitting of valence and conduction band.

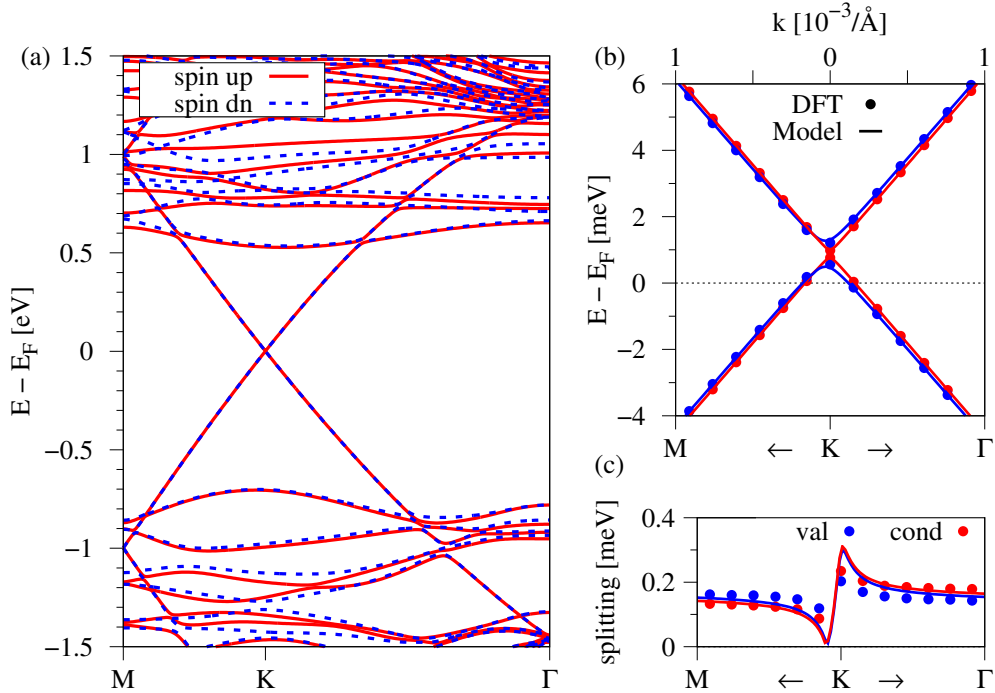

FIG. S40. (a) DFT-calculated band structure of the graphene/FePSe₃ heterostructure along the high-symmetry path M-K-Γ without SOC, when the FePSe₃ is in the antiferromagnetic stripy phase. Red solid (blue dashed) lines correspond to spin up (spin down). (b) Zoom to the DFT-calculated (symbols) low energy Dirac bands near the K point with a fit to the model Hamiltonian (solid line). (c) The splitting of valence and conduction band.

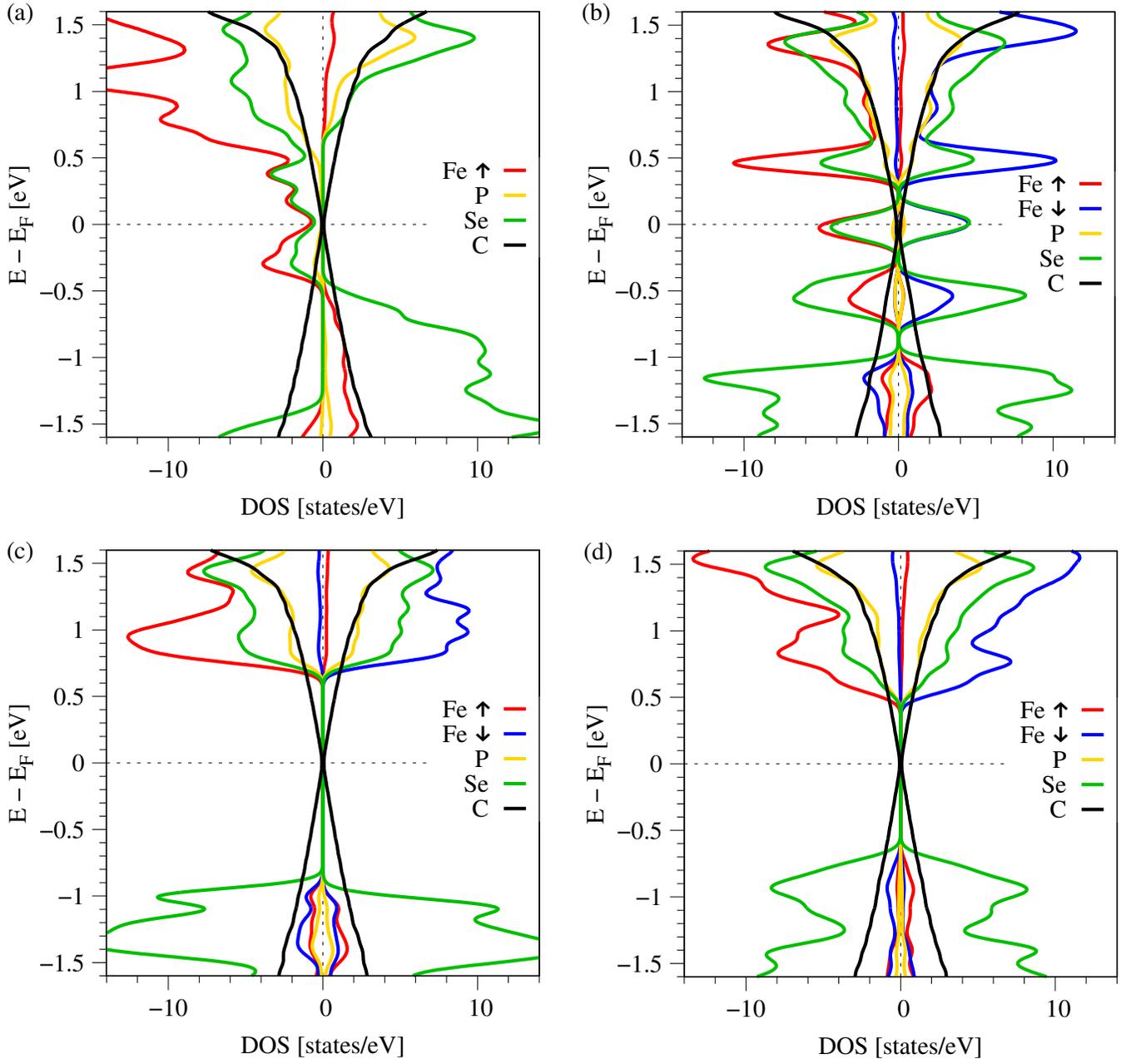

FIG. S41. Spin and atom resolved Density of States (DOS) for graphene on FePSe$_3$ when FePSe$_3$ is in the (a) ferromagnetic, (b) antiferromagnetic Néel, (c) antiferromagnetic zigzag phase, or (d) antiferromagnetic stripy phase.

\clearpage

\end{document}